\documentclass[a4paper,12pt]{article}
\usepackage[a4paper,text={16.8cm,22.4cm}]{geometry}
\usepackage{amsmath,amssymb,bm,braket,psfrag,colortbl,booktabs,latexsym}
\usepackage{graphicx}
\usepackage{cite}

\allowdisplaybreaks 
\newcommand{\Tr}{\mathrm{Tr}}
\newcommand{\Li}{\mathrm{Li}}
\newcommand{\ff}{f\hspace{-0.4em}f}

\newcommand{\nslash}{\rlap{\hspace{0.05em}/}{n}}
\newcommand{\nbslash}{\rlap{\hspace{0.05em}/}{\bar{n}}}
\newcommand{\vslash}{\rlap{\hspace{0.05em}/}{v}}

\newcommand{\dbraket}[1]{\langle\!\langle #1 \rangle\!\rangle}

\newcommand{\gr}{\rowcolor[gray]{0.9}[.88\tabcolsep]}

\def\msbar{{$\overline{\mbox{MS}}$}}

\begin{document}

\begin{titlepage}

  \begin{flushright}
    MZ-TH/10-10 \\
    HD-THEP-10-7 \\
    August 31, 2010 \\
  \end{flushright}
  
  \vspace{5ex}
  
  \begin{center}
    \textbf{\Large Renormalization-Group Improved Predictions for Top-Quark Pair
      Production at Hadron Colliders} \vspace{7ex}
    
    \textsc{Valentin Ahrens$^a$, Andrea Ferroglia$^a$, Matthias Neubert$^{a,b}$,\\
      Ben D. Pecjak$^a$, and Li Lin Yang$^a$} \vspace{2ex}
  
    \textsl{${}^a$Institut f\"ur Physik (THEP), Johannes Gutenberg-Universit\"at\\
      D-55099 Mainz, Germany\\[0.3cm]
      ${}^b$Institut f\"ur Theoretische Physik, Ruprecht-Karls-Universit\"at Heidelberg\\
      Philosophenweg 16, D-69120 Heidelberg, Germany}
  \end{center}

  \vspace{4ex}

  \begin{abstract}
    Precision predictions for phenomenologically interesting observables such as the
    $t\bar t$ invariant mass distribution and forward-backward asymmetry in top-quark pair
    production at hadron colliders require control over the differential cross section in
    perturbative QCD. In this paper we improve existing calculations of the doubly
    differential cross section in the invariant mass and scattering angle by using
    techniques from soft-collinear effective theory to perform an NNLL resummation of
    threshold logarithms, which become large when the invariant mass $M$ of the top-quark
    pair approaches the partonic center-of-mass energy $\sqrt{\hat{s}}$. We also derive an
    approximate formula for the differential cross section at NNLO in fixed-order
    perturbation theory, which completely determines the coefficients multiplying the
    singular plus distributions in the variable $(1-M^2/\hat{s})$. We then match our
    results in the threshold region with the exact results at NLO in fixed-order
    perturbation theory, and perform a numerical analysis of the invariant mass
    distribution, the total cross section, and the forward-backward asymmetry. We argue
    that these are the most accurate predictions available for these observables at
    present. Using MSTW2008NNLO parton distribution functions (PDFs) along with
    $\alpha_s(M_Z)=0.117$ and $m_t=173.1$\,GeV, we obtain for the inclusive production
    cross sections at the Tevatron and LHC the values $\sigma_{\rm Tevatron}=(6.30\pm
    0.19{}_{-0.23}^{+0.31})$\,pb and $\sigma_{\rm LHC}=(149\pm 7\pm 8)$\,pb, where the
    first error results from scale variations while the second reflects PDF uncertainties.
  \end{abstract}
  
\end{titlepage}

\section{Introduction}

The top quark is the heaviest known particle in the Standard Model (SM) of fundamental
interactions. Because of its large mass, it is expected to couple strongly with the fields
responsible for electroweak symmetry breaking, and the detailed study of top-quark
properties is likely to play a key role in elucidating the origin of particle masses. The
measurement of top-quark related observables is therefore one of the main goals of the
Fermilab Tevatron and CERN Large Hadron Collider (LHC). To date, thousands of top-quark
events have been observed by two different experiments at the Tevatron, and the top-quark
mass has been extracted at the percent level \cite{:2009ec}. The experiments at the LHC
are expected to observe millions of top-quark events per year already in the initial
low-luminosity phase, bringing the study of the top-quark properties into the realm of
precision physics. In particular, the total inclusive cross section for top-quark pair
production is expected to be measured with a relative error of $5\%$ to $10\%$ at the LHC
\cite{Bernreuther:2008ju}. In addition to the mass and the total inclusive cross section,
kinematic distributions and charge asymmetries are also of interest. For instance, the
$t\bar{t}$ invariant mass distribution can be used as a complementary method for measuring
$m_t$ \cite{Frederix:2007gi}. The presence of bumps in the smoothly decreasing $t\bar{t}$
invariant mass distribution would be a clear signal of an $s$-channel heavy resonance,
which is predicted in many new physics scenarios \cite{Barger:2006hm, Frederix:2007gi,
  Baur:2008uv}. Such searches have been pursued at the Tevatron \cite{:2007dz, :2007dia,
  Abazov:2008ny}, and results for the top-pair invariant mass distribution were recently
obtained from data collected by the CDF collaboration \cite{Aaltonen:2009iz}. The
forward-backward asymmetry in $t\bar{t}$ production has also been measured recently
\cite{:2007qb, Aaltonen:2008hc, cdf9724} and shows a potential deviation from the
theoretical predictions.

To make optimal use of the experimental measurements requires theoretical predictions of
similar precision. At hadron colliders, a leading-order (LO) prediction is usually
insufficient, and including higher-order corrections in QCD is mandatory. Current
theoretical predictions for $t\bar{t}$ production in fixed-order perturbation theory are
based on next-to-leading order (NLO) calculations of the total cross section
\cite{Nason:1987xz, Beenakker:1988bq, Beenakker:1990maa, Czakon:2008ii}, differential
distributions \cite{Nason:1989zy, Mangano:1991jk, Frixione:1995fj}, and the
forward-backward asymmetry \cite{Kuhn:1998jr, Kuhn:1998kw}. The NLO computations suffer
from theoretical uncertainties larger than $10\%$, both for Tevatron and LHC
center-of-mass energies. These uncertainties are partly due to our imperfect knowledge of
the parton distribution functions (PDFs), and partly to the truncation of the perturbative
series in the strong coupling constant, which introduces a dependence on the unphysical
renormalization and factorization scales into the physical predictions. The latter
uncertainty is typically reduced by including more terms in the perturbative series, and
for this reason the calculation of the partonic cross section to next-to-next-to-leading
order (NNLO) has been an area of active research. This includes studies of the two-loop
virtual corrections \cite{Czakon:2007ej, Czakon:2007wk, Czakon:2008zk, Bonciani:2008az,
  Bonciani:2009nb}, the squared one-loop corrections \cite{Korner:2008bn,
  Anastasiou:2008vd, Kniehl:2008fd}, and the NLO corrections to the final state $t\bar{t}$
+ jet \cite{Dittmaier:2007wz}. However, due to the complexity of the calculations,
complete results for the cross section at NNLO are not yet available.

In the absence of full NNLO results, one way to improve the NLO predictions is to utilize
threshold resummation methods \cite{Sterman:1986aj, Catani:1989ne} to incorporate some
(presumably) dominant contributions from higher orders. For the total cross section, a
common approach is to work in the limit where the top-quark pair is produced nearly at
rest, and to resum logarithms in the velocity $\beta=\sqrt{1-4m_t^2/\hat s}\to 0$, with
$\hat s$ the partonic center-of-mass energy. Such resummations have been carried out at
leading-logarithmic (LL) \cite{Laenen:1991af, Laenen:1993xr, Berger:1995xz, Berger:1996ad,
  Berger:1997gz, Catani:1996dj}, next-to-leading-logarithmic (NLL) \cite{Bonciani:1998vc},
and approximate next-to-next-to-leading-logarithmic (NNLL) order \cite{Moch:2008qy,
  Czakon:2008cx, Langenfeld:2009wd}. Only very recently have the complete expressions at
NNLL order been obtained \cite{Beneke:2009rj, Czakon:2009zw, Beneke:2009ye}. The
calculation of differential cross sections is more involved due to the appearance of
several kinematic variables. A typical approach in that case is to work in the threshold
limit where the parameter $(1-z)=1-M^2/\hat s\to 0$, with $M$ the invariant mass of the
$t\bar t$ pair, and to resum logarithms in $(1-z)$. In that case real gluon emission is
soft, but the parameter $\beta$ is a generic ${\cal O}(1)$ variable. The current frontier
for resummation in this limit is NLL calculations for differential cross sections
\cite{Kidonakis:1996aq, Kidonakis:1997gm, Banfi:2004xa} and the forward-backward asymmetry
\cite{Almeida:2008ug}. In addition, approximate NNLL calculations were performed in
\cite{Kidonakis:2000ui, Kidonakis:2001nj, Kidonakis:2003qe, Kidonakis:2008mu}. Extending
these results to full NNLL order has been made possible by our recent calculation of the
two-loop anomalous-dimension matrices \cite{Ferroglia:2009ep, Ferroglia:2009ii}. We have
presented an approximate NNLO formula for the $t\bar{t}$ invariant mass distribution in
\cite{Ahrens:2009uz}. The goal of the present paper is to derive a renormalization-group
(RG) improved expression for the doubly differential cross section at NNLL order, in which
all threshold-enhanced terms are resummed. We will match this expression with the exact
fixed-order NLO results and study the top-pair invariant mass distribution, the
forward-backward asymmetry, and the total cross section at NLO+NNLL order. An important
part of our analysis is a comparison of the total cross section obtained here with those
obtained in the limit $\beta \to 0$. The outcome of this comparison leads us to conclude
that the predictions obtained in this paper are the most precise available at present.

The paper is organized as follows. In Section~\ref{sec:kin} we review the kinematics and
the structure of factorization in the threshold region. We then derive the factorization
formula for the hard-scattering kernels into products of hard and soft matrices using
soft-collinear effective theory (SCET) in Section~\ref{sec:SCETfact}. In
Section~\ref{sec:Matching} we present the calculation of the hard and soft matrices at
NLO, and describe several checks on our results. Section~\ref{sec:Resummation} deals with
the RG properties of the hard and soft functions. We derive a formula for the resummed
cross section in momentum space using RG methods and describe its evaluation at NNLL
order. We also review the derivation of the approximate NNLO formula, which has been
presented first in \cite{Ahrens:2009uz}. In Section~\ref{sec:Pheno} we perform numerical
studies of the invariant mass distribution, the total cross section, and the
forward-backward asymmetry, utilizing both RG-improved perturbation theory at NNLL order
and the NNLO approximate formula. This section includes the aforementioned comparison of
different methods for obtaining the total cross section. We conclude in
Section~\ref{sec:Conclusions}.

\section{Kinematics and factorization at threshold}
\label{sec:kin}

We consider the process
\begin{align}
  N_1(P_1) + N_2(P_2) \to t(p_3) + \bar{t}(p_4) + X(p_X) \, ,
\end{align}
where $X$ is an inclusive hadronic final state \footnote{Throughout the analysis we treat
  the top-quarks as on-shell partons and neglect their decay. Corrections to this picture
  are suppressed by $\Gamma_t/m_t\ll 1$.}. At Born level this proceeds through the $q\bar
q$ annihilation and gluon-fusion channels
\begin{align}
  q(p_1) + \bar{q}(p_2) &\to t(p_3) + \bar{t}(p_4) \, , \nonumber
  \\
  g(p_1) + g(p_2) &\to t(p_3) + \bar{t}(p_4) \, ,
\end{align}
where $p_1=x_1P_1$ and $p_2=x_2P_2$. We define the kinematic invariants
\begin{align}\label{mandelstam}
  s &= (P_1+P_2)^2 \, , \quad \hat{s}=(p_1+p_2)^2 \, , \quad M^2=(p_3+p_4)^2 \, ,
  \nonumber
  \\
  t_1 &= (p_1-p_3)^2-m_t^2 \, , \quad u_1=(p_2-p_3)^2 -m_t^2 \, ,
\end{align}
and momentum conservation at Born level implies $\hat{s}+t_1+u_1=0$.

In this section we consider the structure of the differential cross section near the
partonic threshold. While the fully differential cross section depends on three kinematic
variables, in this paper we are mainly interested in the doubly differential cross section
expressed in terms of the invariant mass $M$ of the $t\bar{t}$ pair and the scattering
angle $\theta$ between $\vec{p}_1$ and $\vec{p}_3$ in the partonic center-of-mass frame.
To describe this distribution we introduce the variables
\begin{align}\label{betatdef}
  z = \frac{M^2}{\hat{s}} \, , \qquad \tau = \frac{M^2}{s} \, , \qquad \beta_t =
  \sqrt{1-\frac{4m_t^2}{M^2}} \, .
\end{align}
The quantity $\beta_t$ gives the 3-velocity of the top quarks in the $t\bar{t}$ rest
frame. One often considers a related variable $\beta = \sqrt{1-4m_t^2/\hat{s}}$, which
coincides with $\beta_t$ in the limit $z \to 1$.

Before moving on, we would like to clarify our definition of the so-called ``threshold
region''. In general, the definition of this region depends on the measured observable.
For instance, to describe the total inclusive cross section, one often considers the limit
$\beta \to 0$ \cite{Bonciani:1998vc, Moch:2008qy, Czakon:2008cx, Langenfeld:2009wd,
  Beneke:2009rj, Czakon:2009zw, Beneke:2009ye}. To describe the $t\bar{t}$ invariant mass
distribution, on the other hand, we consider the limit $z \to 1$, with $\beta_t$ a generic
$\mathcal{O}(1)$ variable. In that case the top quarks are not produced at rest, nor are
they highly boosted, and the emitted partons in the final state $X$ are constrained to be
soft. This threshold limit is often referred to in the literature as the
pair-invariant-mass kinematics, and the theoretical framework to deal with this situation
was developed in \cite{Kidonakis:1996aq, Kidonakis:1997gm, Kidonakis:1998bk,
  Kidonakis:1998nf}. A closely related framework can be applied to processes where only a
single top-quark is observed \cite{Laenen:1998qw, Catani:1998tm}, but will not be
considered here. Threshold resummation at fixed invariant mass $M$ is completely analogous
to the well-studied case of Drell-Yan or Higgs production, and it becomes more and more
important as $M$ is increased. Therefore, the RG-improved result for $d\sigma/dM$ offers
the best possible prediction for this observable. Moreover, integrating this distribution
over $M$ then yields the best possible prediction for the total cross section. In
particular, we believe that in this way we obtain a better description of the total cross
section than the usual approach of resumming soft gluon and Coulomb terms for $\beta \to
0$. The reason is that at the Tevatron and LHC, the cross section is dominated by events
with $\beta \approx 0.3$--0.6, which is not especially small. We will come back to this
important point in Section~\ref{sec:totcross}.

According to the QCD factorization theorem \cite{Collins:1989gx}, the differential cross
section in $M$ and $\cos\theta$ can be written as
\begin{align}
  \label{eq:genfact}
  \frac{d^2\sigma}{dMd\cos\theta} = \frac{8\pi\beta_t}{3sM} \sum_{i,j} \int_\tau^1
  \frac{dz}{z} \, \ff_{ij}(\tau/z,\mu_f) \, C_{ij}(z,M,m_t,\cos\theta,\mu_f) \, ,
\end{align}
where $\mu_f$ is the factorization scale, and the parton luminosity functions $\ff_{ij}$
are defined by
\begin{align}
  \ff_{ij}(y,\mu_f) = \int_y^1 \frac{dx}{x} \, f_{i/N_1}(x,\mu_f) \, f_{j/N_2}(y/x,\mu_f)
  \, .
\end{align}
The $f_{i/N}$ are universal non-perturbative PDFs of the parton $i$ in the hadron $N$,
which can be extracted from experimental data. The hard-scattering kernels $C_{ij}$ are
related to the partonic cross sections and can be calculated as a power series in
$\alpha_s$. We shall write their expansion as
\begin{align}
  \label{eq:cexp}
  C_{ij} = \alpha_s^2 \left[ C_{ij}^{(0)} + \frac{\alpha_s}{4\pi} C_{ij}^{(1)} + \left(
      \frac{\alpha_s}{4\pi} \right)^2 C_{ij}^{(2)} + \ldots \right] .
\end{align}
At leading order in $\alpha_s$, only $C_{q\bar{q}}$ and $C_{gg}$ are non-zero. They are
proportional to $\delta(1-z)$ and read
\begin{align}
  C_{q\bar{q}}^{(0)} &= \delta(1-z) \, \frac{3}{8N} \, C_F \left( \frac{t_1^2+u_1^2}{M^4}
    + \frac{2m_t^2}{M^2} \right) , \nonumber
  \\
  C_{gg}^{(0)} &= \delta(1-z) \, \frac{3}{8(N^2-1)} \left( C_F \frac{M^4}{t_1u_1} - C_A
  \right) \left[ \frac{t_1^2+u_1^2}{M^4} + \frac{4m_t^2}{M^2} - \frac{4m_t^4}{t_1u_1}
  \right] ,
\end{align}
where $N=3$ is the number of colors in QCD, and $t_1$ and $u_1$ can be expressed in terms
of $M$ and $\cos\theta$ as
\begin{align}
  \label{eq:tu}
  t_1 = -\frac{M^2}{2} ( 1 - \beta_t \cos\theta ) \, , \qquad u_1 = -\frac{M^2}{2} ( 1 +
  \beta_t \cos\theta ) \, .
\end{align}
Note that for the doubly differential cross section the coefficient $C_{\bar{q}q}$ is also
needed. It can be obtained from the expression for $C_{q\bar{q}}$ by replacing $\cos\theta
\to -\cos\theta$, which is a symmetry at tree level but not beyond.

At higher orders in $\alpha_s$ the hard-scattering kernels receive corrections from
virtual loop diagrams and real gluon emissions in the $q\bar{q}$ and $gg$ channels, as
well as from other partonic channels such as $gq \to t\bar{t}q$. The calculation of these
corrections near threshold is greatly simplified. For $z \to 1$ there is no phase-space
available for hard gluon emission, which is thus suppressed by powers of $(1-z)$.
Moreover, contributions from channels such as $gq \to t\bar{t}q$, which involve external
soft-quark fields, are also suppressed. The partonic scattering process is thus dominated
by virtual corrections and the real emission of soft gluons. The phase-space for such
processes is effectively that for a two-body final state, so the hard-scattering kernels
can be written in terms of the kinematic invariants from the Born-level processes.
Therefore, up to corrections of order $(1-z)$, we can rewrite (\ref{eq:genfact}) in the
threshold region as
\begin{align}
  \label{eq:threfact}
  \frac{d^2\sigma}{dMd\cos\theta} &= \frac{8\pi\beta_t}{3sM} \int_\tau^1 \frac{dz}{z} \,
  \Big[ \ff_{gg}(\tau/z,\mu_f) \, C_{gg}(z,M,m_t,\cos\theta,\mu_f) \nonumber
  \\
  &\hspace{-2em} + \ff_{q\bar{q}}(\tau/z,\mu_f) \, C_{q\bar{q}}(z,M,m_t,\cos\theta,\mu_f)+
  \ff_{\bar{q}q}(\tau/z,\mu_f) \, C_{q\bar{q}}(z,M,m_t,-\cos\theta,\mu_f) \Big] \, ,
\end{align}
where $\ff_{q\bar{q}}$ is understood to be summed over all light quark flavors.

In the threshold limit $z\to 1$, the hard-scattering kernels $C_{ij}$ can be factorized
into a product of hard and soft functions according to
\begin{align}
  C_{ij}(z,M,m_t,\cos\theta,\mu_f) = \Tr \left[ \bm{H}_{ij}(M,m_t,\cos\theta,\mu_f) \,
    \bm{S}_{ij}(\sqrt{\hat{s}}(1-z),m_t,\cos\theta,\mu_f) \right] + \mathcal{O}(1-z) \, .
\end{align}
The boldface indicates that the hard functions $\bm{H}_{ij}$ and soft functions
$\bm{S}_{ij}$ are matrices in color space, with respect to which the trace is taken. We
will derive this formula in the next section, using techniques from SCET. (A similar
factorization formula in Mellin moment space was derived in \cite{Kidonakis:1997gm}.) The
hard functions are related to the virtual corrections and are ordinary functions of their
arguments, while the soft functions are related to the real emission of soft gluons and
contain singular distributions in $(1-z)$. In addition to terms proportional to
$\delta(1-z)$, the $n$-th order corrections in $\alpha_s$ also contain plus distributions
of the form
\begin{align}
  \left[ \frac{\ln^m(1-z)}{1-z} \right]_+ ; \qquad m = 0,\ldots,2n-1 \, ,
\end{align}
where 
\begin{align}
  \int_\tau^1 dz \left[ \frac{\ln^m(1-z)}{1-z} \right]_+ g(z) = \int_\tau^1 dz \,
  \frac{\ln^m(1-z)}{1-z} \, [g(z)-g(1)]-g(1)\int_0^\tau dz\,\frac{\ln^m(1-z)}{1-z}
\end{align}
for an arbitrary function $g(z)$. These singular distributions make the perturbative
series badly convergent near threshold and must be resummed to all orders in perturbation
theory. In this paper we perform such a resummation directly in momentum space
\cite{Becher:2006nr}, up to NNLL order. To this end, we extend the procedure for
deep-inelastic scattering \cite{Becher:2006mr}, Drell-Yan process \cite{Becher:2007ty},
Higgs production \cite{Ahrens:2008qu, Ahrens:2008nc}, and direct photon production
\cite{Becher:2009th} to processes with four colored external particles. The formalism will
be described in Section~\ref{sec:Resummation}.

\section{Factorization at threshold in SCET}
\label{sec:SCETfact}

In this section we derive the factorization formula (\ref{eq:threfact}) for the
hard-scattering kernels in the threshold region using SCET \cite{Bauer:2000yr,
  Bauer:2001yt, Beneke:2002ph} and heavy-quark effective theory (HQET) (for a review, see
\cite{Neubert:1993mb}). The derivation is similar to the ones in \cite{Becher:2006mr,
  Becher:2007ty, Becher:2009th}, but is more complicated due to the presence of additional
Dirac and color structures. The derivation of factorization in the effective theory relies
on a two-step matching procedure. In the first step, fluctuations at the hard scale from
virtual corrections are integrated out by matching QCD onto an effective-theory with
collinear and soft degrees of freedom. The Wilson coefficients from this matching step
give the hard function when squaring the amplitude. In the second step, the soft degrees
of freedom are integrated out, giving rise to a soft function, which is defined as the
vacuum expectation value of a Wilson loop operator.

\subsection{Fields and operators}

The scattering amplitude for $t\bar{t}$ production involves several scales, which we
assume to satisfy
\begin{align}
  \hat{s}, M^2, |t_1|, |u_1|, m_t^2 \gg \hat{s}(1-z)^2 \gg \Lambda_{\text{QCD}}^2
\end{align}
in the threshold region. The elements of the first set of scales are taken to be of the
same order and shall be collectively referred to as hard scales, whereas $\hat{s}(1-z)^2$
defines the soft scale. The small quantity $\lambda = (1-z) \ll 1$ then serves as the
expansion parameter in the effective theory. Note that we treat $M$ and $m_t$ as of the
same order, which means that the top quarks are not highly boosted. To describe the
invariant mass spectrum in the region where $M \gg 2m_t$, a more appropriate treatment
would require a different effective theory to separate these two scales, and two jet
functions have to be introduced for the top and anti-top quarks. Such an approach was
adopted in \cite{Fleming:2007xt} for top-quark production in $e^+e^-$ collisions, where
$m_t/M$ was used as a small expansion parameter. However, given that even for $M$ as large
as 1.5\,TeV the ratio $2m_t/M \approx 0.23$ is still a reasonable $\mathcal{O}(1)$
parameter, we see no need to adopt it for the present work.

The formalism for SCET applied to a generic $n$-body scattering process involving both
heavy and light partons was set up in \cite{Bauer:2006mk, Becher:2009kw}. In our case, the
effective theory contains two sets of collinear fields to describe the degrees of freedom
in the incoming hadrons, two sets of HQET fields to describe the outgoing heavy quarks,
and a single set of soft fields describing the final state $X$ and the soft interactions
among particles. In classifying the collinear fields we define two light-like vectors $n$
and $\bar{n}$ in the directions of the colliding partons, which satisfy $n \cdot \bar{n} =
2$. The collinear quark fields are related to the QCD fields by
\begin{align}
  \xi_n(x) = \frac{\nslash\nbslash}{4} \, \psi(x) \, , \qquad \xi_{\bar{n}}(x) =
  \frac{\nbslash\nslash}{4} \, \psi(x) \, .
\end{align}
The collinear gluon fields in a single collinear sector are identical to those in QCD,
with their momenta restricted to be collinear to the given direction. In constructing
operators below, it will be convenient to introduce the manifestly gauge-invariant
combinations of fields \cite{Bauer:2001yt, Hill:2002vw}
\begin{align}
  \chi_n(x) = W_n^\dagger(x) \, \xi_n(x) \, , \qquad \mathcal{A}^\mu_{n\perp}(x) =
  W_n^\dagger(x) \left[ iD^\mu_\perp W_n(x) \right] ,
\end{align}
where the $n$-collinear Wilson line is defined by
\begin{align}
  W_n(x) = \mathcal{P} \exp \left( ig \int_{-\infty}^0 ds \, \bar{n} \cdot A_n(x+s\bar{n})
  \right) ,
\end{align}
and $\mathcal{P}$ denotes path ordering. The corresponding objects for the
$\bar{n}$-collinear fields are obtained by interchanging $n$ and $\bar{n}$. The HQET
fields $h_{v_3}$ and $h_{v_4}$ are labeled by the velocities of the top quark and
anti-quark, which are related to their momenta as
\begin{align}
  p_3^\mu = m_t v_3^\mu + k_3^\mu \, , \qquad p_4^\mu = m_t v_4^\mu + k_4^\mu \, .
\end{align}
The residual momenta $k_i$ scale as soft momenta and are set to zero for on-shell quarks.
In terms of the QCD top-quark fields, the HQET fields are defined as
\begin{align}
  h_{v_i}(x) = \frac{1+\vslash}{2} \, e^{-im_tv_i\cdot x} \, t(x) \, .
\end{align}

A crucial property of the leading-order SCET Lagrangian is that the interactions of soft
gluon fields with collinear and heavy-quark fields are described by eikonal vertices. The
explicit form of the interaction terms for soft gluons with the fermion fields is
\begin{align}
  \label{eq:Lint}
  \mathcal{L}_{\text{int}} &= \bar{\xi}_n(x) \, \frac{\nbslash}{2} \, g \, n \cdot A_s(x)
  \, \xi_n(x) + \bar{\xi}_{\bar{n}}(x) \, \frac{\nslash}{2} \, g\, \bar{n} \cdot A_s(x) \,
  \xi_{\bar{n}}(x) \nonumber
  \\
  &\quad + \bar{h}_{v_3}(x) \, g \, v_3 \cdot A_s(x) \, h_{v_3}(x) + \bar{h}_{v_4}(x) \, g
  \, v_4 \cdot A_s(x) \, h_{v_4}(x) \, ,
\end{align}
and those between collinear and soft gluon fields can be deduced by making the
substitution $A_n\to A_n + n \cdot A_s \, \bar{n}/2$ (and similarly for the
$\bar{n}$-collinear fields) in the Yang-Mills Lagrangian. Such eikonal interactions can be
absorbed into Wilson lines via the field redefinitions \cite{Bauer:2001yt, Becher:2003qh}
\begin{align}
  \label{eq:decoupling}
  \chi_n^a(x) \to \big[ S_n(x) \big]^{ab} \, \chi^{b(0)}_n(x) \, , \quad
  \mathcal{A}_{n\mu}^a(x) \to \big[ S_n^{\text{adj}}(x) \big]^{ab} \,
  \mathcal{A}_{n\mu}^{b(0)}(x) \, , \quad h_{v_3}^a(x) \to \big[ S_{v_3}(x) \big]^{ab} \,
  h^{b(0)}_{v_3}(x) \, ,
\end{align}
with
\begin{align}
  \label{eq:WS}
  \big[ S_n(x) \big]^{ab} &= \mathcal{P} \exp \left( ig \int_{-\infty}^0 dt \, n \cdot
    A_s^c(x+tn) \, t^c_{ab} \right) , \nonumber
  \\
  \big[ S_n^{\text{adj}}(x) \big]^{ab} &= \mathcal{P} \exp \left( ig \int_{-\infty}^0 dt
    \, n \cdot A_s^c(x+tn) \, (-if^{cab}) \right) , \nonumber
  \\
  \big[ S_{v_3}(x) \big]^{ab} &= \mathcal{P} \exp \left( -ig \int_0^{\infty} dt \, v_3
    \cdot A_s^c(x+tv_3) \, t^c_{ab} \right) ,
\end{align}
and similarly for the $\bar{n}$-collinear and $h_{v_4}$ fields. We have used the
superscript ``adj'' to indicate Wilson lines in the adjoint representation. The fields
with the superscript $(0)$ no longer interact with soft gluon fields. Here $t^c$ are
Gell-Mann matrices and $f^{cab}$ are structure constants of QCD.

One must supplement the effective Lagrangian with a set of operators describing the
$(q\bar{q},gg)\to t\bar{t}$ scattering processes. These operators appear in the effective
Hamiltonian in convolutions along light-like directions with perturbative Wilson
coefficients arising from integrating out hard virtual fluctuations. We write this
effective Hamiltonian as
\begin{align}
  \mathcal{H}_{\text{eff}}(x) = \sum_{I,m} \int dt_1 dt_2 \, e^{im_t(v_3+v_4)\cdot x}
  \left[ \tilde{C}^{q\bar{q}}_{Im}(t_1,t_2) \, O^{q\bar{q}}_{Im}(x,t_1,t_2) +
    \tilde{C}^{gg}_{Im}(t_1,t_2) \, O^{gg}_{Im}(x,t_1,t_2) \right] ,
\end{align}
where $I$ labels color structures and $m$ labels Dirac structures. The operators can be
written as
\begin{align}
  \label{eq:ScetOps}
  O^{q\bar{q}}_{Im}(x,t_1,t_2) &= \sum_{\{a\}} \big( c^{q\bar{q}}_I \big)_{\{a\}} \,
  \bar{\chi}^{a_2}_{\bar{n}}(x+t_2n) \, \Gamma'_m \, \chi^{a_1}_n(x+t_1\bar{n}) \;
  \bar{h}^{a_3}_{v_3}(x) \, \Gamma''_m \, h^{a_4}_{v_4}(x) \, , \nonumber
  \\
  O^{gg}_{Im}(x,t_1,t_2) &= \sum_{\{a\}} \big( c^{gg}_I \big)_{\{a\}} \,
  \mathcal{A}^{a_1}_{n\mu\perp}(x+t_1\bar{n}) \,
  \mathcal{A}^{a_2}_{\bar{n}\nu\perp}(x+t_2n) \; \bar{h}^{a_3}_{v_3}(x) \,
  \Gamma^{\mu\nu}_m \, h^{a_4}_{v_4}(x) \, ,
\end{align}
where $\Gamma^{\mu\nu}_m$, $\Gamma'_m$, and $\Gamma''_m$ are combinations of Dirac
matrices and the external vectors $n$, $\bar{n}$, $v_3$, and $v_4$ (note that there can be
contractions of Lorentz indices between $\Gamma'_m$ and $\Gamma''_m$). The
$c^{q\bar{q}}_I$ and $c^{gg}_I$ are tensors in color space, whose indices $\{a\} \equiv
\{a_1,a_2,a_3,a_4\}$ can be in either the fundamental or adjoint representation. For each
channel, they are chosen to be the independent color singlet (e.g.\ gauge-invariant)
structures needed to describe the scattering amplitude. We will choose the color
structures to be in the singlet-octet bases
\begin{gather}
  \big( c^{q\bar{q}}_1 \big)_{\{a\}} = \delta_{a_1a_2} \delta_{a_3a_4} \, , \qquad \big(
  c^{q\bar{q}}_2 \big)_{\{a\}} = t^c_{a_2a_1} t^c_{a_3a_4} \, , \nonumber
  \\
  \big( c^{gg}_1 \big)_{\{a\}} = \delta^{a_1a_2} \delta_{a_3a_4} \, , \qquad \big( c^{gg}_2
  \big)_{\{a\}} = if^{a_1a_2c} \, t^c_{a_3a_4} \, , \qquad \big( c^{gg}_3 \big)_{\{a\}} =
  d^{a_1a_2c} \, t^c_{a_3a_4} \, .
\label{eq:colorstructures}
\end{gather}
When squaring the amplitude and summing over colors, one must evaluate products of the
color structures with their indices contracted. In the absence of soft gluon emissions,
these are of the form $\big(c^{q\bar{q}}_I\big)_{a_1 a_2 a_3 a_4}
\big(c^{q\bar{q}}_J\big)^*_{a_1 a_2 a_3 a_4}$ (and similarly for the gluon fusion channel)
and are equal to an $N$-dependent factor multiplying $\delta_{IJ}$. In this sense the
color structures are orthogonal, but not orthonormal.

Time-ordered products of the operators (\ref{eq:ScetOps}) with the SCET and HQET
Lagrangians describe the collinear and soft contributions to the $(q\bar{q},gg) \to
t\bar{t}X$ scattering amplitudes in QCD, where the final state $X$ contains any number of
soft gluons from real emissions. In the formulation used so far, the final state $X$ is
built up through insertions of the interaction Lagrangian (\ref{eq:Lint}) into the SCET
operators. To account explicitly for soft gluon emission to all orders in the strong
coupling constant, it is convenient to use the decoupling relations (\ref{eq:decoupling})
and represent the soft gluon interactions by Wilson lines. Performing this decoupling and
dropping the superscripts on the new fields, the operators factorize into products of
collinear, heavy-quark, and soft-gluon operators in the form of Wilson loops. The
resulting operators in the $q\bar{q}$ channel read
\begin{align}
  O_{Im}(x,t_1,t_2) = \sum_{\{a\},\{b\}} \big( c_I \big)_{\{a\}} \, \big[ O^h_m(x)
  \big]^{b_3b_4} \, \big[ O^c_m(x,t_1,t_2) \big]^{b_1b_2} \, \big[ O^s(x)
  \big]^{\{a\},\{b\}} \, ,
\end{align}
where
\begin{gather}
  \label{eq:Oscet}
  \big[ O^h_m(x) \big]^{b_3b_4} = \bar{h}^{b_3}_{v_3}(x) \, \Gamma''_m \, h^{b_4}_{v_4}(x)
  \, , \qquad \big[ O^c_m(x,t_1,t_2) \big]^{b_1b_2} = \bar{\chi}^{b_2}_{\bar{n}}(x+t_2n) \,
  \Gamma'_m \, \chi^{b_1}_n(x+t_1\bar{n}) \, , \nonumber
  \\
  \big[ O^s(x) \big]^{\{a\},\{b\}} = \big[ S^\dagger_{v_3}(x) \big]^{b_3a_3} \, \big[
  S_{v_4}(x) \big]^{a_4b_4} \, \big[ S^\dagger_{\bar{n}}(x) \big]^{b_2a_2} \, \big[ S_n(x)
  \big]^{a_1b_1} \, .
\end{gather}
Those in the gluon-fusion channel are obtained by making the obvious replacements in Dirac
structure and the collinear operators, and by changing the Wilson lines $S_n$ and
$S_{\bar{n}}$ to the corresponding ones in the adjoint representation.

In the form shown above, the different sectors no longer interact with each other. After
squaring the amplitude, this property leads to the factorized form (\ref{eq:threfact}) for
the hard-scattering kernels. A complication is that the color indices on the Wilson lines
representing the soft-gluon interactions act on the color structures
(\ref{eq:colorstructures}) and can mix them into each other. Consider, for example, the
calculation of an $\mathcal{O}(\alpha_s)$ correction due to soft gluon exchange between
partons $1$ and $3$ in the squared amplitude in the $q\bar q$ channel. This could either
be from the product of diagrams involving the real emission of one soft gluon, or from a
virtual diagram with a soft loop (which would be scaleless for on-shell quarks, but
appears in the calculation of the anomalous-dimension matrix). After summing over colors
one must evaluate contractions of the form
\begin{align}
  \big( c^{q\bar{q}}_I \big)_{\{b_1 a_2 b_3 a_4\}} t^c_{b_1a_1} t^c_{a_3b_3} \big(
  c^{q\bar{q}}_J \big)^*_{\{a_1 a_2 a_3 a_4\}} \, .
\end{align}
In general, this contraction is not proportional to $\delta_{IJ}$, and must be worked out
case by case. This mixing of the different color structures due to soft gluon exchange is
responsible for the non-trivial matrix structure of the hard and soft functions. To
organize this color algebra, it is convenient to use the color-space formalism
\cite{Catani:1996jh, Catani:1996vz}. Before moving on to the calculation of the
differential cross section in Section \ref{sec:dsigma}, we briefly pause to review this
formalism.

\subsection{Color-space formalism}

Consider the on-shell scattering amplitudes for $(q\bar{q},gg) \to t\bar{t}$, for a given
color configuration of the external particles. We write this in the quark channel as
\begin{align}
  \mathcal{M}^{q\bar{q}}_{\{a\}} = \Braket{t^{a_3}(p_3) \, \bar{t}^{a_4}(p_4) |
    \mathcal{H}_{\text{eff}}(0) | q^{a_1}(p_1) \, \bar{q}^{a_2}(p_2)} \, ,
\end{align}
and also define the object $\mathcal{M}^{gg}_{\{a\}}$ in the obvious way. In what follows,
we will drop the superscript indicating the channel, and work with a single amplitude
which can represent either $\mathcal{M}^{q\bar{q}}$ or $\mathcal{M}^{gg}$. As in
\cite{Catani:1996jh, Catani:1996vz}, we introduce an orthonormal basis of vectors
$\{\Ket{a_1,a_2,a_3,a_4}\}$, where the indices $\{a\}$ refer to the colors of the external
particles. The amplitude can then be written as
\begin{align}
  \mathcal{M}_{\{a\}} = \Braket{a_1,a_2,a_3,a_4 | \mathcal{M}} \, ,
\end{align}
where the object $\Ket{\mathcal{M}}$ is an abstract vector in color space. Since we only
consider color-singlet amplitudes, we can decompose the QCD amplitude into the set of
color structures (\ref{eq:colorstructures}). In the color-space formalism, this is done by
writing
\begin{align}
  \Ket{\mathcal{M}}=\sum_I \mathcal{M}_I \sum_{\{a\}}\big(c_I\big)_{\{a\}} \Ket{\{a\}}
  \equiv \mathcal{M}_I \Ket{c_I} \, ,
\end{align}
where the coefficients $\mathcal{M}_I$ are combinations of Dirac matrices, external
vectors, spinors, and polarization vectors. Note that with this definition the basis
vectors $\Ket{c_I}$ are orthogonal but not normalized, so to project out the
$\mathcal{M}_I$ one must use
\begin{align}
  \mathcal{M}_I = \frac{1}{\Braket{c_I|c_I}} \Braket{c_I|\mathcal{M}} \, .
\end{align}
The square of the amplitude summed over colors is then given by the inner product of
$\Ket{\mathcal{M}}$:
\begin{align}
  \sum_{\rm colors}|\mathcal{M}|^2 = \Braket{\mathcal{M}|\mathcal{M}}=\sum_I
  \mathcal{M}_I^*\mathcal{M}_I \Braket{c_I|c_I}= \sum_I \sum_{\{a\}}
  \mathcal{M}_I^*\mathcal{M}_I \big(c_I\big)^*_{a_1 a_2 a_3 a_4} \big(c_I\big)_{a_1 a_2
    a_3 a_4} \, .
\end{align}

Following \cite{Catani:1996jh, Catani:1996vz}, we introduce color generators $\bm{T}_i$ to
describe the color algebra associated with the emission of a soft gluon from parton
$i=1,2,3,4$. These matrices act on the color indices of the $i$-th parton as
\begin{align}
  \label{eq:genact}
  \bm{T}^c_i \ket{\ldots,a_i,\ldots} = (\bm{T}^c_i)_{b_i a_i} \ket{\ldots,b_i,\ldots} \, .
\end{align}
If the $i$-th parton is a final-state quark or an initial-state anti-quark we set
$(\bm{T}^c_i)_{ba} = t^c_{ba}$, for a final-state anti-quark or an initial-state quark we
have $(\bm{T}^c_i)_{ba} = -t^c_{ab}$, and for a gluon we use $(\bm{T}^c_i)_{ba} =
if^{abc}$. We also use the notation $\bm{T}_i \cdot \bm{T}_j \equiv \bm{T}^c_i \,
\bm{T}^c_j$, and $\bm{T}_i^2$ denotes the quadratic Casimir operator in the representation
of the $i$-th parton, with eigenvalues $C_F$ for quarks and $C_A$ for gluons. Since we
consider color-singlet amplitudes, color conservation implies the relation
\begin{align}
  ( \bm{T}_1 + \bm{T}_2 + \bm{T}_3 + \bm{T}_4 ) \Ket{\mathcal{M}} = 0 \, .
\end{align}

We will be particularly interested in products of color generators acting on the
scattering amplitudes, which appear in the calculation of perturbative corrections to the
differential cross section. We write such products as, for instance,
\begin{align}
  \label{eq:CCrules}
  \Braket{\mathcal{M}|\bm{T}_2\cdot \bm{T}_4|\mathcal{M}}= \mathcal{M}^*_{a_1 b_2 a_3 b_4}
  (\bm{T}^c_2)_{b_2a_2} (\bm{T}^c_4)_{b_4a_4} \mathcal{M}_{a_1a_2a_3a_4} \, .
\end{align}
Rather than evaluating such expressions for each amplitude $\mathcal{M}$, it is more
convenient to work out how the products $\bm{T}_i \cdot \bm{T}_j$ act on the basis vectors
$\Ket{c_I}$:
\begin{align}
  \bm{T}_i \cdot \bm{T}_j \Ket{c_J} = \left[ \bm{T}_i \cdot \bm{T}_j \right]_{IJ}
  \Ket{c_I} \, ,
\end{align}
where the matrix elements are given by
\begin{align}
  \label{eq:matrixdef}
  \left[ \bm{T}_i \cdot \bm{T}_j \right]_{IJ} = \frac{1}{\Braket{c_I|c_I}}
  \Braket{c_I|\bm{T}_i \cdot \bm{T}_j|c_J} \, .
\end{align}
It is worth emphasizing that while the generators themselves act in the abstract color
space, on the left-hand side of the above equation they are just labels to identify a
matrix acting in the space of color-singlet structures. This matrix is thus a 2$\times$2
matrix for the $q\bar q$ channel, and a 3$\times$3 matrix for the $gg$ channel.

We now consider the SCET representation of the amplitude $\mathcal{M}_{\{a\}}$. For the
$q\bar{q}$ channel, this is equal to
\begin{align}
  \sum_{I,m} & \int dt_1 dt_2 \, \tilde{C}_{Im}(t_1,t_2) \, \big(c_I\big)_{\{a\}}
  \nonumber
  \\[-3mm]
  & \times \Braket{t^{b_3}(p_3) \, \bar{t}^{b_4}(p_4) |\big[ O^h_m\big]^{b_3b_4} \big[
    O^c_m \big]^{b_1b_2} \big[ O^s \big]^{\{a\},\{b\}}(0,t_1,t_2) | q^{b_1}(p_1) \,
    \bar{q}^{b_2}(p_2)} \, ,
\end{align}
where no summation over the set of indices $\{b\}$ is performed. We have made clear that
for the amplitude to be non-zero, the colors of the heavy-quark and collinear fields must
coincide with those of the external partons. The reason is that after the decoupling of
soft gluons, the heavy-quark fields are effectively free fields, while collinear exchanges
take place only within $O_m^c$ itself and are diagonal in color space. We can therefore
suppress the color indices on $O^h_m$ and $O^c_m$ as well as on the external states, and
keep in mind that we shall sum over the color indices of $O^s$ when we square the
amplitude. Color correlations such as (\ref{eq:CCrules}) are mediated by the exchange of
soft gluons, which are represented by the Wilson lines in the soft operator $O_s$. To
describe these exchanges in the color-space formalism we use the operator
\begin{align}
  \bm{O}_s(x) = \big[ \bm{S}_n \bm{S}_{\bar{n}}^\dagger \bm{S}_{v_3}^\dagger \bm{S}_{v_4}
  \big] (x) \, ,
\end{align}
where the Wilson lines $\bm{S}_i$ are defined as in (\ref{eq:WS}), with the color
generators promoted to the abstract ones $\bm{T}_i$ according to the rules stated below in
(\ref{eq:genact}). The decoupling relations (\ref{eq:decoupling}) are of the same form for
quarks and gluons when expressed in terms of the $\bm{S}_i$ \cite{Becher:2009qa}, so this
operator is used for both the $q\bar{q}$ and $gg$ channels. Its action on the basis
vectors $\Ket{c_I}$ is then defined according to (\ref{eq:CCrules}). We define the full
SCET operator as $\bm{O}_m = O^h_m O^c_m \bm{O}^s$.

To evaluate the partonic matrix elements of the collinear fields we use
\begin{align}
  \Braket{0 | \big(\chi_n\big)^a_\alpha(t\bar{n}) | p_i;a_i,s_i} &= \delta_{aa_i} \,
  e^{-it\bar{n} \cdot p} \, u_\alpha(p_i,s_i) \, , \nonumber
  \\
  \Braket{0 | \big(\mathcal{A}_{n\perp}\big)^a_\alpha(t\bar n) | p_i;a_i,s_i} &=
  \delta_{aa_i} \, e^{-it\bar{n} \cdot p} \, \epsilon_\alpha(p_i,s_i) \, .
\end{align}
The heavy-quark fields are always taken to be on-shell, so their partonic matrix elements
are equal to HQET spinors multiplied by Kronecker delta symbols in the spin and color
indices. Upon taking the partonic matrix elements, the integrals over $t_1$ and $t_2$
produce the Fourier-transformed Wilson coefficients
\begin{align}
  \label{eq:MomC}
  C_{Im}(M,m_t,\cos\theta,\mu) = \int dt_1 dt_2 \, e^{-it_1 \bar{n} \cdot p_1 - it_2 n
    \cdot p_2} \, \tilde{C}_{Im}(t_1,t_2) \, ,
\end{align}
where we have made the full dependence of the momentum space Wilson coefficients on the
kinematic variables explicit. Defining a vector of Wilson coefficients as
\begin{align}
  \Ket{C_m} \equiv \sum_I C_{Im} \Ket{c_I} \, ,
\end{align}
and introducing the symbol $\dbraket{\ldots}$ for partonic matrix elements as in
\cite{Ferroglia:2009ii}, i.e.,
\begin{align}
  \dbraket{\bm{O}_m} = \Braket{t(p_3) \, \bar{t}(p_4) | \bm{O}_m(0,0,0) | q(p_1) \,
    \bar{q}(p_2)} \, ,
\end{align} 
we can write the color-space representation of the SCET scattering amplitude as
\begin{align}
  \ket{\mathcal{M}} = \sum_m \dbraket{\bm{O}_m} \ket{C_m} \, .
\end{align}

\subsection{The differential cross section}
\label{sec:dsigma}

We now return to the derivation of the factorization formula (\ref{eq:threfact}) by
calculating the partonic cross sections. The hadronic cross section is then obtained by
convoluting these results with the PDFs. Below we will discuss the $q\bar{q}$ case in
detail; the $gg$ channel can be analyzed in an analogous way.

The differential cross section is given by the phase-space integral of the squared
amplitude
\begin{align}
  d\hat \sigma &= \frac{1}{2\hat{s}} \frac{d^3\vec{p}_3}{(2\pi)^32E_3}
  \frac{d^3\vec{p}_4}{(2\pi)^32E_4} \sum_{X_s} (2\pi)^4 \delta^{(4)}(p_s+p_3+p_4-p_1-p_2)
  \nonumber
  \\[-2mm]
  &\quad \times \frac{1}{4d_R^2}\,\bigg| \sum_m \Braket{t(p_3)\,\bar{t}(p_4)\,X_s(p_s) |
      \bm{O}_m(0) | q(p_1)\,\bar q(p_2)} \Ket{C_m} \bigg|^2 ,
\end{align}
where $\bm{O}_m(x) \equiv \bm{O}_m(x,0,0)$, and we have used translational invariance to
write the result in terms of the Fourier-transformed coefficients $C_m$ from
(\ref{eq:MomC}). The cross section is implicitly summed over the external colors and
spins, and the factor $d_R=N$ for quarks and $d_R=N^2-1$ for gluons arises from averaging
over the colors of the initial-state partons. The factor 1/4 accounts for the averaging
over the polarizations of the initial-state partons. Since the different types of
effective-theory fields do not interact with each other, we can factorize the matrix
element into soft, heavy-quark, and collinear pieces. The partonic matrix elements of the
heavy-quark and collinear pieces just give the usual products of spinors, which we combine
into the tree-level matrix element
\begin{align}
  \dbraket{O_m}_{\text{tree}} \equiv \dbraket{O^h_m(0) O^c_m(0)}_{\text{tree}} =
  \Braket{t(p_3) \, \bar{t}(p_4) | O^h_m(0) O^c_m(0) | q(p_1) \,
    \bar{q}(p_2)}_{\text{tree}} \, .
\end{align}
Summed over spins, these give rise to the usual Dirac traces. The matrix element of the
soft operator is taken using the vacuum as the initial state and is of the form
$|\Braket{X_s(p_s)|\bm{O}_s(0)|0}|^2$. This can be evaluated directly, but we prefer
instead to sum over the final states $X_s$ and convert it into a forward matrix element
using the formalism explained in Appendix~C of \cite{Becher:2007ty}. Written in this form,
the differential cross section reads
\begin{align}
  \label{eq:Forward}
  d\hat\sigma &= \frac{1}{2\hat{s}} \frac{d^3\vec{p}_3}{(2\pi)^32E_3}
  \frac{d^3\vec{p}_4}{(2\pi)^32E_4} \int d^4x \, e^{i(p_1+p_2-p_3-p_4) \cdot x} \nonumber
  \\
  &\quad \times \frac{1}{4d_R^2}\sum_{m,m'} \bigg[ \, \dbraket{O_m}^\dagger_{\text{tree}}
  \, \dbraket{O_{m'}}_{\text{tree}} \times \Bra{C_m} \Braket{0 | \bar{\bm{\mathrm{T}}} [
    \bm{O}^{s\dagger}(x) ] \, \bm{\mathrm{T}} [ \bm{O}^s(0) ] | 0} \Ket{C_{m'}} \bigg] \,
  ,
\end{align}
where $\bm{\mathrm{T}}$ and $\bar{\bm{\mathrm{T}}}$ represent time and anti-time ordering
\cite{Becher:2007ty}. Since we treat the soft scale as perturbative, the soft fields can
be integrated out by evaluating the vacuum matrix element. We now define a hard matrix and
a position-space soft function as
\begin{align}
  \label{eq:hard}
  \bm{H}(M,m_t,\cos\theta,\mu) &= \frac{3}{8} \, \frac{1}{(4\pi)^2} \, \frac{1}{4d_R}
  \sum_{m,m'} \dbraket{O_{m'}}_{\text{tree}} \, \Ket{C_{m'}} \Bra{C_m} \,
  \dbraket{O_m}^\dagger_{\text{tree}} \, ,
  \nonumber
  \\
  \bm{W}(x,\mu) &= \frac{1}{d_R} \Braket{0 | \bar{\bm{\mathrm{T}}} [ \bm{O}^{s\dagger}(x) ]
    \, \bm{\mathrm{T}} [ \bm{O}^s(0) ] | 0} \, ,
\end{align}
where we have chosen the prefactors to match the overall normalization of
(\ref{eq:threfact}). The elements of these matrices in the chosen color basis are defined
as
\begin{align}
  \label{eq:hardsoftmat}
  H_{IJ} \equiv \frac{1}{\braket{c_I|c_I}\braket{c_J|c_J}} \Braket{c_I | \bm{H} | c_J} \, ,
  \qquad W_{IJ} \equiv \Braket{c_I | \bm{W} | c_J} \, ,
\end{align}
so that the term in square brackets in the second line of (\ref{eq:Forward}) is
proportional to
\begin{align}
  \Tr \big[ \bm{H} \, \bm{W} \big]=\sum_{I,J}H_{IJ}W_{JI} \, .
\end{align}

In order to compute the invariant mass spectrum for the $t\bar{t}$ pair, we define
$q=p_3+p_4$ and insert
\begin{align}
  1 = \int d^4q \, dM \, \delta^{(4)}(q-p_3-p_4) \, 2M \, \delta(M^2-q^2)
\end{align}
into (\ref{eq:Forward}). After performing the $\vec{p}_4$ integral using the first
$\delta$-function, the $q^0$ integration using the second, and carrying out the trivial
angular integration in the $\vec{p_3}$ integral, we arrive at
\begin{align}\label{sigma_parton}
  \frac{d^2\hat \sigma}{dMd\cos\theta} &= \frac{16M}{3\hat{s}} \, \frac{1}{(2\pi)^3} \int
  \frac{d^3\vec{q}}{2q^0} \int dE_3 |\vec{p}_3| \, \, \delta(M^2 - 2 q \cdot p_3)
  \nonumber
  \\
  &\quad \times \int d^4x \, e^{i(p_1+p_2-q) \cdot x} \, \Tr \big[
  \bm{H}(M,m_t,\cos\theta,\mu_f) \, \bm{W}(x,\mu_f) \big] \, ,
\end{align}
where $|\vec{p}_3|=\sqrt{E_3^2-m_t^2}$ and $q_0=\sqrt{M^2+\vec q^2}$. In the partonic
center-of-mass frame, we have $|\vec{q}|=\mathcal{O}(\sqrt{\hat s}(1-z))$, so we can set
$q_0=M$ and drop $\vec{q}\,$ in the $\delta$-function. Then the integral over $\vec{q}\,$
produces a factor $\delta^3(\vec{x})$ from the exponential, and after a few manipulations
we find
\begin{align}
  \frac{d^2\hat\sigma}{dMd\cos\theta} &= \frac{8\pi\beta_t}{3\hat{s}M} \Tr \big[
  \bm{H}(M,m_t,\cos\theta,\mu_f) \, \bm{S}(\sqrt{\hat{s}}(1-z),M,m_t,\cos\theta,\mu_f)
  \big] \, ,
\end{align}
where the momentum-space soft function is defined by \cite{Becher:2007ty}
\begin{align}
  \label{eq:softmom}
  \bm{S}(\sqrt{\hat{s}}(1-z),M,m_t,\cos\theta,\mu) = \sqrt{\hat{s}} \int \frac{dx_0}{4\pi}
  \, e^{i\sqrt{\hat{s}}(1-z)x_0/2} \, \bm{W}(x_0,\vec{x}=0,\mu) \, .
\end{align}

The hadronic cross section (\ref{eq:genfact}) is now obtained by convoluting the partonic
cross sections in (\ref{sigma_parton}) with the parton luminosities. Comparing with
(\ref{eq:threfact}), we finally arrive at the factorized form of the hard-scattering
kernel
\begin{align}
  C(z,M,m_t,\cos\theta,\mu_f) = \Tr \big[ \bm{H}(M,m_t,\cos\theta,\mu_f) \,
  \bm{S}(\sqrt{\hat{s}}(1-z),M,m_t,\cos\theta,\mu_f) \big] \, ,
\end{align}
where we have set $\hat{s}=M^2$ everywhere except in the first argument of the soft
function.

\section{The hard and soft functions at NLO}
\label{sec:Matching}

In this section we describe the calculation of the soft and hard matrices up to NLO in
perturbation theory.

\subsection{Hard functions}

The hard functions are related to products of Wilson coefficients, as shown in
(\ref{eq:hard}). To obtain the Wilson coefficients $C_{Im}$, one matches renormalized
Green's functions in QCD with those in SCET. The matching can be done with any choice of
external states and infrared (IR) regulators. It is by far simplest to use on-shell
partonic states for $(q\bar{q},gg) \to t\bar{t}$ scattering and dimensional regularization
in $d=4-2\epsilon$ dimensions to regularize both the ultraviolet (UV) and IR divergences.
With this choice, the loop graphs in SCET are scaleless and vanish, so the
effective-theory matrix elements are equal to their tree-level expressions multiplied by a
UV renormalization matrix $\bm{Z}$. The matrix elements in QCD, on the other hand, are
just the virtual corrections to the $(q\bar{q},gg) \to t\bar{t}$ scattering amplitudes.
The matching condition then reads \cite{Becher:2009qa, Becher:2009kw, Ferroglia:2009ii}
\begin{align}
  \label{eq:matching}
  \lim_{\epsilon \to 0} \bm{Z}^{-1}(\epsilon,M,m_t,\cos\theta,\mu)
  \Ket{\mathcal{M}(\epsilon,M,m_t,\cos\theta)} = \sum_m \dbraket{O_m}_{\text{tree}}
  \Ket{C_m(M,m_t,\cos\theta,\mu)} \, ,
\end{align}
where $\mathcal{M}$ is the UV-renormalized virtual QCD amplitude expressed in terms of
$\alpha_s$ with $n_l=5$ active flavors. We have moved the SCET renormalization matrix
$\bm{Z}$ to act on the QCD amplitude, so that both sides of the equation are finite in the
limit $\epsilon \to 0$. The explicit results for the matrix elements $\bm{Z}_{IJ}$ in our
color basis for the $q\bar q$ and $gg$ channels can be found in \cite{Ferroglia:2009ii}.

In practice, we are not interested in the Wilson coefficients themselves, but rather the
hard matrix $H_{IJ}$. To calculate this, we first define
\begin{align}
  \label{eq:renM}
  \Ket{\mathcal{M}_{\text{ren}}} \equiv \lim_{\epsilon \to 0} \bm{Z}^{-1}(\epsilon)
  \Ket{\mathcal{M}(\epsilon)} = 4\pi\alpha_s \left[ \Ket{\mathcal{M}_{\text{ren}}^{(0)}} +
    \frac{\alpha_s}{4\pi} \Ket{\mathcal{M}_{\text{ren}}^{(1)}} + \ldots \right] ,
\end{align}
and expand the hard function as
\begin{align}
  \bm{H} = \alpha_s^2 \, \frac{3}{8d_R} \left( \bm{H}^{(0)} + \frac{\alpha_s}{4\pi}
    \bm{H}^{(1)} + \ldots \right) .
\end{align}
Using (\ref{eq:matching}) and (\ref{eq:renM}) to express the SCET matrix element in terms
of the finite, IR-subtracted QCD amplitudes in the definition of the hard function
(\ref{eq:hard}), the matrix elements (\ref{eq:hardsoftmat}) can be written as
\begin{align}
  \label{eq:hform}
  H^{(0)}_{IJ} &= \frac{1}{4} \, \frac{1}{\braket{c_I|c_I}\braket{c_J|c_J}} \Braket{c_I |
    \mathcal{M}_{\text{ren}}^{(0)}} \Braket{\mathcal{M}_{\text{ren}}^{(0)} | c_J} \, ,
  \nonumber
  \\
  H^{(1)}_{IJ} &= \frac{1}{4} \, \frac{1}{\braket{c_I|c_I}\braket{c_J|c_J}} \bigg[
  \Braket{c_I | \mathcal{M}_{\text{ren}}^{(0)}} \Braket{\mathcal{M}_{\text{ren}}^{(1)} |
    c_J} + \Braket{c_I | \mathcal{M}_{\text{ren}}^{(1)}}
  \Braket{\mathcal{M}_{\text{ren}}^{(0)} | c_J} \bigg] \, .
\end{align}
The leading-order result for the $q\bar q$ channel follows from a simple calculation and
reads
\begin{align}
  \bm{H}_{q\bar{q}}^{(0)} =
  \begin{pmatrix}
    0 & 0
    \\
    0 & 2
  \end{pmatrix}
  \Bigg[ \frac{t_1^2 + u_1^2}{M^4} + \frac{2m_t^2}{M^2} \Bigg] \, ,
\end{align}
while that for the $gg$ channel is  
\begin{align}
  \bm{H}_{gg}^{(0)}  =
  \begin{pmatrix}
    \frac{1}{N^2} & \frac{1}{N}\,\frac{t_1-u_1}{M^2} & \frac{1}{N}
    \\
    \frac{1}{N}\,\frac{t_1-u_1}{M^2} & \frac{(t_1-u_1)^2}{M^4} & \frac{t_1-u_1}{M^2}
    \\
    \frac{1}{N} & \frac{t_1-u_1}{M^2} & 1
  \end{pmatrix}
  \frac{M^4}{2t_1u_1} \Bigg[ \frac{t_1^2+u_1^2}{M^4} + \frac{4m_t^2}{M^2} -
  \frac{4m_t^4}{t_1u_1} \Bigg] \, .
\end{align}

To calculate the NLO hard function requires the one-loop virtual corrections to the
partonic scattering amplitudes, decomposed into the singlet-octet basis. Although results
for the NLO virtual corrections interfered with the Born-level amplitudes exist in the
literature \cite{Nason:1987xz, Beenakker:1988bq, Beenakker:1990maa}, results for the
one-loop amplitude decomposed into our color basis are not available and must be
calculated from scratch. For this purpose we use in-house routines written in the computer
algebraic system FORM \cite{Vermaseren:2000nd}. The results are rather lengthy (especially
for the $gg$ channel) and are in a Mathematica package which can be obtained from the
authors upon request.

We have been able to perform several checks on our results. First, we have verified that
applying the renormalization factor $\bm{Z}$ to the tree-level amplitude indeed absorbs
the IR poles in the UV-renormalized QCD amplitudes at one-loop order. Second, we have
checked that inserting the results for the products of one-loop hard functions and
tree-level soft functions, given in (\ref{eq:Stree}) below, into the formula for the
differential cross section, we reproduce the results of \cite{Nason:1987xz,
  Beenakker:1988bq, Beenakker:1990maa}, as required. Finally, using the one-loop hard
functions, we were able to calculate the IR singularities of the two-loop amplitudes in
\cite{Ferroglia:2009ii}, which agree with all the available results in the literature
\cite{Czakon:2007ej, Czakon:2007wk, Czakon:2008zk}.

\subsection{Soft functions}

The soft functions are given by the vacuum expectation values of the soft Wilson-loop
operators, as defined in (\ref{eq:hard}). In what follows we will calculate the one-loop
corrections to these objects directly in position space. When performing the resummation
in the next section, it will be more convenient to work with the Laplace-transformed
functions. They are defined as
\begin{align}
  \label{eq:stilde}
  \tilde{\bm{s}}(L,M,m_t,\cos\theta,\mu) &= \frac{1}{\sqrt{\hat s}} \int_0^\infty d\omega
  \, \exp \left( -\frac{\omega}{e^{\gamma_E}\mu e^{L/2}} \right)
  \bm{S}(\omega,M,m_t,\cos\theta,\mu) \nonumber
  \\[-2mm]
  &= \bm{W} \bigg( x_0 = \frac{-2i}{e^{\gamma_E}\mu e^{L/2}}, \mu \bigg) \, ,
\end{align}
where the second equality was shown in \cite{Becher:2007ty} and follows from the
functional form of position-space Wilson loops \cite{Korchemsky:1993uz}.

%%%%%%%%%%%%%%%%%%%%%%%%%%%%%%%%%%%%%%%%%%%%%%%%%%%%%%%%%%%%%%%%%%%%%%%%%%%%%%%%
\begin{figure}[t]
\centering
\includegraphics[width=0.79\textwidth]{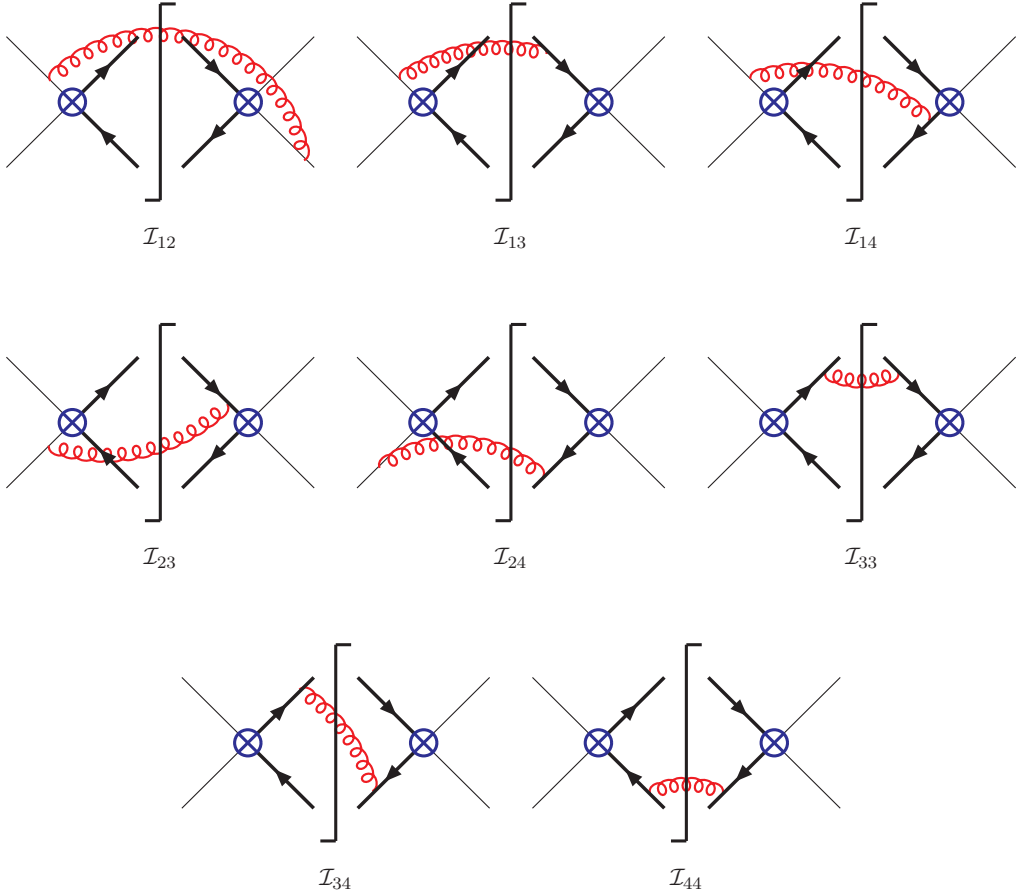}
\caption{\label{fig:NLOsoft} Diagrams contributing to the soft functions at NLO. The thick
  lines represent Wilson lines in the time-like directions $v_3$ and $v_4$, the thin lines
  Wilson lines in the light-like directions $n$ and $\bar{n}$, and the cut curly lines
  represent the cut gluon propagator (\ref{eq:cutprop}).}
\end{figure}
%%%%%%%%%%%%%%%%%%%%%%%%%%%%%%%%%%%%%%%%%%%%%%%%%%%%%%%%%%%%%%%%%%%%%%%%%%%%%%%%

We expand the soft functions in power of $\alpha_s$ as
\begin{align}
  \tilde{\bm{s}}=\tilde{\bm{s}}^{(0)}+\frac{\alpha_s}{4\pi}\tilde{\bm{s}}^{(1)}
  +\left(\frac{\alpha_s}{4\pi}\right)^2\tilde{\bm{s}}^{(2)} + \ldots \, .
\end{align}
At leading order, the Wilson loop is just the unit matrix, so $\tilde{s}^{(0)}_{IJ} =
\Braket{c_I|c_J}/d_R$, and it is easy to show that
\begin{align}
  \label{eq:Stree}
  \tilde{\bm{s}}_{q\bar{q}}^{(0)} =
  \begin{pmatrix}
    N & 0
    \\
    0 & \frac{C_F}{2}
  \end{pmatrix}
  , \qquad \tilde{\bm{s}}_{gg}^{(0)} =
  \begin{pmatrix}
    N & 0 & 0
    \\
    0 & \frac{N}{2} & 0
    \\
    0 & 0 & \frac{N^2-4}{2N}
  \end{pmatrix}
  .
\end{align}
At NLO, the soft functions receive contributions from the diagrams depicted in Figure
\ref{fig:NLOsoft}. The calculation is similar to that in \cite{Korchemsky:1993uz}. To
evaluate the diagrams we associate an eikonal factor $v_i^{\mu}/{k\cdot v_i}$ multiplied
by a color generator $\bm{T}_i$ for each attachment of a gluon to a particle with velocity
$v_i$ (we define $v_1=n$ and $v_2=\bar n$), and contract with the cut gluon propagator in
position space, which in Feynman gauge reads
\begin{equation}
  \label{eq:cutprop}
  D^{\mu\nu}_+(x)=-g^{\mu\nu} \int \frac{d^dk}{(2\pi)^d}\,e^{-ik\cdot x}\,(2\pi)   \,
  \delta(k^2) \, \theta(k^0) \, .
\end{equation}
We can then write the bare soft function in position space as
\begin{align}
  \bm{W}^{(1)}_{\text{bare}}(\epsilon,x_0,\mu) = \sum_{i,j} \, \bm{w}_{ij} \,
  \mathcal{I}_{ij}(\epsilon,x_0,\mu) \, ,
\end{align}
where the matrices $\bm{w}_{ij}$ are related to products of color generators and will be
given in (\ref{eq:qTT}) and (\ref{eq:gTT}) below. The integrals $\mathcal{I}_{ij}$ are
defined as
\begin{align}
  \label{eq:softintegrals}
  \mathcal{I}_{ij}(\epsilon,x_0,\mu) = -\frac{(4\pi\mu^2)^\epsilon}{\pi^{2-\epsilon}} \,
  v_i \cdot v_j \int d^dk \, \frac{e^{-ik^0x_0}}{v_i \cdot k \, v_j \cdot k} \, (2\pi) \,
  \delta(k^2) \, \theta(k^0) \, ,
\end{align}
which are obviously symmetric in the indices $i$ and $j$. The integrals
$\mathcal{I}_{11}=\mathcal{I}_{22}=0$, and the non-vanishing integrals are
\begin{align}
  \mathcal{I}_{12} &= -\left( \frac{2}{\epsilon^2} + \frac{2}{\epsilon} L_0 + L_0^2 +
    \frac{\pi^2}{6}\right) , \nonumber
  \\
  \mathcal{I}_{33} = \mathcal{I}_{44} &= \frac{2}{\epsilon} + 2L_0 - \frac{2}{\beta_t} \ln
  x_s \, , \nonumber
  \\[1mm]
  \mathcal{I}_{34} &= -\frac{1+x_s^2}{1-x_s^2} \left[ \left( \frac{2}{\epsilon} + 2L_0
    \right) \ln x_s - \ln^2x_s + 4\ln x_s \ln(1-x_s) + 4\Li_2(x_s) - \frac{2\pi^2}{3}
  \right] , \nonumber
  \\[1mm]
  \mathcal{I}_{13} = \mathcal{I}_{24} &= -\left[\frac{1}{2} \left( L_0 -
      \ln\frac{(1+y_t)^2x_s}{(1+x_s)^2} \right)^2 + \frac{\pi^2}{12} +
    2\Li_2\left(\frac{1-x_s y_t}{1+x_s}\right) +
    2\Li_2\left(\frac{x_s-y_t}{1+x_s}\right)\right] , \nonumber
  \\[2mm]
  \mathcal{I}_{14} = \mathcal{I}_{23} &= \mathcal{I}_{13}(y_t \to z_u) \, ,
\end{align}
where $x_s=(1-\beta_t)/(1+\beta_t)$, $y_t=-t_1/m_t^2-1$, $z_u=-u_1/m_t^2-1$, and
\begin{align}
  L_0 = \ln \bigg( -\frac{\mu^2x_0^2e^{2\gamma_E}}{4} \bigg) \,.
\end{align}
The renormalized soft functions $\bm{W}^{(1)}$ can then be obtained by subtracting the
divergent part from $\bm{W}^{(1)}_{\text{bare}}$. Later on we will need the
Laplace-transformed function $\tilde{\bm{s}}$, which according to (\ref{eq:stilde}) is
obtained by replacing $L_0 \to -L$. To finish the calculation, we must also determine the
matrix elements of
\begin{align}
  \big( \bm{w}_{ij} \big)_{IJ} = \frac{1}{d_R} \, \Braket{c_I | \bm{T}_i \cdot \bm{T}_j |
    c_J} \, .
\end{align}
For the $q\bar{q}$ channel, the results are
\begin{align}
  \label{eq:qTT}
  \bm{w}_{12}^{q\bar{q}} = \bm{w}_{34}^{q\bar{q}} &= -\frac{C_F}{4N}
  \begin{pmatrix}
    4N^2 & 0
    \\
    0 & -1
  \end{pmatrix}
  , \nonumber
  \\
  \bm{w}_{33}^{q\bar{q}} = \bm{w}_{44}^{q\bar{q}} &= \frac{C_F}{2}
  \begin{pmatrix}
    2N & 0
    \\
    0 & C_F
  \end{pmatrix}
  , \nonumber
  \\
 \bm{w}_{13}^{q\bar{q}} = \bm{w}_{24}^{q\bar{q}}
 &=- \frac{C_F}{2}
  \begin{pmatrix}
    0 & 1
    \\
    1 & 2C_F - \frac{N}{2}
  \end{pmatrix}
  , \nonumber
  \\
  \bm{w}_{14}^{q\bar{q}} = \bm{w}_{23}^{q\bar{q}} &= -\frac{C_F}{2N}
  \begin{pmatrix}
    0 & -N
    \\
    -N & 1
  \end{pmatrix}
  ,
\end{align}
while for the $gg$ channel we obtain
\begin{align}
  \label{eq:gTT}
  \bm{w}_{12}^{gg}&=- \frac{1}{4}
  \begin{pmatrix}
    4N^2 & 0 & 0
    \\
    0 & N^2 & 0
    \\
    0 & 0 & N^2-4
  \end{pmatrix}
  , \nonumber
  \\
  \bm{w}_{34}^{gg}&=-
  \begin{pmatrix}
    C_FN & 0 & 0
    \\
    0 & -\frac{1}{4} & 0
    \\
    0 & 0 & -\frac{N^2-4}{4N^2}
  \end{pmatrix}
  , \nonumber
  \\
  \bm{w}_{33}^{gg}=\bm{w}_{44}^{gg}&= \frac{C_F}{2N}
  \begin{pmatrix}
    2N^2 & 0 & 0
    \\
    0 & N^2 & 0
    \\
    0 & 0 & N^2-4
  \end{pmatrix}
  , \nonumber
  \\
 \bm{w}_{13}^{gg}= \bm{w}_{24}^{gg}&= -\frac{1}{8}
  \begin{pmatrix}
    0 & 4N & 0
    \\
    4N & N^2 & N^2-4
    \\
    0 & N^2-4 & N^2-4
  \end{pmatrix}
  , \nonumber
  \\
  \bm{w}_{14}^{gg}= \bm{w}_{23}^{gg}&= -\frac{1}{8}
  \begin{pmatrix}
    0 & -4N & 0
    \\
    -4N & N^2 & -(N^2-4)
    \\
    0 & -(N^2-4) & N^2-4
  \end{pmatrix}
  .
\end{align}

The NLO contributions to the cross section from real emissions in the soft limit have been
known for some time \cite{Mangano:1991jk}. Transforming our results to momentum space, we
have checked that the integrals (\ref{eq:softintegrals}) are consistent with those given
in Appendix~A of that paper (after taking into account some misprints in
\cite{Mangano:1991jk} later corrected in \cite{Frixione:1993dg}). As another check, we
have used our results along with the one-loop hard functions from the previous section to
calculate the total partonic cross sections at NLO in the limit $\beta\to 0$ (see
Section~\ref{sec:totcross}), reproducing the analytic expressions from
\cite{Czakon:2008ii}. Finally, using the RG invariance of the cross section we will derive
the RG equation for $\tilde{\bm{s}}$ in the next section. We have checked that our
one-loop result satisfies this equation, which also justifies our procedure of simply
subtracting the $1/\epsilon$ poles in the bare function to get the renormalized results.

\section{Threshold resummation in SCET}
\label{sec:Resummation}

In the region where the cross section is dominated by the threshold terms, one needs to
resum the leading singular terms in $(1-z)$ to all orders in perturbation theory. This is
accomplished by deriving and solving RG equations for the hard and soft functions in the
effective theory, which will be described in what follows. Since these equations contain
information on the logarithmic structure of the hard-scattering kernels at higher-orders
in perturbation theory, they can also be used to derive an approximate NNLO formula for
the differential cross section in the threshold region. We discuss this further in
Section~\ref{sec:NNLOapprox}.

\subsection{RG evolution and resummation at NNLL}

The hard function satisfies the evolution equation
\begin{align}
  \label{eq:Hev}
  \frac{d}{d\ln\mu} \bm{H}(M,m_t,\cos\theta,\mu) &= \bm{\Gamma}_H(M,m_t,\cos\theta,\mu) \,
  \bm{H}(M,m_t,\cos\theta,\mu) \nonumber
  \\
  &\quad\mbox{}+ \bm{H}(M,m_t,\cos\theta,\mu) \,
  \bm{\Gamma}_H^\dagger(M,m_t,\cos\theta,\mu) \, .
\end{align}
Using (\ref{eq:hardsoftmat}), we can write the above equation in a matrix form, where the
matrix elements of $\bm{\Gamma}_H$ are defined according to (\ref{eq:matrixdef}), and can
be obtained from the matrices $\bm{\Gamma}_{q\bar{q}}$ or $\bm{\Gamma}_{gg}$ in
\cite{Ferroglia:2009ii}. The form of the evolution equation follows from (\ref{eq:renM})
and (\ref{eq:hform}), along with the defining relation
\begin{align}
  \bm{Z}^{-1}\frac{d}{d\ln\mu}\bm{Z}=-\bm{\Gamma}_H
\end{align}
for the anomalous dimension. The explicit results to two-loop order are
\begin{align}
  \label{eq:qqmatrix}
  \bm{\Gamma}_{q\bar{q}} &= \left[ C_F \, \gamma_{\text{cusp}}(\alpha_s) \, \left(
      \ln\frac{M^2}{\mu^2} - i\pi \right) + C_F \,
    \gamma_{\text{cusp}}(\beta_{34},\alpha_s) + 2\gamma^q(\alpha_s) + 2\gamma^Q(\alpha_s)
  \right] \bm{1} \nonumber
  \\
  &\quad\mbox{} + \frac{N}{2} \left[ \gamma_{\text{cusp}}(\alpha_s) \, \left(
      \ln\frac{t_1^2}{M^2m_t^2} + i\pi \right) - \gamma_{\text{cusp}}(\beta_{34},\alpha_s)
  \right]
  \begin{pmatrix}
    0 & 0
    \\
    0 & 1
  \end{pmatrix}
  \nonumber
  \\
  &\quad\mbox{} + \gamma_{\text{cusp}}(\alpha_s) \, \ln\frac{t_1^2}{u_1^2} \left[
    \begin{pmatrix}
      0 & \frac{C_F}{2N}
      \\
      1 & -\frac{1}{N}
    \end{pmatrix}
    + \frac{\alpha_s}{4\pi} \, g(\beta_{34})
    \begin{pmatrix}
      0 & \frac{C_F}{2}
      \\
      -N & 0
    \end{pmatrix}
  \right] , 
\end{align}
and 
\begin{align}
  \label{eq:ggmatrix}
  \bm{\Gamma}_{gg} &= \left[ N \, \gamma_{\text{cusp}}(\alpha_s) \, \left(
      \ln\frac{M^2}{\mu^2} -i\pi \right) + C_F \,
    \gamma_{\text{cusp}}(\beta_{34},\alpha_s) + 2\gamma^g(\alpha_s) + 2\gamma^Q(\alpha_s)
  \right] \bm{1} \nonumber
  \\
  &\quad\mbox{} + \frac{N}{2} \left[ \gamma_{\text{cusp}}(\alpha_s) \, \left(
      \ln\frac{t_1^2}{M^2m_t^2} + i\pi\right) - \gamma_{\text{cusp}}(\beta_{34},\alpha_s)
  \right]
  \begin{pmatrix}
    0 & 0 & 0
    \\
    0 & 1 & 0
    \\
    0 & 0 & 1
  \end{pmatrix}
  \nonumber
  \\
  &\quad\mbox{} + \gamma_{\text{cusp}}(\alpha_s) \, \ln\frac{t_1^2}{u_1^2} \left[
    \begin{pmatrix}
      0 & \frac{1}{2} & 0
      \\
      1 & -\frac{N}{4} & \frac{N^2-4}{4N}
      \\
      0 & \frac{N}{4} & -\frac{N}{4}
    \end{pmatrix} 
    + \frac{\alpha_s}{4\pi}\,g(\beta_{34})
    \begin{pmatrix}
      0 & \frac{N}{2} & 0
      \\
      -N & 0 & 0 \\
      0 & 0 & 0
    \end{pmatrix}
  \right] , 
\end{align}
where the various anomalous-dimension functions can be found in the Appendix, and the cusp
angle $\beta_{34}= i\pi - \ln(1+\beta_t)/(1-\beta_t)$. The solution to the evolution
equation can be written as
\begin{align}
  \bm{H}(M,m_t,\cos\theta,\mu) = \bm{U}(M,m_t,\cos\theta,\mu_h,\mu) \,
  \bm{H}(M,m_t,\cos\theta,\mu_h) \, \bm{U}^\dagger(M,m_t,\cos\theta,\mu_h,\mu) \, ,
\end{align}
where the unitary matrix $\bm{U}$ satisfies the equation
\begin{align}
  \label{eq:Uev}
  \frac{d}{d\ln\mu} \bm{U}(M,m_t,\cos\theta,\mu_h,\mu) =
  \bm{\Gamma}_H(M,m_t,\cos\theta,\mu) \, \bm{U}(M,m_t,\cos\theta,\mu_h,\mu) \, .
\end{align}
The matching scale $\mu_h$ must be chosen of order a typical hard scale, so that the
matching condition for the hard function is free of large logarithms. With the help of the
evolution matrix $\bm{U}$, the hard function can then be evolved to an arbitrary scale
$\mu$. The formal solution to this equation is
\begin{align}
  \bm{U}(M,m_t,\cos\theta,\mu_h,\mu) = \mathcal{P} \exp \int\limits_{\mu_h}^{\mu}
  \frac{d\mu'}{\mu'} \, \bm{\Gamma}_H(M,m_t,\cos\theta,\mu') \, ,
\end{align}
where the path-ordering is necessary because $\bm{\Gamma}_H$ is a matrix. To evaluate the
path-ordered exponential, it is convenient to separate the explicit logarithmic dependence
on the scale $\mu$, which is related to Sudakov double logarithms, from the remaining
piece, which is related to single logarithmic evolution. We thus write the anomalous
dimension as
\begin{align}
  \label{eq:gammaH}
  \bm{\Gamma}_H(M,m_t,\cos\theta,\mu) = \Gamma_{\text{cusp}}(\alpha_s) \left(
    \ln\frac{M^2}{\mu^2} - i\pi \right) \bm{1} + \bm{\gamma}^h(M,m_t,\cos\theta,\alpha_s)
  \, ,
\end{align}
where $\Gamma_{\text{cusp}}$ is equal to $C_F \, \gamma_{\text{cusp}}$ for $q\bar{q}$ and
$N \, \gamma_{\text{cusp}}$ for $gg$, and the matrices $\bm{\gamma}^h$ are defined through
a comparison with (\ref{eq:qqmatrix}) and (\ref{eq:ggmatrix}). Since the term proportional
to $\Gamma_{\text{cusp}}$ multiplies the unit matrix, we can factor this piece out of the
path-ordering and evaluate it using standard techniques. The result for the evolution
matrix is then
\begin{align}
  \bm{U}(M,m_t,\cos\theta,\mu_h,\mu) &= \exp \bigg[ 2S(\mu_h,\mu) - a_\Gamma(\mu_h,\mu)
  \left( \ln\frac{M^2}{\mu_h^2} - i\pi \right) \bigg] \,
  \bm{u}(M,m_t,\cos\theta,\mu_h,\mu) \, .
\end{align}
The RG exponents in the square brackets of the exponential factor are given by
\begin{align}
  \label{eq:Sa}
  S(\mu_h,\mu) = -\int\limits_{\alpha_s(\mu_h)}^{\alpha_s(\mu)} \!d\alpha\,
  \frac{\Gamma_{\text{cusp}}(\alpha)}{\beta(\alpha)} \int\limits_{\alpha_s(\mu_h)}^\alpha
  \!\frac{d\alpha'}{\beta(\alpha')} \, , \qquad a_\Gamma(\mu_h,\mu) =
  -\int\limits^{\alpha_s(\mu)}_{\alpha_s(\mu_h)}\!
  d\alpha\,\frac{\Gamma_{\text{cusp}}(\alpha)}{\beta(\alpha)} \, ,
\end{align}
where $\beta(\alpha_s)=d\alpha_s/d\ln\mu$ is the QCD $\beta$-function. The quantity
$\bm{u}$ contains the non-trivial matrix evolution due to $\bm{\gamma}^h$ and reads
\begin{align}
  \label{eq:offdiagu}
  \bm{u}(M,m_t,\cos\theta,\mu_h,\mu) = \mathcal{P} \exp
  \int\limits_{\alpha_s(\mu_h)}^{\alpha_s(\mu)} \!\frac{d\alpha}{\beta(\alpha)} \,
  \bm{\gamma}^h(M,m_t,\cos\theta,\alpha) \, .
\end{align}
The perturbative solutions to the above equations are reviewed in the Appendix.

We now turn to the evolution of the soft function. We derive its evolution equation by
using the RG invariance of the cross section,
\begin{align}
  \label{eq:RGinv}
  \frac{d}{d\ln\mu} \Tr\left[\bm{H}\bm{S}\right] \otimes \ff = 0 \, ,
\end{align}
along with the evolution equations for the hard function and parton luminosities. The
evolution equation for the hard function was given above, and the parton luminosity
functions satisfy the DGLAP equations \cite{Gribov:1972ri, Altarelli:1977zs,
  Dokshitzer:1977sg}. While the full DGLAP equations involve flavor mixing, what we need
here is the $x \to 1$ limit of them, which is flavor-diagonal and can be written as
\begin{align}
  \frac{d}{d\ln\mu} \ff(y,\mu) = 2\int_y^1 \frac{dx}{x} \, P(x) \, \ff(y/x,\mu) \, ,
\end{align}
where $P(x)$ is given by
\begin{align}
  \label{eq:SplittingF}
  P(x) = \frac{2\Gamma_{\text{cusp}}(\alpha_s)}{(1-x)_+} + 2\gamma^\phi(\alpha_s) \,
  \delta(1-x) \, .
\end{align}
The evolution for the momentum-space soft function is then
\begin{align}
  \frac{d}{d\ln\mu} \bm{S}(\omega,M,m_t,\cos\theta,\mu) &= - \left[
    2\Gamma_{\text{cusp}}(\alpha_s) \ln\frac{\omega}{\mu} +
    \bm{\gamma}^{s\dagger}(M,m_t,\cos\theta,\alpha_s) \right]
  \bm{S}(\omega,M,m_t,\cos\theta,\mu) \nonumber
  \\
  &\hspace{-4em} - \bm{S}(\omega,M,m_t,\cos\theta,\mu) \left[
    2\Gamma_{\text{cusp}}(\alpha_s) \ln\frac{\omega}{\mu} +
    \bm{\gamma}^{s}(M,m_t,\cos\theta,\alpha_s) \right] \nonumber
  \\
  &\hspace{-4em} - 4\Gamma_{\text{cusp}}(\alpha_s) \int_0^\omega d\omega' \,
  \frac{\bm{S}(\omega',M,m_t,\cos\theta,\mu)-\bm{S}(\omega,M,m_t,\cos\theta,\mu)}
  {\omega-\omega'} \, ,
\end{align}
where we have defined
\begin{align}
  \bm{\gamma}^s(M,m_t,\cos\theta,\alpha_s) = \bm{\gamma}^h(M,m_t,\cos\theta,\alpha_s) +
  2\gamma^{\phi}(\alpha_s) \, \bm{1} \, .
\end{align}
As in \cite{Becher:2007ty}, the non-local evolution equation for the soft function can be
turned into a local one by the Laplace transformation (\ref{eq:stilde}). The evolution
equation for the Laplace-transformed function reads
\begin{align}
  \label{eq:Sev}
  \frac{d}{d\ln\mu} \, \tilde{\bm{s}}
  &\left(\ln\frac{M^2}{\mu^2},M,m_t,\cos\theta,\mu\right) = \nonumber
  \\
  &- \left[ \Gamma_{\text{cusp}}(\alpha_s) \ln\frac{M^2}{\mu^2} +
    \bm{\gamma}^{s\dagger}(M,m_t,\cos\theta,\alpha_s) \right] \tilde{\bm{s}}
  \left(\ln\frac{M^2}{\mu^2},M,m_t,\cos\theta,\mu\right) \nonumber
  \\
  &- \tilde{\bm{s}} \left(\ln\frac{M^2}{\mu^2},M,m_t,\cos\theta,\mu\right) \left[
    \Gamma_{\text{cusp}}(\alpha_s) \ln\frac{M^2}{\mu^2} +
    \bm{\gamma}^s(M,m_t,\cos\theta,\alpha_s) \right] .
\end{align}
This can be solved using the same methods as for the hard function. Transforming the
results back to momentum space, we find
\begin{align}
  \bm{S}(\omega,M,m_t,\cos\theta,\mu_f) &= \sqrt{\hat s}\, \exp \left[ -4S(\mu_s,\mu_f) +
    4a_{\gamma^\phi}(\mu_s,\mu_f) \right] \nonumber
  \\
  &\hspace{-9em} \times \bm{u}^\dagger(M,m_t,\cos\theta,\mu_f,\mu_s) \,
  \tilde{\bm{s}}(\partial_\eta,M,m_t,\cos\theta,\mu_s) \,
  \bm{u}(M,m_t,\cos\theta,\mu_f,\mu_s) \, \frac{1}{\omega}
  \left(\frac{\omega}{\mu_s}\right)^{2\eta} \frac{e^{-2\gamma_E\eta}}{\Gamma(2\eta)} \, ,
\end{align}
where $\eta=2a_\Gamma(\mu_s,\mu_f)$. The soft scale $\mu_s$ should be chosen such that the
contribution from the soft function to the cross section is perturbatively well-behaved,
and will be discussed in detail in Section~\ref{sec:Pheno}.

Combining the results for the hard and soft functions, our final resummed expression for
the hard-scattering kernel is
\begin{align}
  \label{eq:MasterFormula}
  C(z,M,m_t,\cos\theta,\mu_f) &= \exp \big[ 4a_{\gamma^{\phi}}(\mu_s,\mu_f) \big]
  \nonumber
  \\
  &\hspace{-2em} \times \Tr \Bigg[ \bm{U}(M,m_t,\cos\theta,\mu_h,\mu_s) \,
  \bm{H}(M,m_t,\cos\theta,\mu_h) \, \bm{U}^\dagger(M,m_t,\cos \theta,\mu_h,\mu_s)
  \nonumber
  \\
  &\hspace{-2em} \times \tilde{\bm{s}}
  \left(\ln\frac{M^2}{\mu_s^2}+\partial_\eta,M,m_t,\cos\theta,\mu_s\right) \Bigg]
  \frac{e^{-2\gamma_E \eta}}{\Gamma(2\eta)} \frac{z^{-\eta}}{(1-z)^{1-2\eta}} \, .
\end{align}
For values $\mu_s<\mu_f$ the parameter $\eta<0$, and one must use a subtraction at $z=1$
and analytic continuation to express integrals over $z$ in terms of star (or plus)
distributions \cite{Bosch:2004th}. Formula~(\ref{eq:MasterFormula}) can be evaluated
order-by-order in RG-improved perturbation theory, using the standard counting $\ln
\mu_h/\mu_s\sim \ln (1-z)\sim 1/\alpha_s$. The perturbative solutions for the RG factors
needed to evaluate the evolution matrix $\bm{U}$ to NLO in this counting scheme are given
in (\ref{asol}), (\ref{eq:Ssol}), and (\ref{eq:usol}) of the Appendix. The correspondence
between this counting and the standard counting of logarithms (e.g.\ NLL, NNLL), along
with the accuracy of the anomalous dimensions and matching functions needed at a given
order, can be summarized as follows:

\vspace{1ex}
\begin{center}
  \begin{tabular}{|c|c|c|c|c|}
    \hline
    RG-improved PT & log accuracy & $\Gamma_{\text{cusp}}$ & $\bm{\gamma}^h$, $\gamma^\phi$ & $\bm{H}$,
    $\tilde{\bm{s}}$
    \\ \hline
    LO & NLL & 2-loop & 1-loop & tree-level
    \\ \hline       
    NLO & NNLL & 3-loop & 2-loop & 1-loop
    \\ \hline
  \end{tabular}
\end{center}
\vspace{1ex}
\noindent
In the remainder of the paper we will use the logarithmic counting (e.g.\ NNLL) when
referring to the resummed results obtained in this section. These results are valid for
the leading-order term in the threshold expansion in $(1-z)$, whereas the full result at
NLO in fixed-order perturbation theory also contains information on subleading terms. In
phenomenological applications we can match the resummed results with the NLO fixed-order
results to achieve an NLO+NNLL precision. The method for doing this is described in
Section~\ref{sec:Pheno}.

\subsection{Approximate NNLO results}
\label{sec:NNLOapprox}

In the previous subsection we derived a formula for the resummed differential cross
section, which is valid up to NNLL order. Starting from (\ref{eq:MasterFormula}), it is
also possible to obtain expressions for the differential cross section which are valid in
fixed-order perturbation theory \cite{Ahrens:2009uz}. Indeed, our results allow one to
obtain analytic expression for all of the coefficients multiplying singular plus
distributions in the variable $(1-z)$ appearing in the hard-scattering kernels up to NNLO.
With the same method, which is outlined below, it is also possible to determine
analytically, up to ${\mathcal O}(\alpha_s^4)$, the scale-dependent parts of the
coefficient multiplying $\delta(1-z)$.

In order to derive fixed-order formulas from (\ref{eq:MasterFormula}), we first set
$\mu_h=\mu_s=\mu_f=\mu$. In that case the evolution matrix $\bm{U}$ is equal to unity, and
$\eta = 2 a_\Gamma(\mu_f, \mu_s) \to 0$. The formula for the hard-scattering kernels then
becomes
\begin{align}
  \label{eq:FixedOrder}
  C(z,M,m_t,\cos\theta,\mu) = \tilde{c}(\partial_\eta,M,m_t,\cos\theta,\mu) \left(
    \frac{M}{\mu} \right)^{2\eta} \frac{e^{-2\gamma_E \eta}}{\Gamma(2\eta)}
  \frac{z^{-\eta}}{(1-z)^{1-2\eta}} \Bigg|_{\eta=0} \, ,
\end{align}
where 
\begin{align}
  \label{eq:ctilde}
  \tilde{c}(L,M,m_t,\cos\theta,\mu) = \Tr \Big[ \bm{H}(M,m_t,\cos\theta,\mu)\,
  \tilde{\bm{s}}(L,M,m_t,\cos\theta,\mu) \Big] \, .
\end{align}
By using (\ref{eq:Hev}) and (\ref{eq:Sev}) in combination with the analytic expressions
for the hard and soft functions at NLO, it is possible to determine all terms proportional
to $\ln\mu$ in the two-loop hard function $\bm{H}^{(2)}(M,m_t,\cos\theta,\mu)$, as well as
all terms proportional to $L$ in the two-loop soft function
$\tilde{\bm{s}}^{(2)}(L,M,m_t,\cos\theta,\mu)$. This information allows us to derive an
approximate expression for $\tilde{c}$ at NNLO. By inserting that formula for $\tilde{c}$
into (\ref{eq:FixedOrder}), we obtain the corresponding NNLO expression for the
hard-scattering kernel $C$. The results are conventionally written in terms of the plus
distributions
\begin{align}
  P_n(z) = \left[ \frac{\ln^n(1-z)}{1-z} \right]_+ .
\end{align}
However, the right-hand side of (\ref{eq:FixedOrder}) is more conveniently expressed in
terms of the distributions
\begin{align}
  P'_n(z) = \left[\frac{1}{1-z} \ln^n\left(\frac{M^2 (1-z)^2}{\mu^2 z}\right) \right]_+ .
\end{align}
It is possible to show that taking the derivatives with respect to $\eta$ and the limit
$\eta\to 0$ in (\ref{eq:FixedOrder}) is equivalent to making the following set of
replacements in $\tilde{c}(L,M,m_t,\cos\theta,\mu)$:
\begin{align}
  1 &\to \delta(1-z) \, , \nonumber
  \\
  L &\to 2 P'_0(z) + \delta(1-z) \ln\left(\frac{M^2}{\mu^2} \right) , \nonumber
  \\
  L^2 &\to 4 P'_1(z) + \delta(1-z) \ln^2\left(\frac{M^2}{\mu^2} \right) , \nonumber
  \\
  L^3 &\to 6 P'_2(z) - 4 \pi^2 P'_0(z) + \delta(1-z) \left[ \ln^3 \left(\frac{M^2}{\mu^2}
    \right) + 4 \zeta_3 \right] , \nonumber
  \\
  L^4 &\to 8 P'_3(z) -16 \pi^2 P'_1(z) + 128 \zeta_3 P'_0(z) +\delta(1-z) \left[\ln^4
    \left( \frac{M^2}{\mu^2} \right) + 16 \zeta_3 \ln \left( \frac{M^2}{\mu^2} \right)
  \right] .
\end{align}
In order to translate the $P'_n$ into the conventional $P_n$ distributions, we employ the
general relation
\\
\begin{align}
  P'_n(z) &= \sum_{k=0}^n \binom{n}{k} \ln^{n-k}\left(\frac{M^2}{\mu^2}\right) \Bigg[ 2^k
  P_k(z)
  \\
  &\quad\mbox{} + \sum_{j=0}^{k-1} \binom{k}{j} 2^j (-1)^{k-j} \left(
    \frac{\ln^j(1-z)\ln^{k-j}z}{1-z} - \delta(1-z) \int_0^1 dx
    \frac{\ln^j(1-x)\ln^{k-j}x}{1-x} \right) \Bigg] \, .\nonumber
\end{align}

The final result for the hard-scattering kernels at NNLO can be written as
\begin{align}
  \label{eq:C2}
  C^{(2)}(z,M,m_t,\cos\theta,\mu) &= D_3 \left[\frac{\ln^3 (1-z)}{1-z} \right]_+ + D_2
  \left[\frac{\ln^2(1-z)}{1-z} \right]_+ \nonumber
  \\
  &\quad + D_1 \left[\frac{\ln(1-z)}{1-z}\right]_+ + D_0 \left[\frac{1}{1-z} \right]_+ +
  C_0\,\delta(1-z) + R(z) \, .
\end{align}
The coefficients $D_0,\ldots, D_3$ and $C_0$ are functions of the variables $M,m_t,\cos
\theta,$ and $\mu$. The analytic expression for $D_i$ ($i=1,2,3$) were first derived in
\cite{Kidonakis:2003qe} starting from resummed formulas in Mellin moment space. In
\cite{Ahrens:2009uz} the coefficients $D_0,\dots, D_3$ were completely determined by
following the procedure outlined above. $D_0$ can be calculated in this way because the
process-dependent anomalous-dimension matrices in (\ref{eq:qqmatrix}) and
(\ref{eq:ggmatrix}) are now known up to NNLO. With the same method, it was possible to
calculate the scale dependence of $\delta$-function coefficient $C_0$. The function $R(z)$
is finite for $z \to 1$. The computation of the scale-independent part of $C_0$ requires
the knowledge of the hard and soft functions at two-loop order. As long as these are
missing, there is an ambiguity in $C_0$, associated with the fact that it is possible to
normalize the scale-dependent logarithms in an arbitrary way, i.e.\
\begin{align}
  \ln\left({\frac{\mu_0^2}{\mu^2}}\right) = \ln\left(\frac{\mu_1^2}{\mu^2} \right)
  +\ln\left(\frac{\mu_0^2}{\mu^2_1} \right) . \label{eq:shift}
\end{align}
Scale-independent terms proportional to the second logarithm on the right-hand side can be
absorbed into the unknown $\mu$-independent part of the coefficient function $C_0$.
Therefore, our numerical results for $C_0$ depend on the choice of the second mass scale,
which appears in the scale-dependent logarithms; we indicate this second scale by $\mu_0$.
In the next section we shall consider two different choices for $\mu_0$: $\mu_0 = M$
(scheme A), and $\mu_0 = m_t$ (scheme B). The situation is summarized in the following
formula:
\begin{align}
  C_0 = 
  \begin{cases}
    \displaystyle \sum_{i=0}^4 c^{\text{A}}_n \ln^n \frac{M^2}{\mu^2} & \text{(scheme A)}
    \\
    \displaystyle \sum_{i=0}^4 c^{\text{B}}_n \ln^n \frac{m_t^2}{\mu^2} & \text{(scheme
      B)}
  \end{cases}
\end{align}
The coefficients $c^{\text{A}}_n$ and $c^{\text{B}}_n$ ($n =1,2,3$) are known, while
$c^{\text{A}}_0$ and $c^{\text{B}}_0$ are unknown. In \cite{Ahrens:2009uz}, the explicit
expressions for the coefficients $D_0,\ldots,D_3$ and $C_0$ were collected in a {\tt
  Mathematica} file, which can be downloaded from the arXiv version of that work.

\section{Phenomenological applications}
\label{sec:Pheno}

In this section we perform a numerical study of the invariant mass distribution,
forward-backward asymmetry, and total cross section at the Tevatron and LHC. The main
purpose is to investigate the impact of the NNLL resummation and the approximate NNLO
corrections on the central values and perturbative uncertainties in these observables. The
higher-order corrections computed in this paper are limited to leading order in the
threshold expansion, whereas the exact NLO results in fixed order also contain subleading
terms in $(1-z)$. To make optimal use of our results, we match them onto the NLO
fixed-order expressions in such a way that these subleading corrections are fully taken
into account. For the resummed results, NLO+NNLL accuracy can be achieved by evaluating
differential cross sections according to
\begin{align}
  \label{eq:FixedMatching}
  d\sigma^{\text{NLO+NNLL}} &\equiv d\sigma^{\text{NNLL}} \Big|_{\mu_h,\mu_s,\mu_f} +
  d\sigma^{\text{NLO, subleading}} \Big|_{\mu_f} \nonumber
  \\
  &\equiv d\sigma^{\text{NNLL}} \Big|_{\mu_h,\mu_s,\mu_f} + \left( d\sigma^{\text{NLO}}
    \Big|_{\mu_f} - d\sigma^{\text{NLO, leading}} \Big|_{\mu_f} \right) ,
\end{align}
where $d\sigma^{\text{NLO}}$ is the exact result in fixed order, and
$d\sigma^{\text{NLO,leading}}|_{\mu_f} \equiv d\sigma^{\text{NNLL}}|_{\mu_h=\mu_s=\mu_f}$
captures the leading singular terms in the threshold limit $z \to 1$ at NLO. The term
$d\sigma^{\text{NLO, subleading}}$ is of subleading order in $(1-z)$ and ensures that the
total result reduces to the exact fixed-order result when all the scales are set equal. It
also makes the result invariant under variations of the factorization scale, up to terms
at NNLO in the perturbative expansion, even at subleading order in $(1-z)$. To obtain
approximate NNLO results in fixed order, we simply add the NNLO correction onto the exact
NLO results, i.e.,
\begin{equation}
  \label{eq:NNLOapprox}
  d\sigma^{\text{NNLO, approx}} = d\sigma^{\text{NLO}} + d\sigma^{\text{(2), approx}} \, ,
\end{equation}
where $d\sigma^{\text{(2), approx}}$ is the NNLO correction to the differential cross
section obtained using the coefficient function (\ref{eq:C2}), in either scheme~A or B
described at the end of the previous section.

\begin{figure}
\begin{center}
\begin{tabular}{lr}
\psfrag{x}[][][1][90]{$d\sigma/dM$ [fb/GeV]}
\psfrag{y}[]{$M$ [GeV]}
\psfrag{z}[][][0.85]{$\sqrt{s}=1.96$\,TeV}
\includegraphics[width=0.43\textwidth]{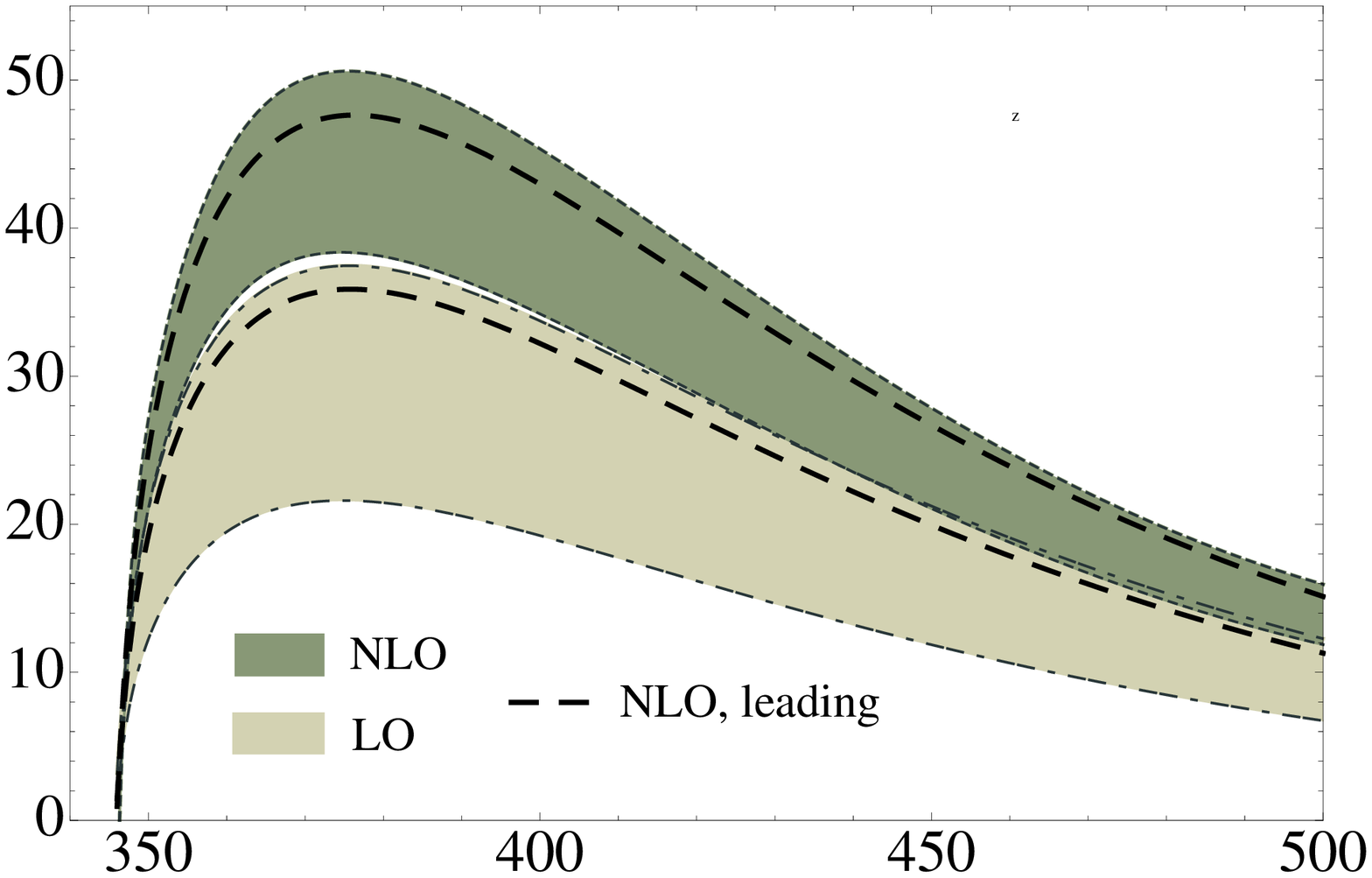}
&
\psfrag{x}[][][1][90]{$d\sigma/dM$ [pb/TeV]}
\psfrag{y}[]{$M$ [GeV]}
\psfrag{z}[][][0.85]{$\sqrt{s}=7$\,TeV}
\includegraphics[width=0.434\textwidth]{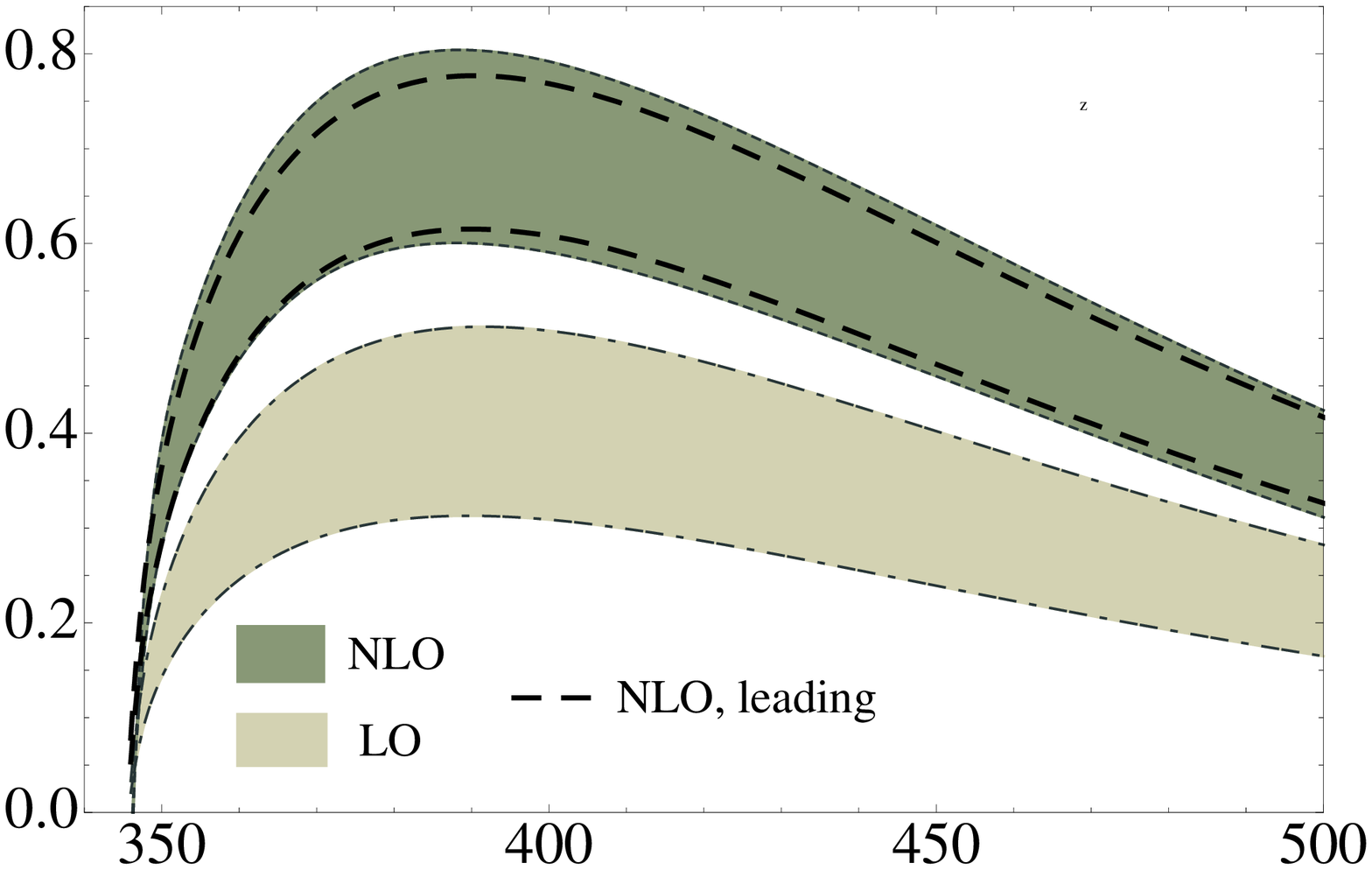}
\end{tabular}
\end{center}
\vspace{-1ex}
\caption{\label{fig:subl_terms} Fixed-order predictions for the invariant mass spectrum at
  LO (light bands) and NLO (dark bands) for the Tevatron (left) and LHC (right). We use
  MSTW2008NLO PDFs \cite{Martin:2009bu} with $\alpha_s(M_Z)=0.120$. The width of the bands
  reflects the uncertainty of the spectrum under variations of the matching and
  factorization scales. The dashed lines refer to the leading terms in the threshold
  expansion.}
\end{figure}

It is important to examine the relative size of the leading and subleading terms in
(\ref{eq:FixedMatching}). If the subleading terms were comparable in size to the leading
ones, it would make little sense to resum the logarithms in $(1-z)$, which we have
discussed so far, or to construct approximate fixed-order formulas at NNLO which capture
only the effects of the singular terms. The naive expectation would be that the singular
terms are dominant only when $\tau=M^2/s\to 1$, since then the integrand in
(\ref{eq:genfact}) is needed only in the $z \to 1$ limit and the less singular terms are
clearly subleading. However, the most interesting region for phenomenology ranges from
$M\sim 2m_t$ to around 1\,TeV at the Tevatron and up to several TeV at the LHC, which
corresponds to $\tau<0.3$ (at most). For the leading-order singular terms to be dominant,
it is necessary that the parton luminosity functions $\ff_{ij}(\tau/z,\mu)$ fall off
sufficiently fast for $\tau/z\to 1$ that only the largest values of $z$ give significant
contributions to the integrand, an effect referred to in \cite{Becher:2007ty,
  Ahrens:2008qu, Ahrens:2008nc} as dynamical threshold enhancement. The study of the
invariant mass spectrum in \cite{Ahrens:2009uz} indicated that such an enhancement
actually does take place in the range of invariant mass from near the peak of the
distribution at $M\sim 380$\,GeV up to much higher values. This is illustrated in
Figure~\ref{fig:subl_terms}, which shows the invariant mass distributions at the Tevatron
and LHC predicted using different approximations in fixed-order perturbation theory. The
difference between the boundaries of the dark NLO bands and the dashed lines is due to the
small contributions from the subleading terms $d\sigma^{\rm NLO, subleading}$ in
(\ref{eq:FixedMatching}). The fact that, even at these relatively low values of $M$, the
leading terms provide a very good approximation to the full NLO result provides a strong
motivation to study within our formalism higher-order corrections to integrated quantities
such as the total cross section and forward-backward asymmetry, which receive their
dominant contributions from low values of the invariant mass.

%%%%%%%%%%%%%%%%%%%%%%%%%%%%%%%%%%%%%%%%%%%%%%%%%%%%%%%%%%%%%%%%%%%%%%%%%%%%%%%%
%%%%%%%%%%%%%%%%%%%%%%%%%%%%%%%%%%%%%%%%%%%%%%%%%%%%%%%%%%%%%%%%%%%%%%%%%%%%%%%%

We will always do the matching onto fixed-order results as in (\ref{eq:FixedMatching}) and
(\ref{eq:NNLOapprox}), when the goal is to provide quantitative phenomenological
predictions. Such a matching is straightforward for integrated quantities such as the
total cross section and forward-backward asymmetry, since the NLO results in fixed order
are available in analytic form. For the invariant mass distribution, on the other hand,
the fixed-order NLO results are available in the form of Monte Carlo programs such as MCFM
\cite{Campbell:2000bg}. This makes it difficult to get accurate values of the top-quark
pair invariant mass spectrum at high $M$, where the differential cross section is small,
and makes it impractical to calculate the spectrum with the scale choice $\mu_f=M$ used in
the next section, since doing so would require to run the program separately at each point
in $\mu_f$. (Monte Carlo programs generate the invariant mass spectrum by first producing
a set of events for a given $\mu_f$, and then grouping them into bins in $M$). When we
study certain aspects of the invariant mass distribution in
Section~\ref{subsec:InvMassSyst}, we will take the NLO correction in the threshold
approximation, so that (\ref{eq:FixedMatching}) and (\ref{eq:NNLOapprox}) are evaluated
with $d\sigma^{\text{NLO}} \to d\sigma^{\text{NLO, leading}}$. This is still a good
approximation to the full NLO result, and allows us to study the qualitative behavior of
the invariant mass spectrum with $\mu_f=M$ over a large range of $M$, as well as PDF
uncertainties, in a simple way. For this purpose, we also define an NNLO approximation
which includes only the singular terms at threshold in the NLO correction:
\begin{align}
  \label{eq:NNLOthresh}
  d\sigma^{\text{NNLO, leading}} = d\sigma^{\text{NLO, leading}} + d\sigma^{\text{(2),
      approx}} \, .
\end{align}

\subsection{Invariant mass distribution: Systematic studies}
\label{subsec:InvMassSyst}

%%%%%%%%%%%%%%%%%%%%%%%%%%%%%%%%%%%%%%%%%%%%%%%%%%%%%%%%%%%%%%%%%%%%%%%%%%%%%%%%
\begin{figure}[t]
\begin{center}
\begin{tabular}{lr}
\psfrag{x}[][][1][90]{$\alpha_s$ correction}
\psfrag{y}[]{$\mu_s/M$}
\psfrag{z}[][][0.85]{$\sqrt{s}=1.96$\,TeV}
\includegraphics[width=0.433\textwidth]{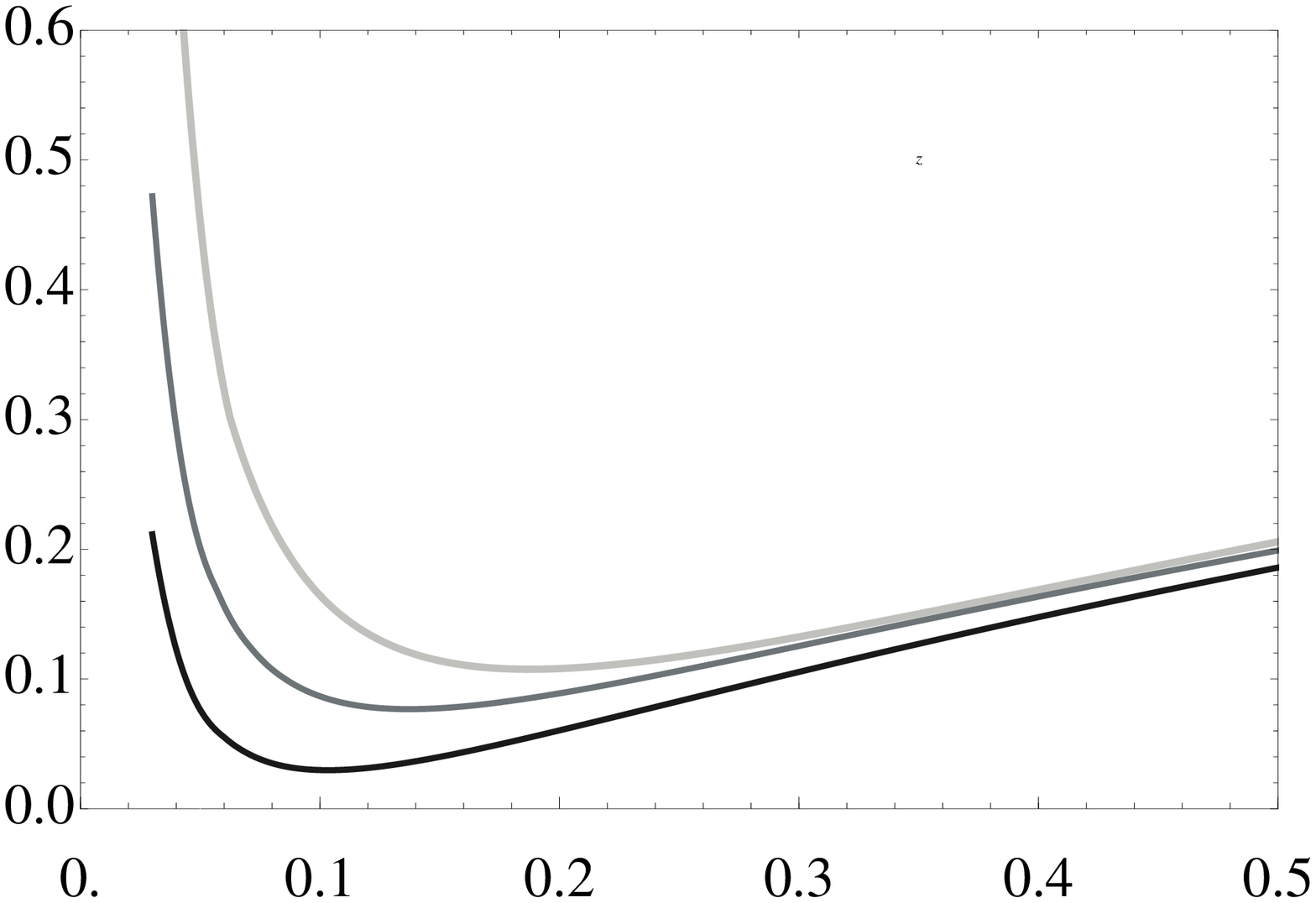}
& 
\psfrag{x}[][][1][90]{$\mu_s/M$}
\psfrag{y}[]{$M$ [GeV]}
\psfrag{z}[][][0.85]{$\sqrt{s}=1.96$\,TeV}
\includegraphics[width=0.427\textwidth]{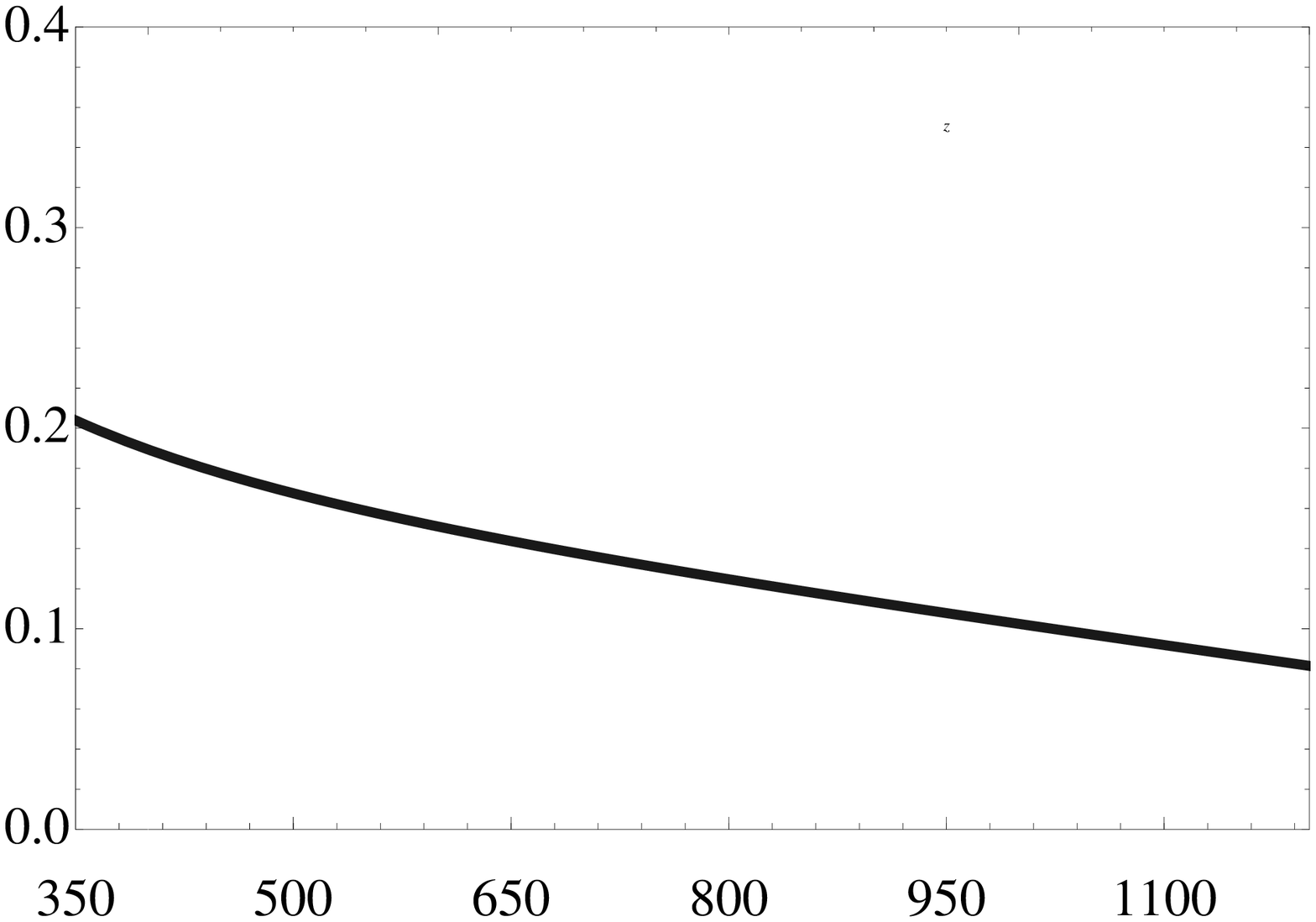}
\\[3mm]
\psfrag{x}[][][1][90]{$\alpha_s$ correction}
\psfrag{y}[]{$\mu_s/M$}
\psfrag{z}[][][0.85]{$\sqrt{s}=7$\,TeV}
\includegraphics[width=0.433\textwidth]{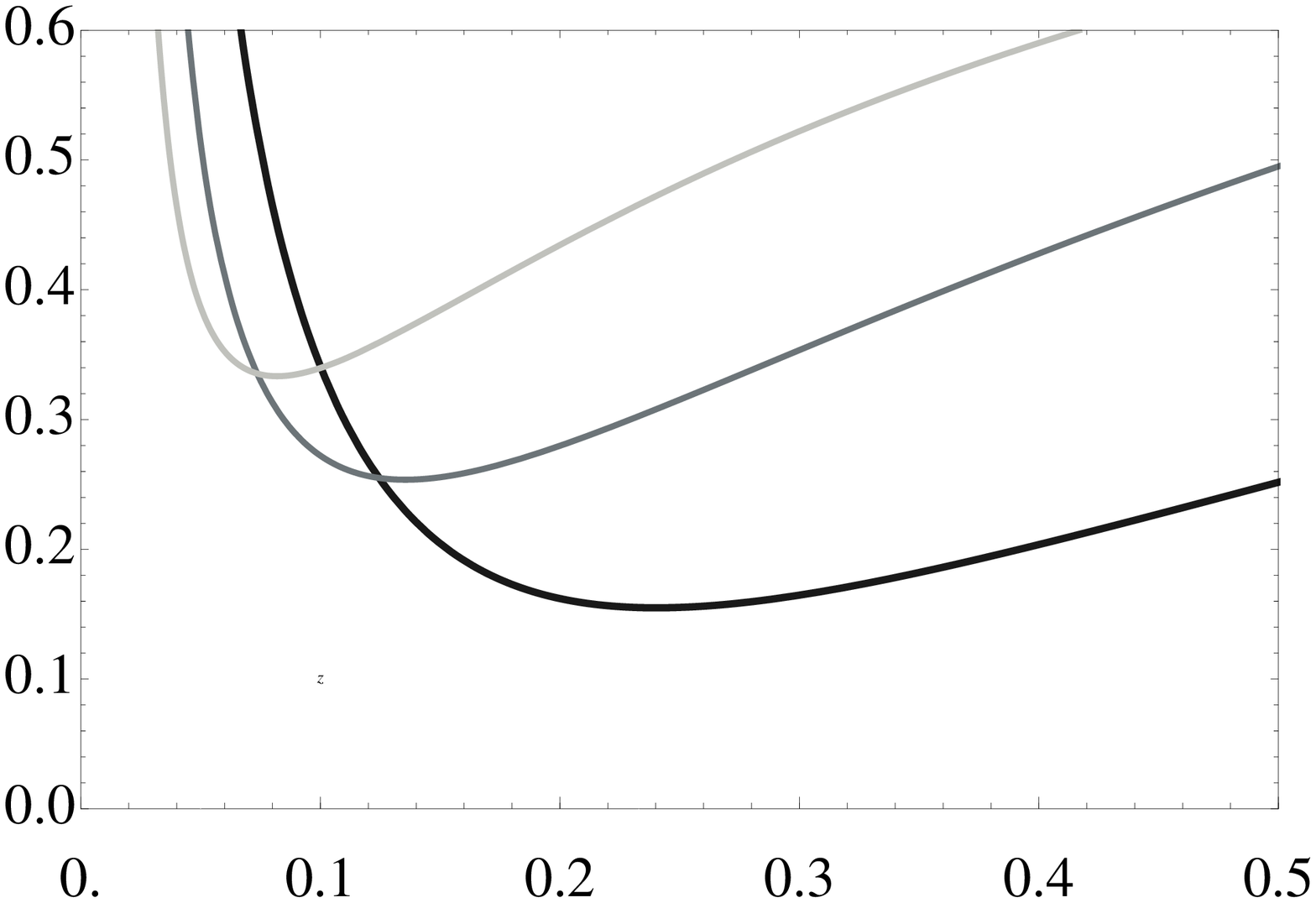}
& 
\psfrag{x}[][][1][90]{$\mu_s/M$}
\psfrag{y}[]{$M$ [GeV]}
\psfrag{z}[][][0.85]{$\sqrt{s}=7$\,TeV}
\includegraphics[width=0.427\textwidth]{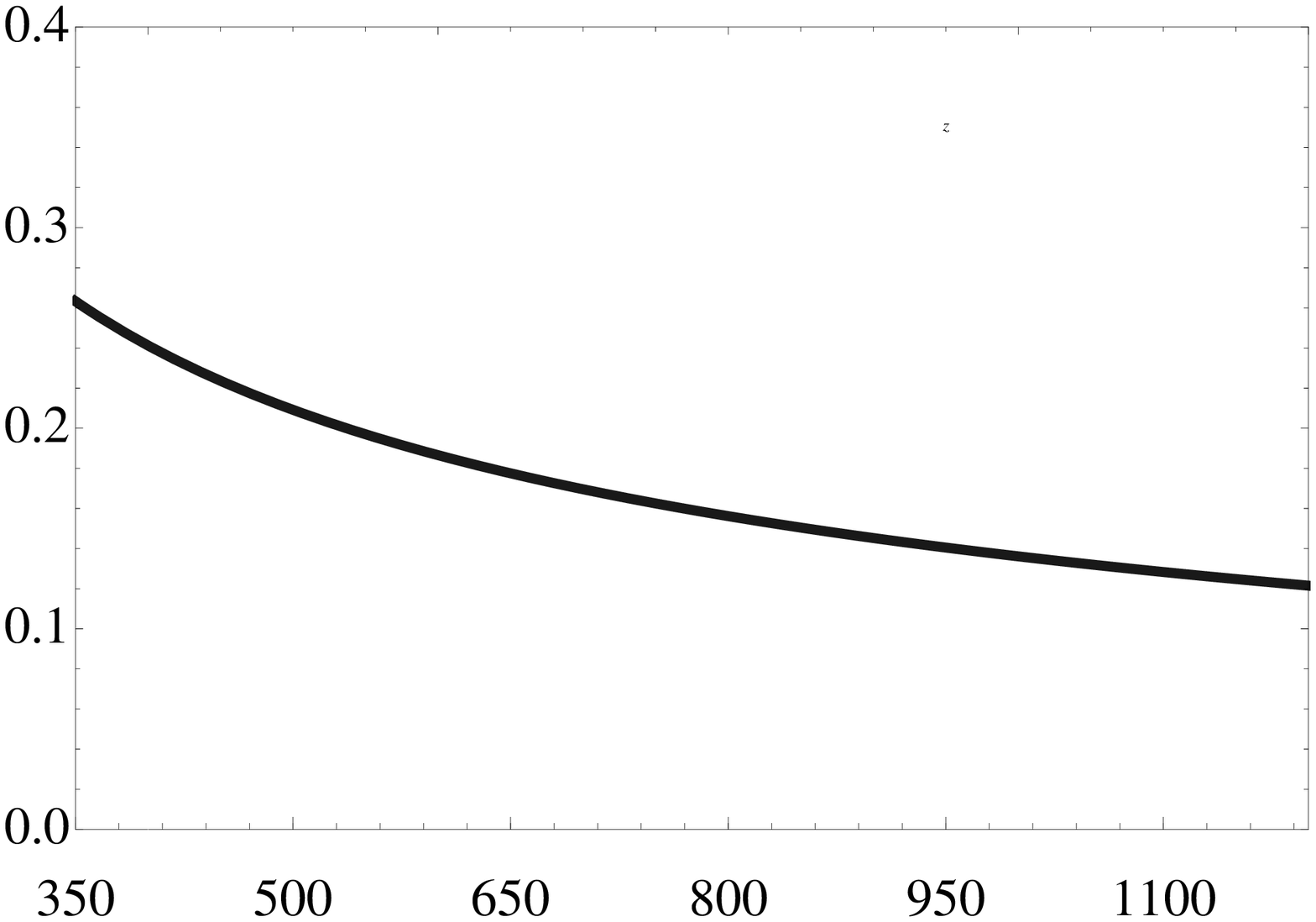} 
\end{tabular}
\end{center}
\vspace{-1ex}
\caption{\label{fig:SoftScales} Top left: Ratio of the one-loop correction from the soft
  function over the leading-order result for top pair production at the Tevatron, as a
  function of $\mu_s/M$, for $M=400$\,GeV (dark), $M=700$\,GeV (medium), and $M=1000$\,GeV
  (light). Top right: The scale $\mu_s/M$ determined by the point where the one-loop
  correction from the soft function is minimal, as a function of the invariant mass $M$.
  Bottom: Analogous plots for the LHC, but with $M=400$\,GeV (dark), $M=1000$\,GeV
  (medium), and $M=2000$\,GeV (light).}
\end{figure}
%%%%%%%%%%%%%%%%%%%%%%%%%%%%%%%%%%%%%%%%%%%%%%%%%%%%%%%%%%%%%%%%%%%%%%%%%%%%%%%%

The invariant mass distribution is obtained by integrating the doubly differential rate
over the range $-1< \cos\theta <1$. The resummed results (\ref{eq:MasterFormula}) depend
on the three scales $\mu_s,\mu_h$, and $\mu_f$, and to give a numerical result we must
first specify how to choose them. In the similar cases of Drell-Yan \cite{Becher:2007ty}
and Higgs production \cite{Ahrens:2008qu, Ahrens:2008nc} at threshold, the soft and hard
scales were chosen by examining the contributions of the one-loop soft and hard matching
coefficients as functions of the scales $\mu_s$ and $\mu_h$, and then choosing default
values of the scales in such a way as to minimize these corrections. We shall use this
approach here, a small complication being the extra dependence on the kinematic variable
$\cos\theta$ in (\ref{eq:MasterFormula}). For the analysis of this section we use the
default set of MSTW2008NNLO PDFs \cite{Martin:2009bu}, take $\alpha_s(M_Z)=0.117$ with
three-loop running in the $\overline{\rm{MS}}$ scheme with five active flavors, and employ
the value $m_t=173.1$\,GeV for the top-quark mass defined in the pole scheme. Using a
fixed set of PDFs helps to elucidate more clearly the behavior of the perturbative
expansion of the hard-scattering kernels in higher orders of perturbation theory. When
presenting our phenomenological results in
Sections~\ref{subsec:InvMassPheno}--\ref{sec:Afb}, we will change the PDF sets according
to the order of perturbation theory employed, and we will also study the theoretical
uncertainties related to the parameterization of the PDFs. Finally, in
Section~\ref{sec:topmass}, we will explore different schemes for the definition of the
top-quark mass and study their impact on the phenomenological results.

%%%%%%%%%%%%%%%%%%%%%%%%%%%%%%%%%%%%%%%%%%%%%%%%%%%%%%%%%%%%%%%%%%%%%%%%%%%%%%%%
\begin{figure}
\begin{center}
\begin{tabular}{lr}
\psfrag{x}[][][1][90]{$\alpha_s$ correction}
\psfrag{y}[]{$\mu_h/M$}
\psfrag{z}[][][0.85]{$\sqrt{s}=1.96$\,TeV}
\includegraphics[width=0.43\textwidth]{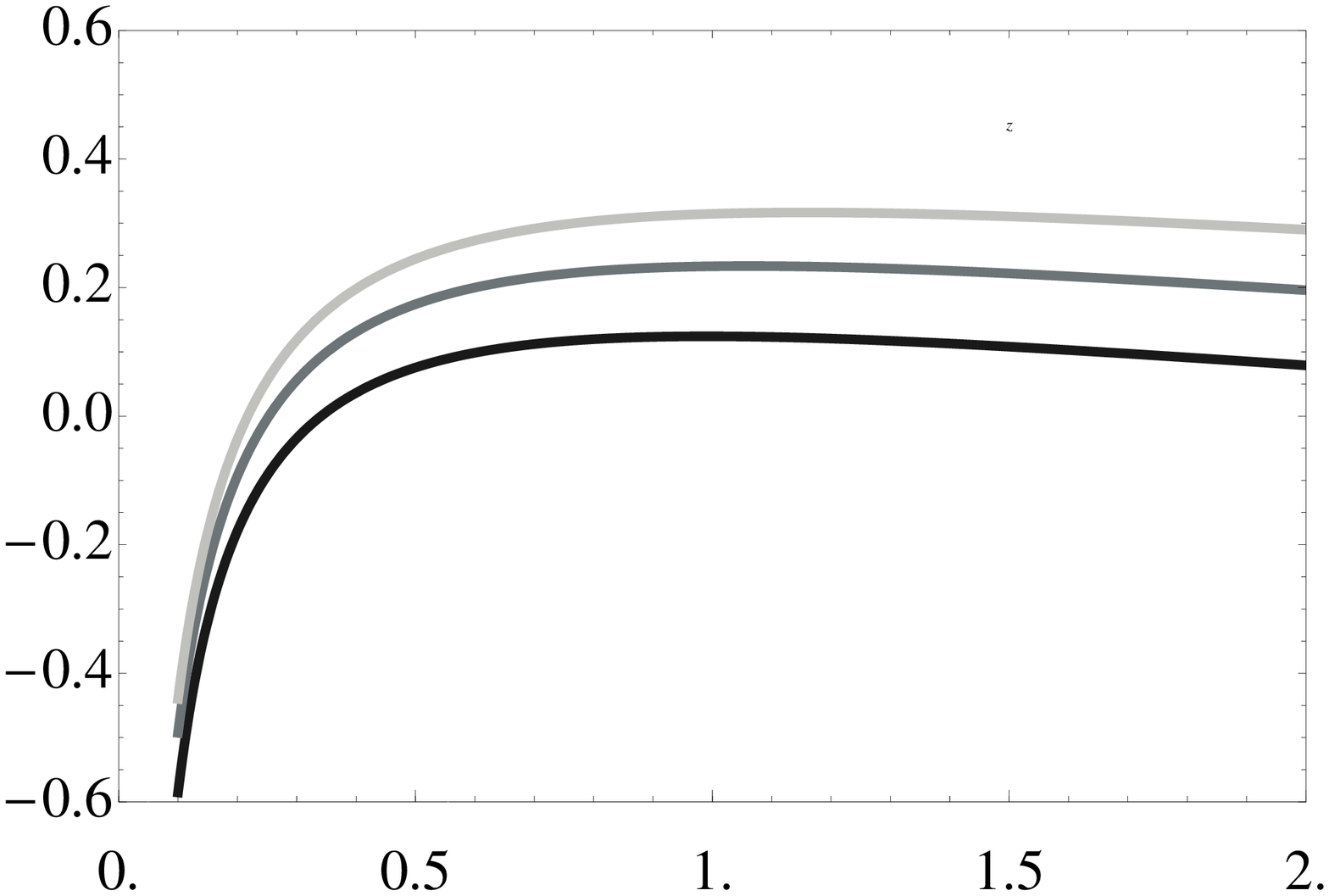}
& 
\psfrag{x}[][][1][90]{$\alpha_s$ correction}
\psfrag{y}[]{$\mu_h/M$}
\psfrag{z}[][][0.85]{$\sqrt{s}=7$\,TeV}
\includegraphics[width=0.43\textwidth]{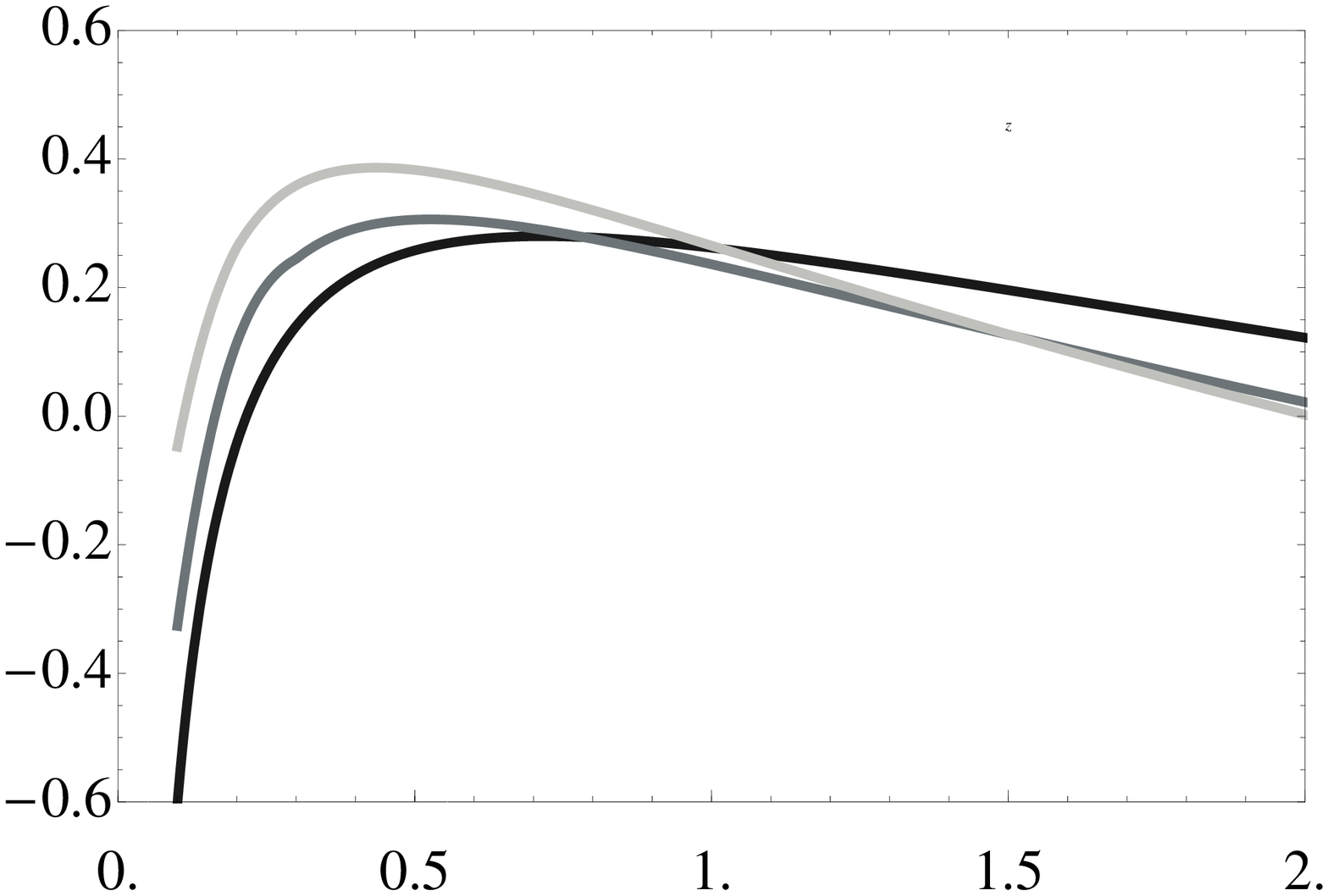}
\end{tabular}
\end{center}
\vspace{-1ex}
\caption{\label{fig:HardScales} Left: Ratio of the one-loop correction the from hard
  function over the leading-order result at the Tevatron, as a function of $\mu_h/M$, at
  $M=400$\,GeV (dark), $M=700$\,GeV (medium), and $M=1000$\,GeV (light). Right: Analogous
  plot for the LHC, with $M=400$\,GeV (dark), $M=1000$\,GeV (medium), and $M=2000$\,GeV
  (light).}
\end{figure}
%%%%%%%%%%%%%%%%%%%%%%%%%%%%%%%%%%%%%%%%%%%%%%%%%%%%%%%%%%%%%%%%%%%%%%%%%%%%%%%%

\subsubsection{Determination of the matching and factorization scales}

We begin by examining the corrections from the NLO soft matching coefficient as a function
of $\mu_s$. We isolate this contribution by picking out the piece of the NNLL
approximation to (\ref{eq:MasterFormula}) proportional to $\widetilde{\bm{s}}^{(1)}$,
evaluating the differential cross section using only this piece, and dividing the result
by that at NLL, for the choice $\mu_f=\mu_h=M$. The results are shown in the left-hand
plots of Figure~\ref{fig:SoftScales} for the Tevatron and the LHC with $\sqrt{s}=7$\,TeV,
for several different values of $M$. We note that the corrections are larger at the LHC
than at the Tevatron, especially at high values of $M$. This behavior appears to be a
property of the gluon channel, which gives the dominant contribution at the LHC. The
correction is generally at its minimum between $M/4$ and $M/10$, and moves to lower values
of $\mu_s$ at higher values of $M$. The exact position of the minimum as a function of $M$
is shown in the right-hand plots of Figure~\ref{fig:SoftScales}. To a good approximation,
the numerical results for $\mu_s$ can be fitted by the function
\begin{equation}
  \label{eq:SoftScale}
  \mu^{\rm def}_s=\frac{M(1-\tau)}{(a+b \, \tau^{1/4})^c}
\end{equation}
with $a=-33$, $b=150$, and $c=0.46$ for the Tevatron, and $a=-1.3$, $b=23$, and $c=0.98$
for the LHC at $\sqrt{s}=7$\,TeV. In Section~\ref{sec:totcross} we will also study the
total cross section at the LHC with $\sqrt{s}=10$, 14\,TeV, and in those cases we use
$a=0.95$, $b=6.7$, and $c=1.6$.

%%%%%%%%%%%%%%%%%%%%%%%%%%%%%%%%%%%%%%%%%%%%%%%%%%%%%%%%%%%%%%%%%%%%%%%%%%%%%%%%
\begin{figure}[t]
\begin{center}
\begin{tabular}{lr}
\psfrag{x}[][][1][90]{$d\sigma/dM$ [fb/GeV]}
\psfrag{y}[]{$\mu_f$ [GeV]}
\psfrag{z}[][][0.8]{$\sqrt{s}=1.96$\,TeV; $M=400$\,GeV}
\includegraphics[width=0.43\textwidth]{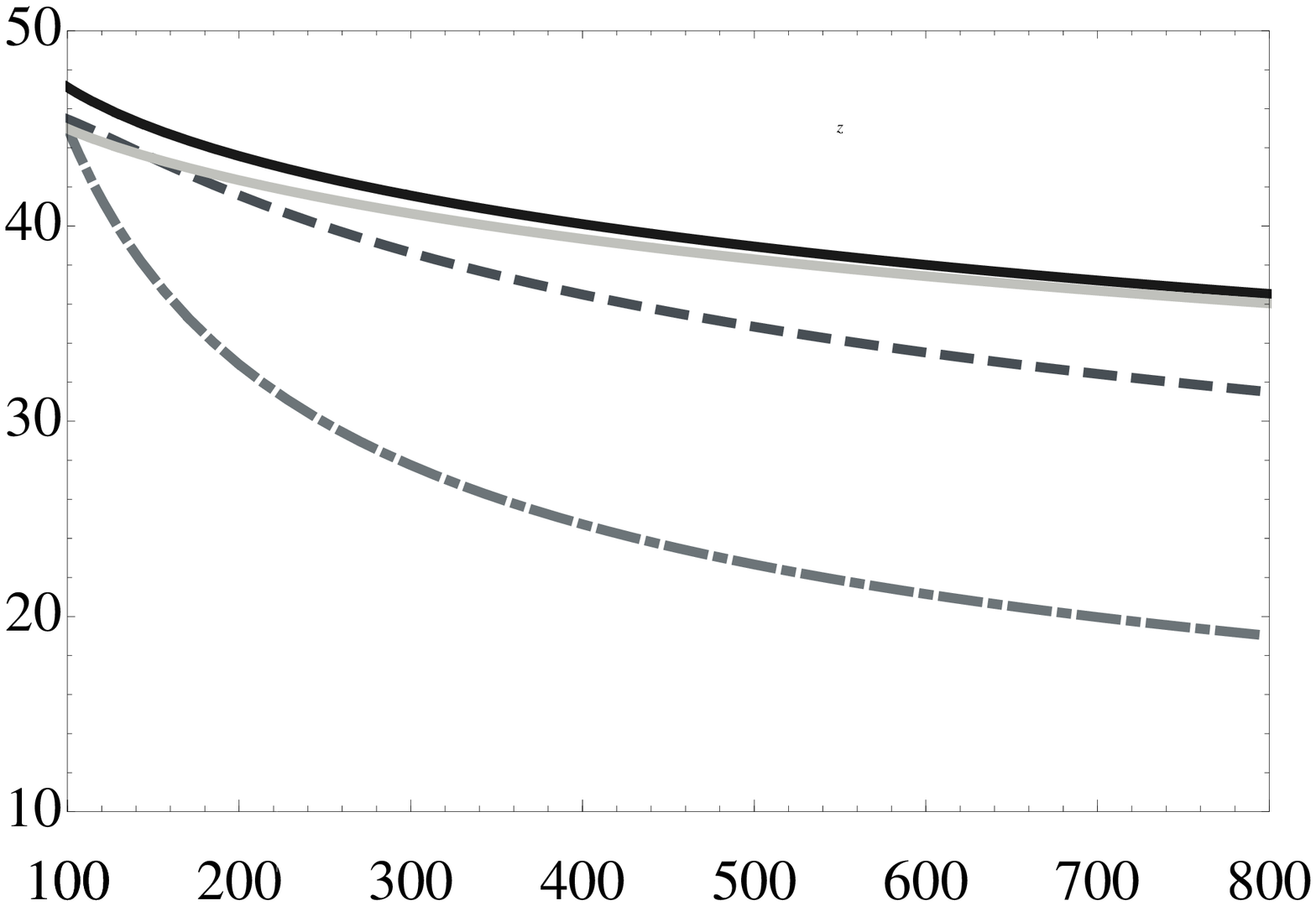}
& 
\psfrag{x}[][][1][90]{$d\sigma/dM$ [fb/GeV]}
\psfrag{y}[]{$\mu_f$ [GeV]}
\psfrag{z}[][][0.8]{$\!\!\!\!\!\!\!\!\sqrt{s}=1.96$\,TeV; $M=1$\,TeV}
\includegraphics[width=0.45\textwidth]{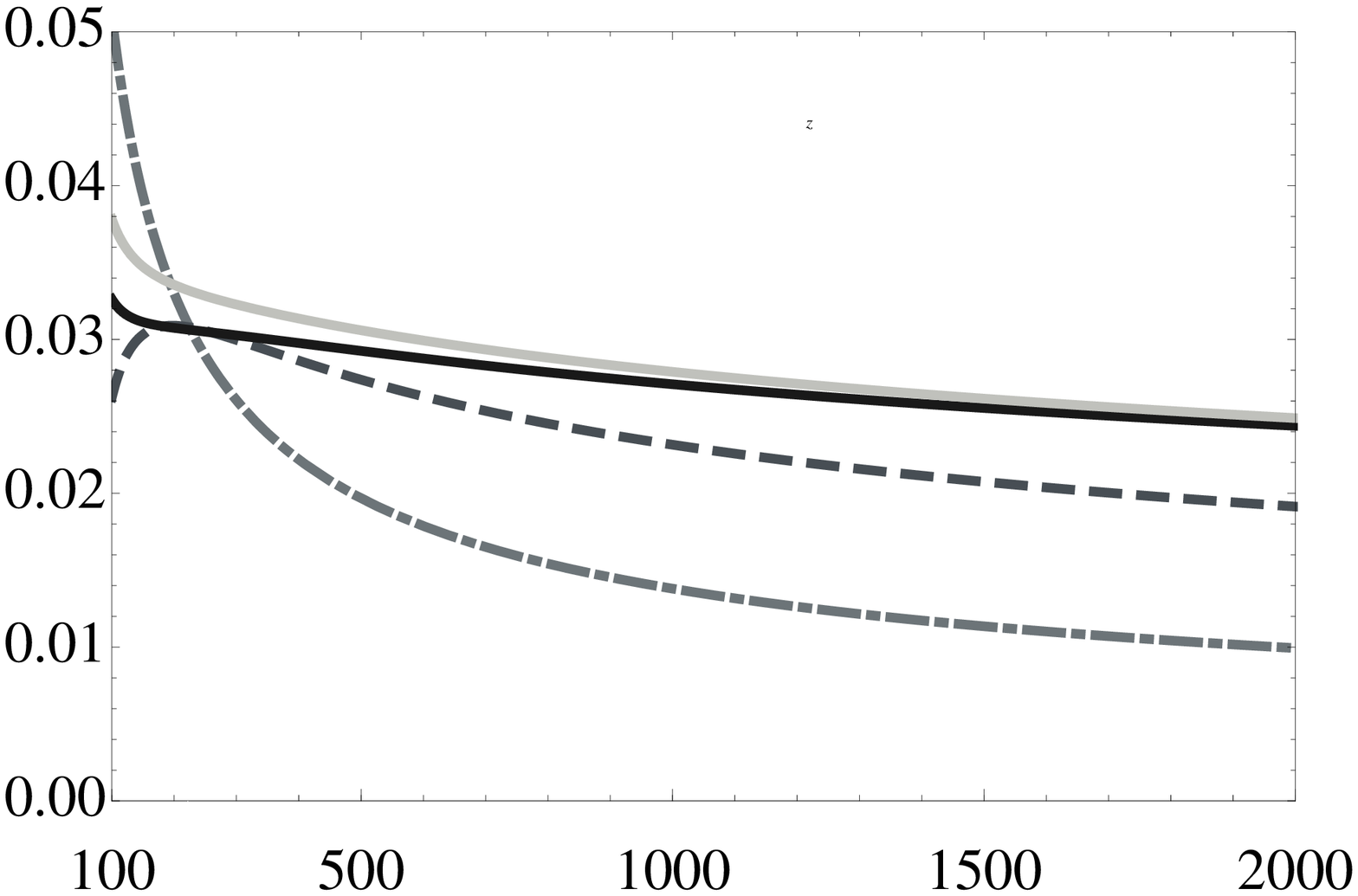}
\\[3mm]
\psfrag{x}[][][1][90]{$d\sigma/dM$ [fb/GeV]}
\psfrag{y}[]{$\mu_f$ [GeV]}
\psfrag{z}[][][0.8]{$\sqrt{s}=7$\,TeV; $M=400$\,GeV}
\includegraphics[width=0.43\textwidth]{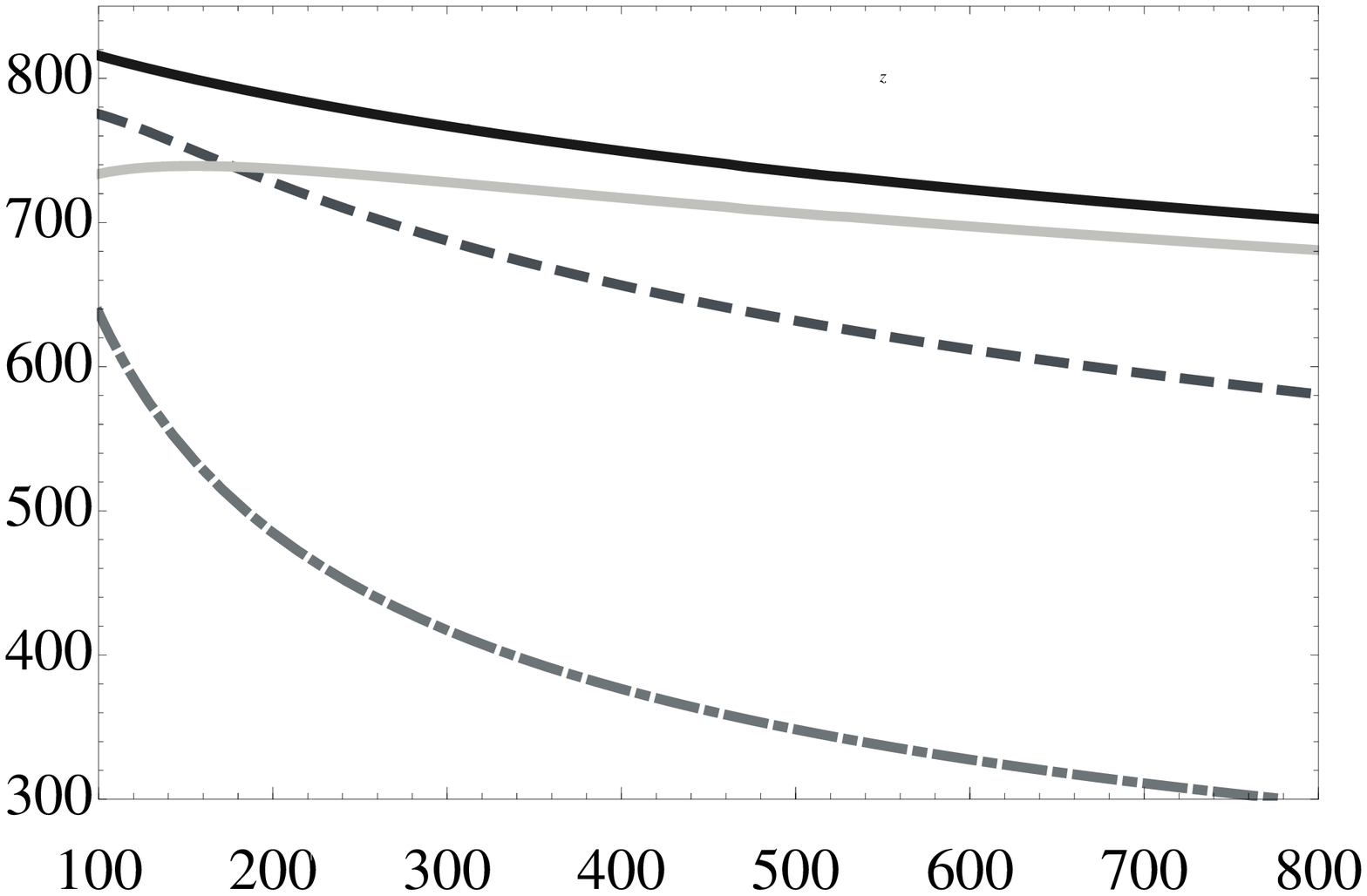}
& 
\psfrag{x}[][][1][90]{$d\sigma/dM$ [fb/GeV]}
\psfrag{y}[]{$\mu_f$ [GeV]}
\psfrag{z}[][][0.8]{$\sqrt{s}=7$\,TeV; $M=1$\,TeV}
\includegraphics[width=0.43\textwidth]{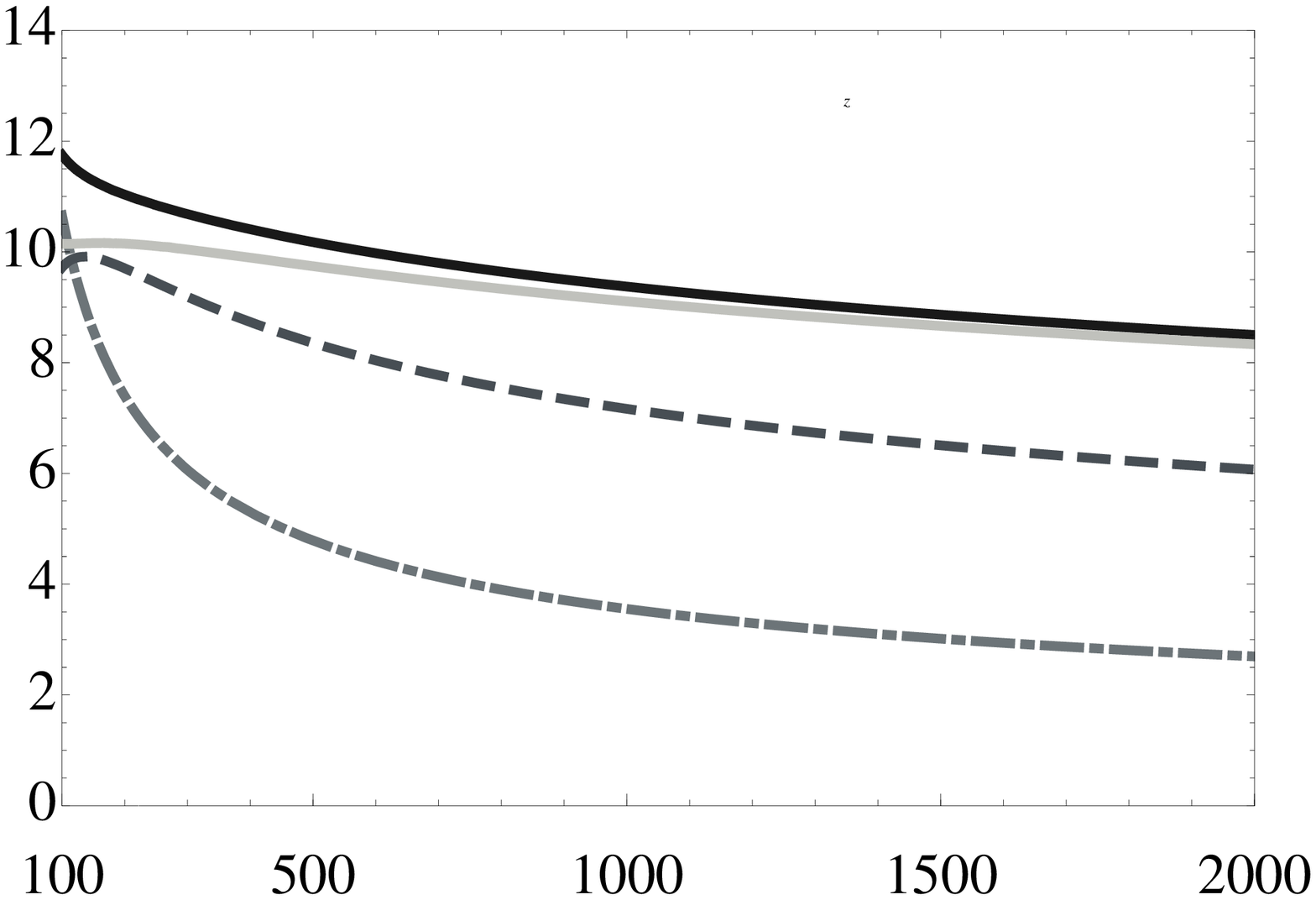}
\end{tabular}
\end{center}
\vspace{-1ex}
\caption{\label{fig:ScaleVarFixed} Dependence of $d\sigma/dM$ at the Tevatron (left) and
  LHC (right) on the scale $\mu_f$ in fixed-order perturbation theory. The dashed-dotted
  lines show $d\sigma^{\text{LO}}$, the dashed lines $d\sigma^{\text{NLO, leading}}$, and
  the dark (light) solid lines the approximate threshold expansion (\ref{eq:NNLOthresh})
  at NNLO in scheme~A (scheme~B).}
\end{figure}
%%%%%%%%%%%%%%%%%%%%%%%%%%%%%%%%%%%%%%%%%%%%%%%%%%%%%%%%%%%%%%%%%%%%%%%%%%%%%%%%

%%%%%%%%%%%%%%%%%%%%%%%%%%%%%%%%%%%%%%%%%%%%%%%%%%%%%%%%%%%%%%%%%%%%%%%%%%%%%%%%
\begin{figure}[t]
\begin{center}
\begin{tabular}{ll}
\psfrag{x}[][][1][90]{$d\sigma/dM$ [fb/GeV]}
\psfrag{y}[]{$\mu_f$ [GeV]}
\psfrag{z}[][][0.85]{$M=400$\,GeV}
\includegraphics[width=0.43\textwidth]{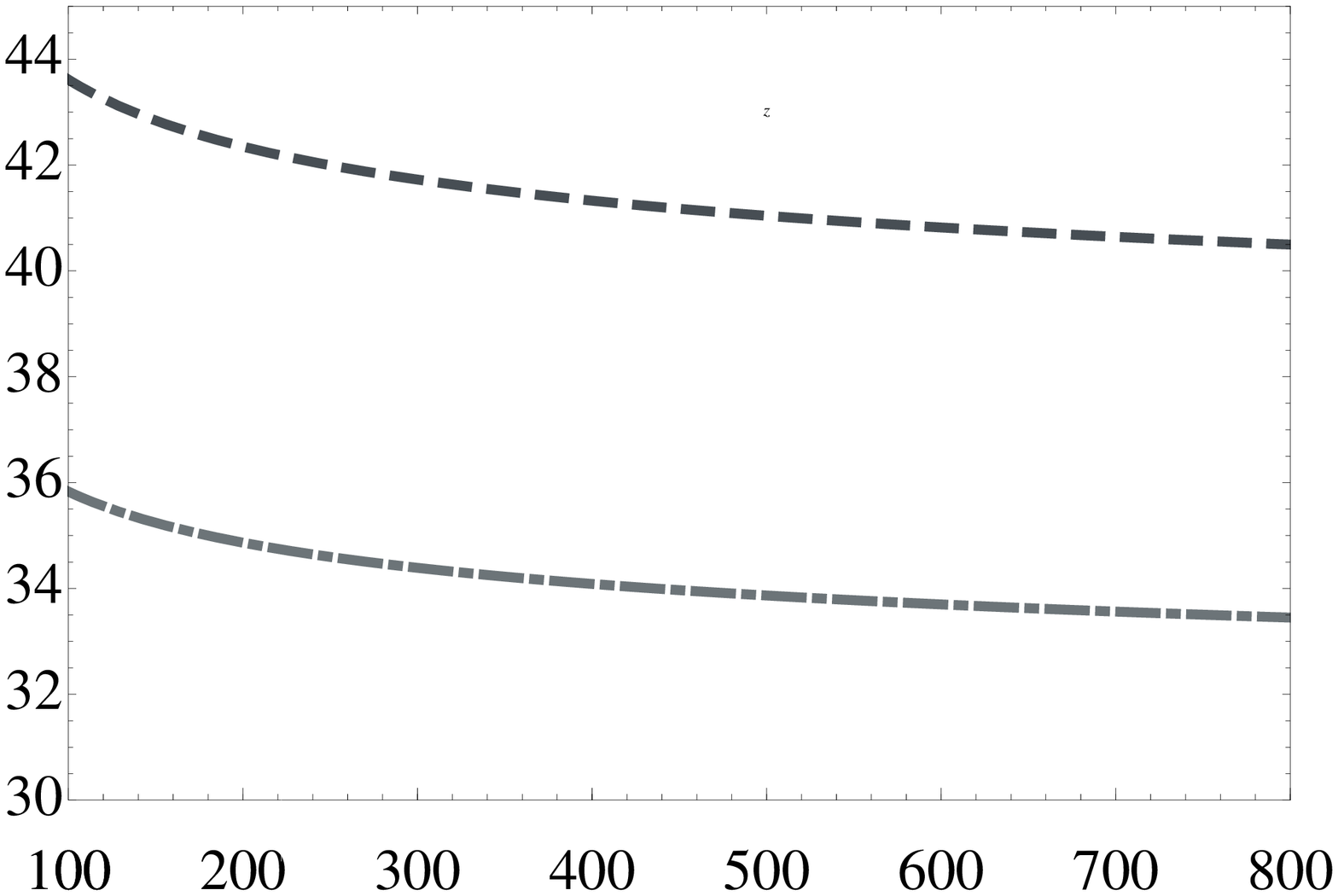}
& 
\psfrag{x}[][][1][90]{$d\sigma/dM$ [fb/GeV]}
\psfrag{y}[]{$\mu_f$ [GeV]}
\psfrag{z}[][][0.85]{$M=1$\,TeV}
\includegraphics[width=0.43\textwidth]{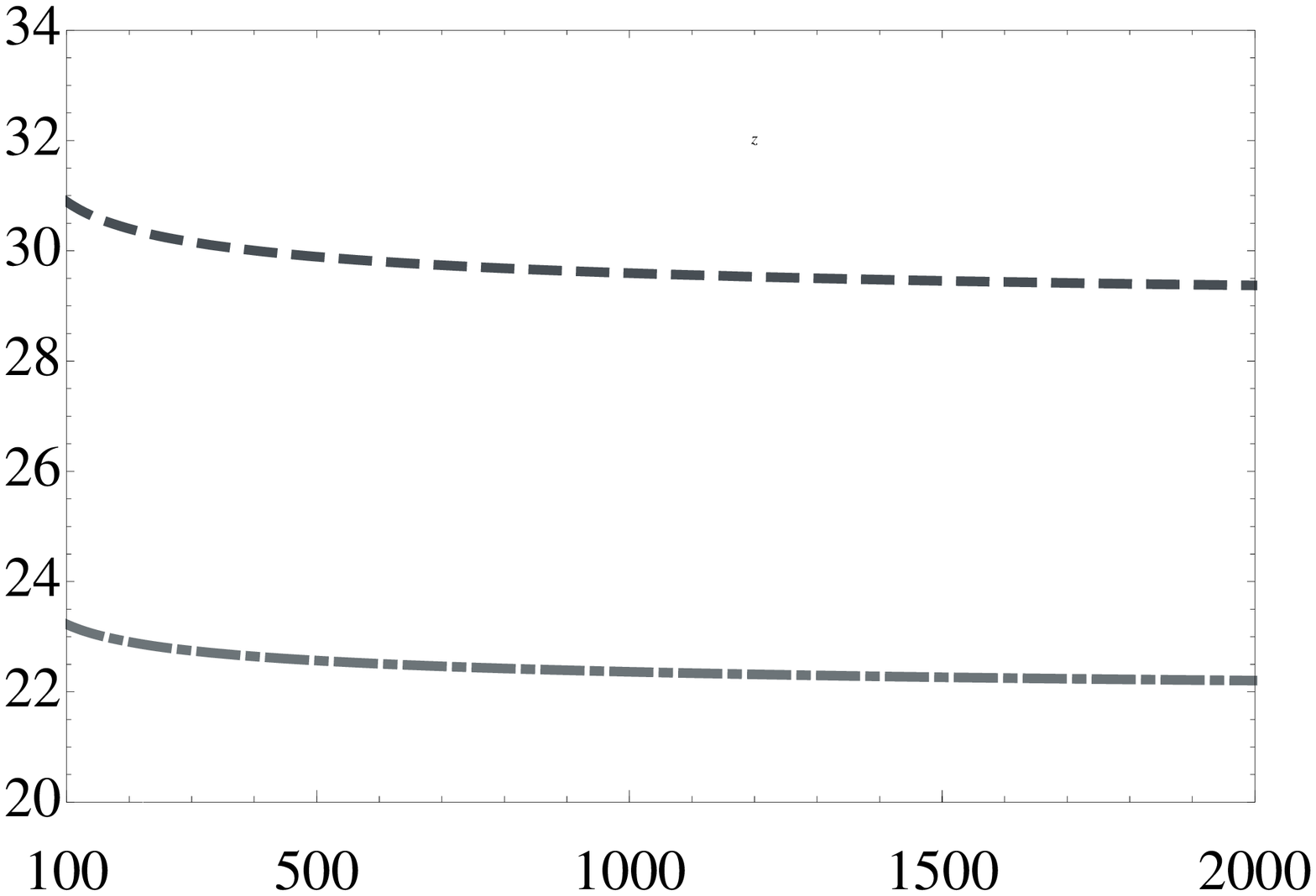}
\\[3mm]
\psfrag{x}[][][1][90]{$d\sigma/dM$ [fb/GeV]}
\psfrag{y}[]{$\mu_h$ [GeV]}
\psfrag{z}[][][0.85]{$M=400$\,GeV}
\includegraphics[width=0.43\textwidth]{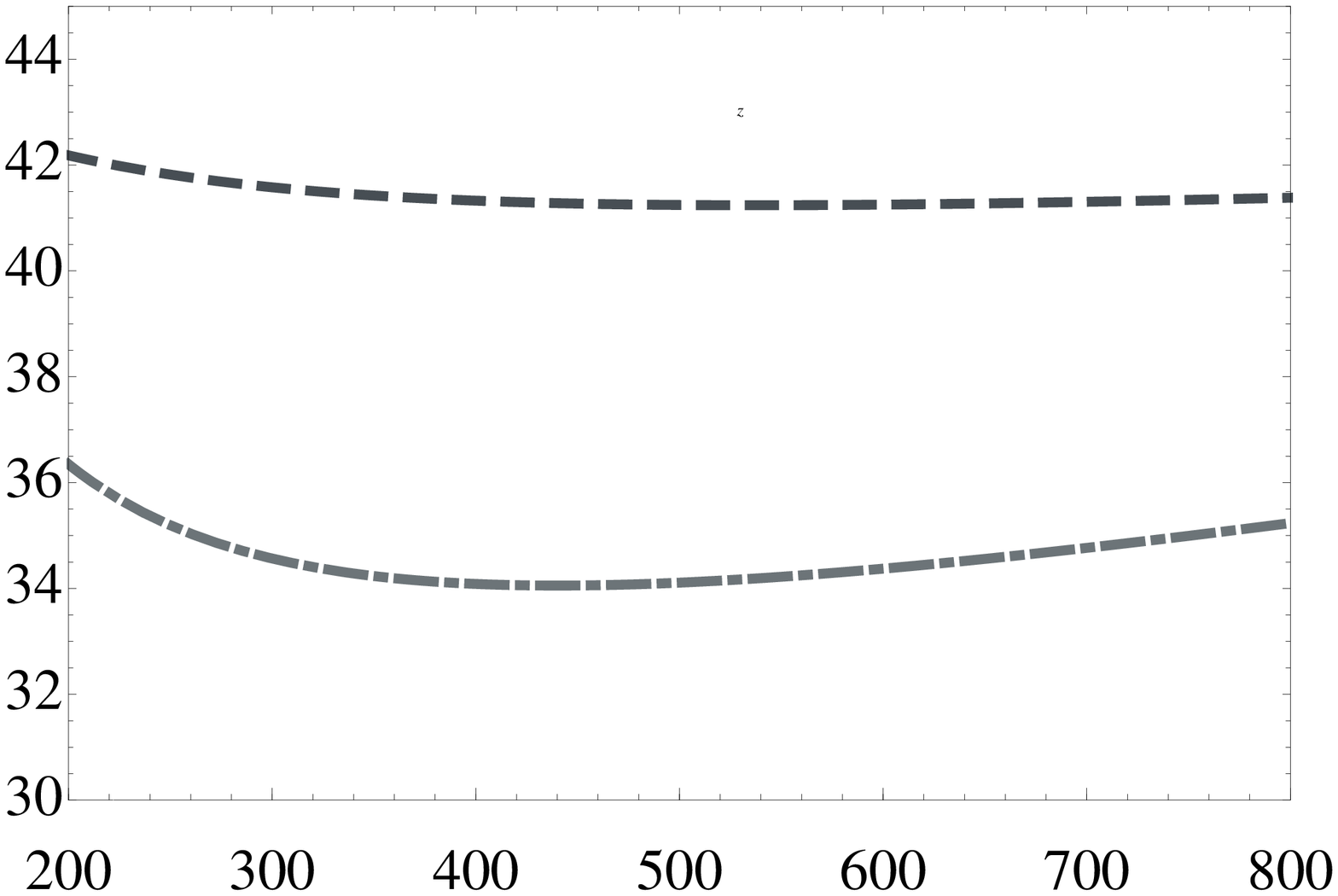}
& 
\psfrag{x}[][][1][90]{$d\sigma/dM$ [fb/GeV]}
\psfrag{y}[]{$\mu_s$ [GeV]}
\psfrag{z}[][][0.85]{$M=400$\,GeV}
\includegraphics[width=0.43\textwidth]{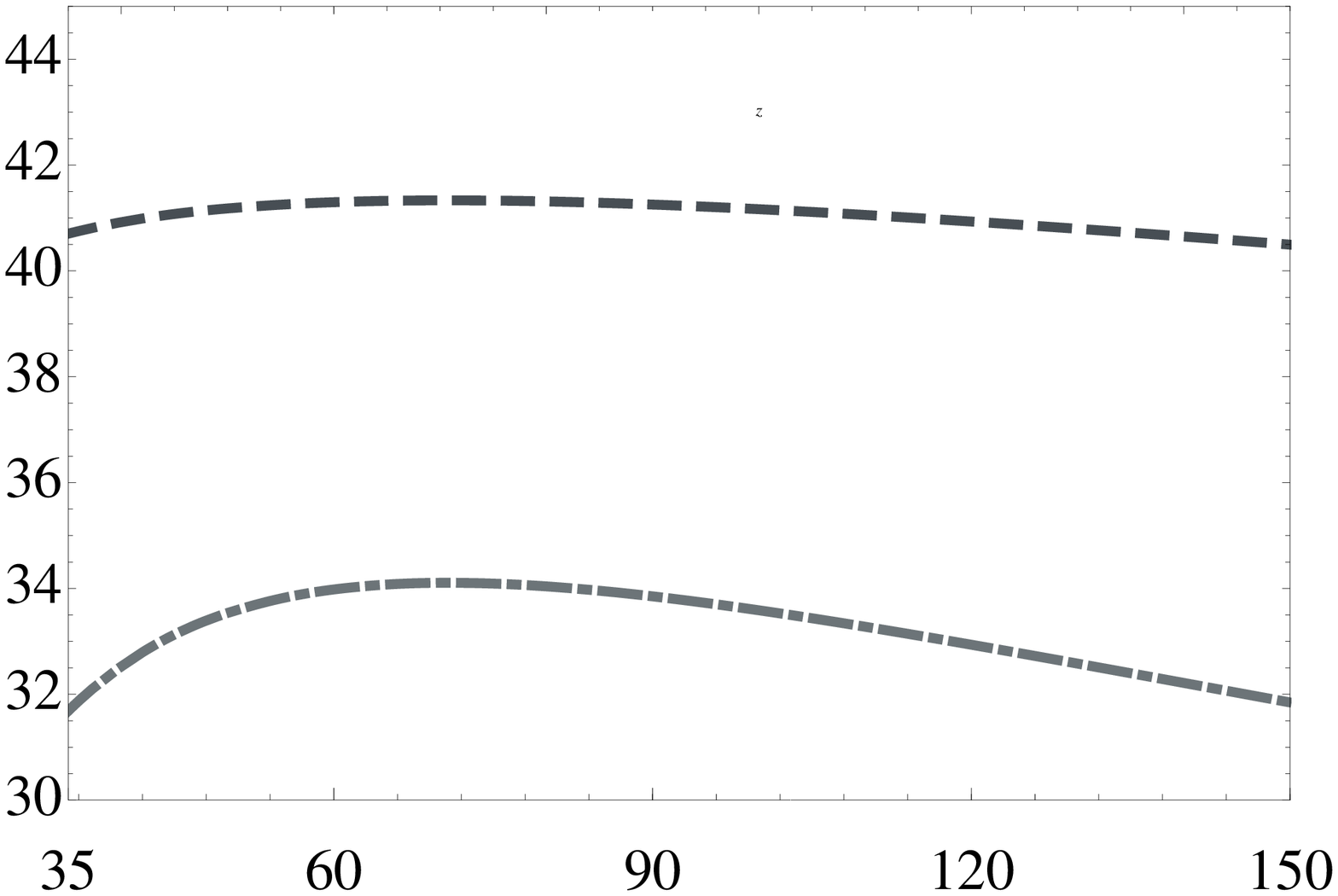}
\end{tabular}
\end{center}
\vspace{-1ex}
\caption{\label{fig:ScaleVarTeV} Dependence of $d\sigma/dM$ at the Tevatron on the scales
  $\mu_h$, $\mu_s$, and $\mu_f$ in resummed perturbation theory. The default choices are
  $\mu_h=\mu_f=M$, and $\mu_s$ according to (\ref{eq:SoftScale}). The dashed-dotted lines
  refer to NLL, the dashed to NNLL.}
\end{figure}
%%%%%%%%%%%%%%%%%%%%%%%%%%%%%%%%%%%%%%%%%%%%%%%%%%%%%%%%%%%%%%%%%%%%%%%%%%%%%%%%

The most appropriate choice of the hard scale $\mu_h$ is not immediately apparent, since
the invariant mass spectrum depends on the two hard scales $m_t$ and $M$. As a guide to an
appropriate choice we look at the size of the correction from the hard matching function
for different choices of $\mu_h$. We show in Figure~\ref{fig:HardScales} the correction
obtained by isolating the contribution of $\bm{H}^{(1)}$ to the differential cross section
at NNLL, and dividing it by the NLL result, for the choice $\mu_f=M$ and $\mu_s$
determined according to (\ref{eq:SoftScale}). We see that at lower values of $\mu_h$
closer to $m_t$ the correction typically gets smaller and can even become negative. In
this lower range of $\mu_h$, however, the correction depends very strongly on the scale.
The results are more stable in the range $M/2 <\mu_h <2M$, where the correction is
generally below $30$\% at the Tevatron and between $20-40$\% at the LHC. In what follows
we shall choose $\mu_h=M$ by default, in order to avoid the instability at lower $\mu_h$.
In the case of Higgs production, a negative hard scale squared $\mu_h^2 \sim
-m_H^2-i\epsilon$ was chosen to minimize the logarithms arising from time-like kinematics
\cite{Ahrens:2008qu, Ahrens:2008nc}. In the $t\bar{t}$ case, however, there are both
time-like and space-like momentum invariants, and it is not straightforward to tell which
point in the complex plane should be chosen to minimize the logarithms. We have thus
investigated the choice $\mu_h^2=M^2e^{i\phi_h}$ with $\phi_h$ varied between $-\pi$ and
$\pi$. The results show that at the LHC, the correction is smallest for $\phi_h \sim
-\pi$, and is about 10\% compared to 20\% at $\phi_h \sim 0$. At the Tevatron, however,
the minimal correction is obtained for values close to $\phi_h \sim 0$. In view of this,
and since the corrections are in any case not very large, we will not go into this
complication in our numerical analyses below.

Finally, we must choose a default value for the factorization scale $\mu_f$ in both the
resummed and fixed-order results. To do so, we study the behavior of the cross sections as
a function of this scale. For the fixed-order results, the invariant mass spectrum as a
function of $\mu_f$ at $M=400$\,GeV and $M=1$\,TeV is shown in
Figure~\ref{fig:ScaleVarFixed}. For the moment we do {\em not\/} match the results onto
fixed-order perturbation theory at NLO, using instead the threshold expansion
$d\sigma^{\text{NLO, leading}}$ and $d\sigma^{\text{NNLO, leading}}$. As a result, our
predictions are not strictly independent of the scale $\mu_f$, but a slight scale
dependence enters via subleading terms in $(1-z)$. At $M=400$\,GeV the approximate NNLO
formulas differ from each other less at $\mu_f \sim 400$\,GeV than at $\mu_f \sim m_t$.
The same is true at $M=1$\,TeV, but in this case the results become very unstable at
$\mu_f \sim m_t$. It therefore seems more appropriate to make the choice $\mu_f\sim M$
when studying the invariant mass spectrum. The resummed results at $M=400$\,GeV and
$M=1$\,TeV as a function of $\mu_f$, with $\mu_h=M$ and $\mu_s$ as in
(\ref{eq:SoftScale}), are shown in the upper two plots of Figure~\ref{fig:ScaleVarTeV},
for the case of the Tevatron (plots for the LHC would look very similar). Again, the
results at $\mu_f \sim M$ are more stable than at $\mu_f \sim m_t$, although compared to
the fixed-order results the difference is less pronounced. We will thus make the choice
$\mu_f=M$ by default in the resummed result.

%%%%%%%%%%%%%%%%%%%%%%%%%%%%%%%%%%%%%%%%%%%%%%%%%%%%%%%%%%%%%%%%%%%%%%%%%%%%%%%%
\begin{figure}[t]
\begin{center}
\begin{tabular}{ll}
\psfrag{x}[][][1][90]{$K$}
\psfrag{y}[]{$M$ [GeV]}
\psfrag{z}[][][0.85]{$\sqrt{s}=1.96$\,TeV}
\includegraphics[width=0.45\textwidth]{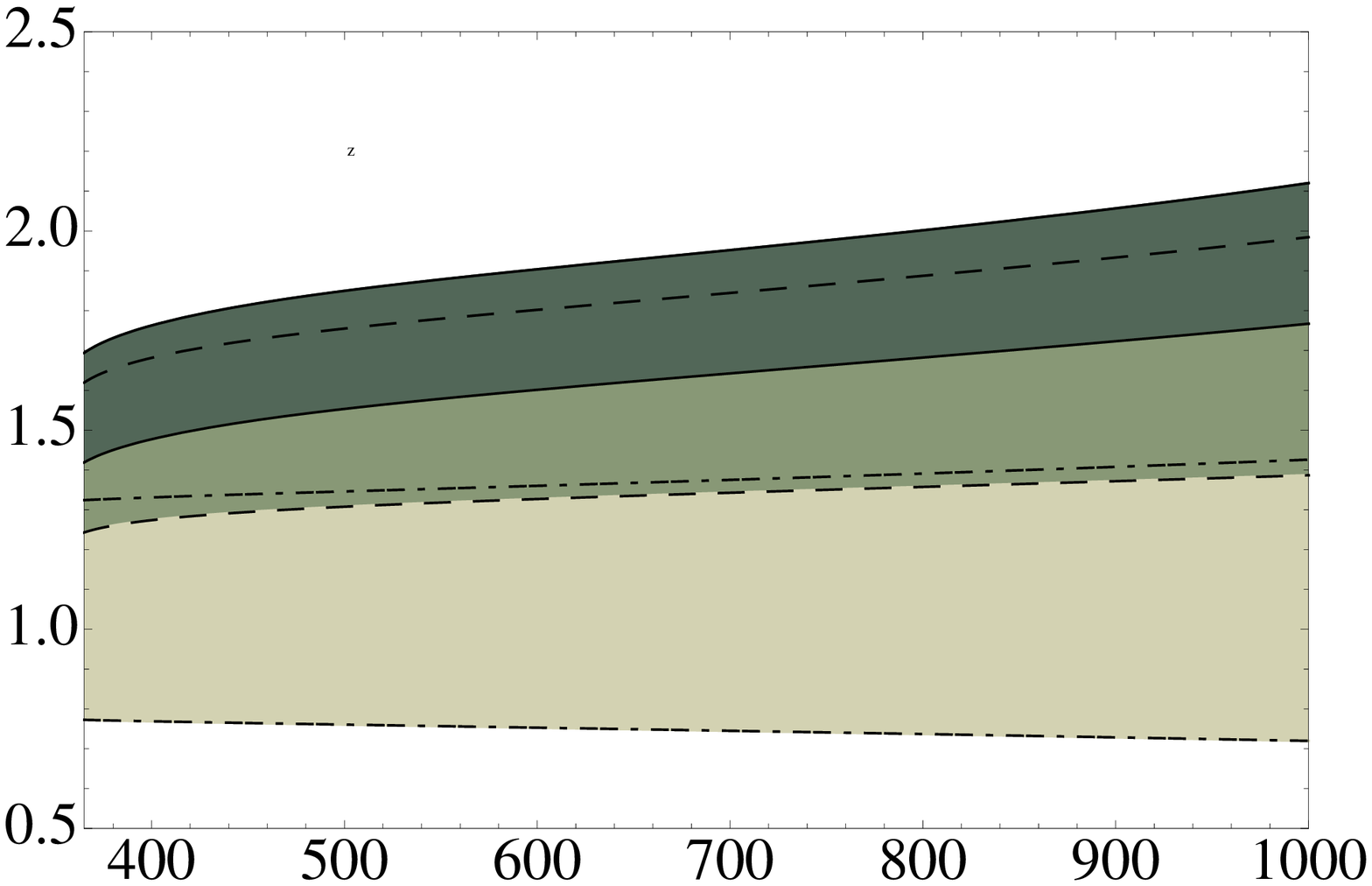}
&
\psfrag{x}[][][1][90]{$K$}
\psfrag{y}[]{$M$ [GeV]}
\psfrag{z}[][][0.85]{$\sqrt{s}=1.96$\,TeV}
\includegraphics[width=0.45\textwidth]{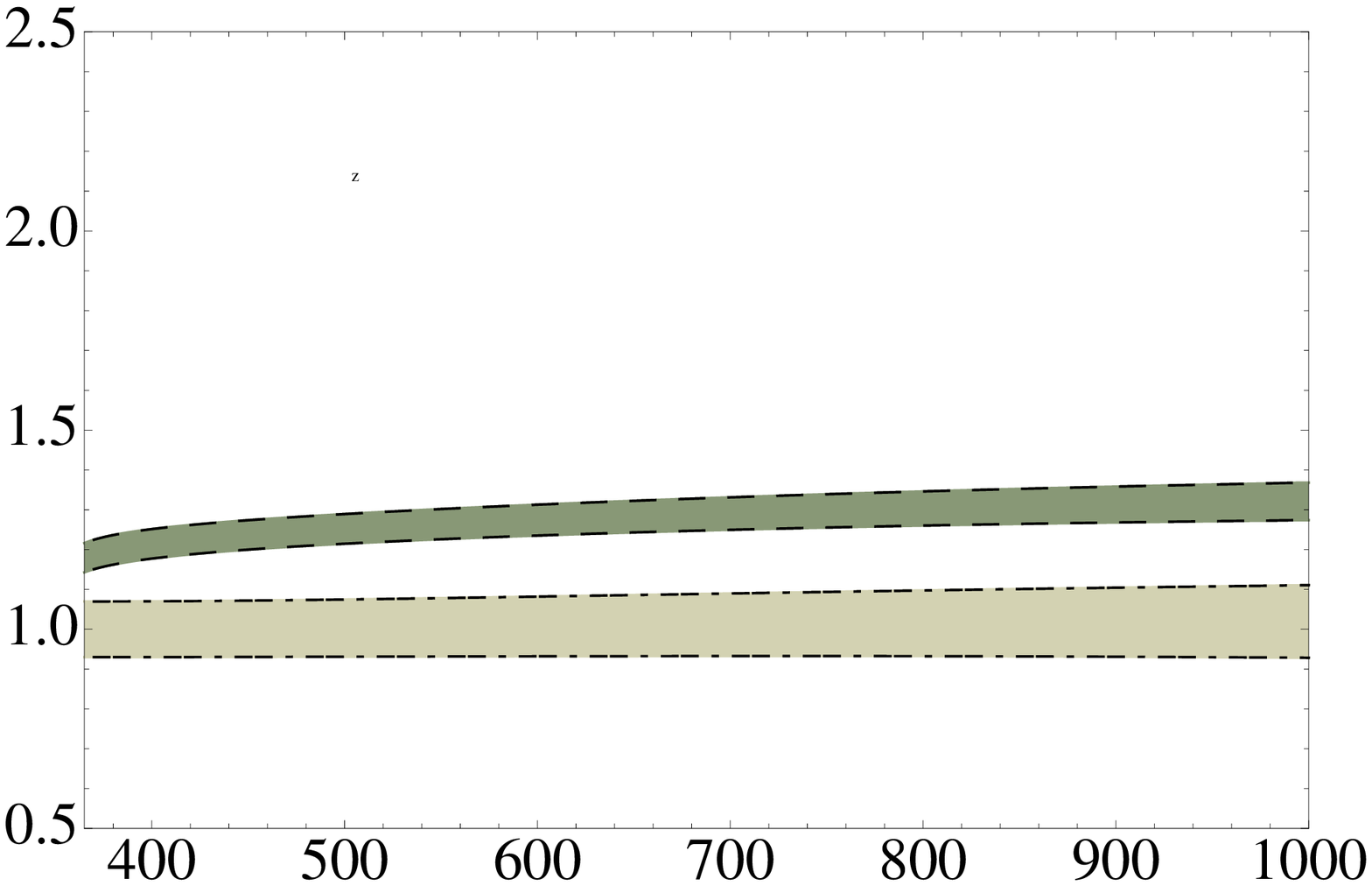}
\\[3mm]
\psfrag{x}[][][1][90]{$K$}
\psfrag{y}[]{$M$ [GeV]}
\psfrag{z}[][][0.85]{$\sqrt{s}=7$\,TeV}
\includegraphics[width=0.43\textwidth]{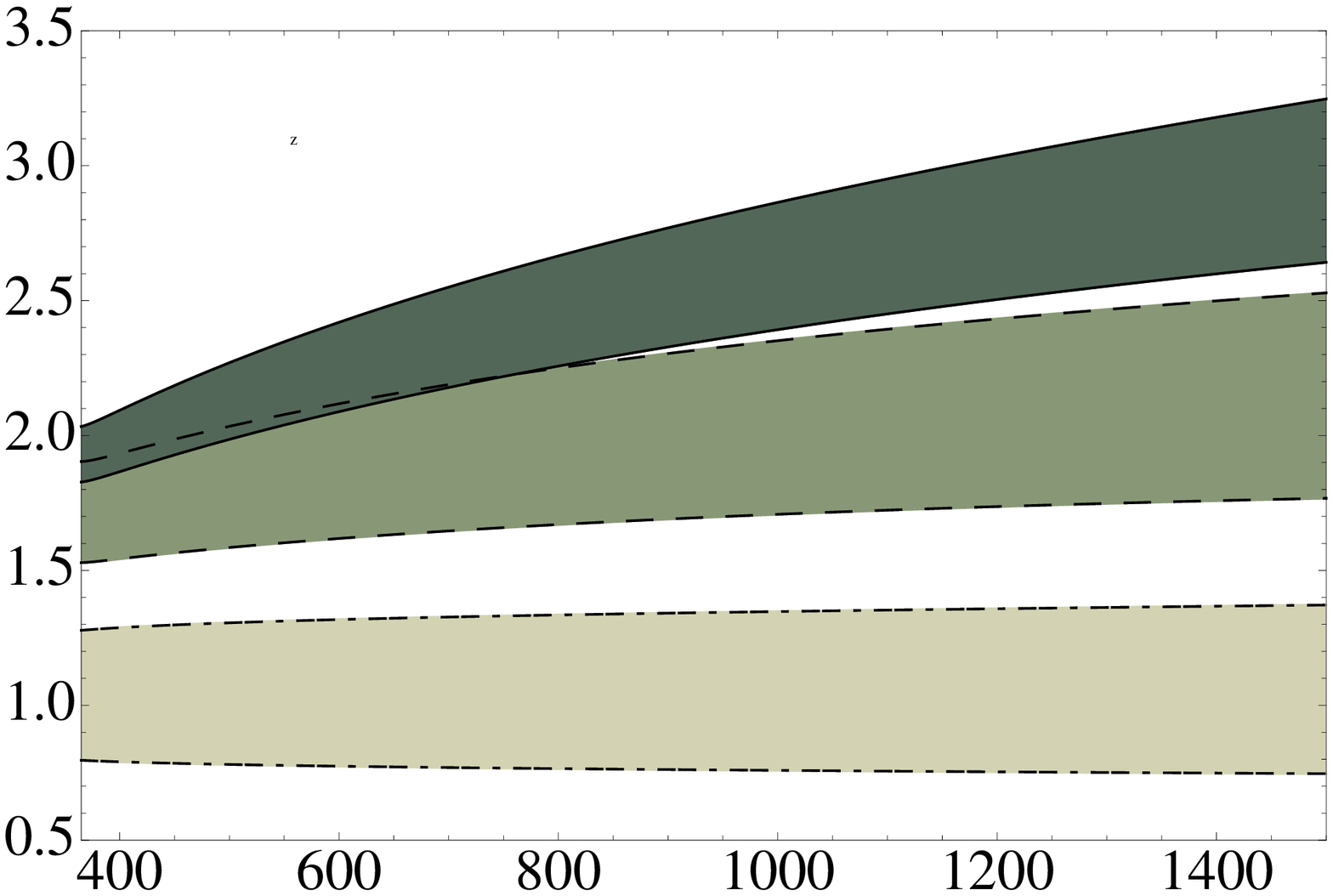}
&
\psfrag{x}[][][1][90]{$K$}
\psfrag{y}[]{$M$ [GeV]}
\psfrag{z}[][][0.85]{$\sqrt{s}=7$\,TeV}
\includegraphics[width=0.43\textwidth]{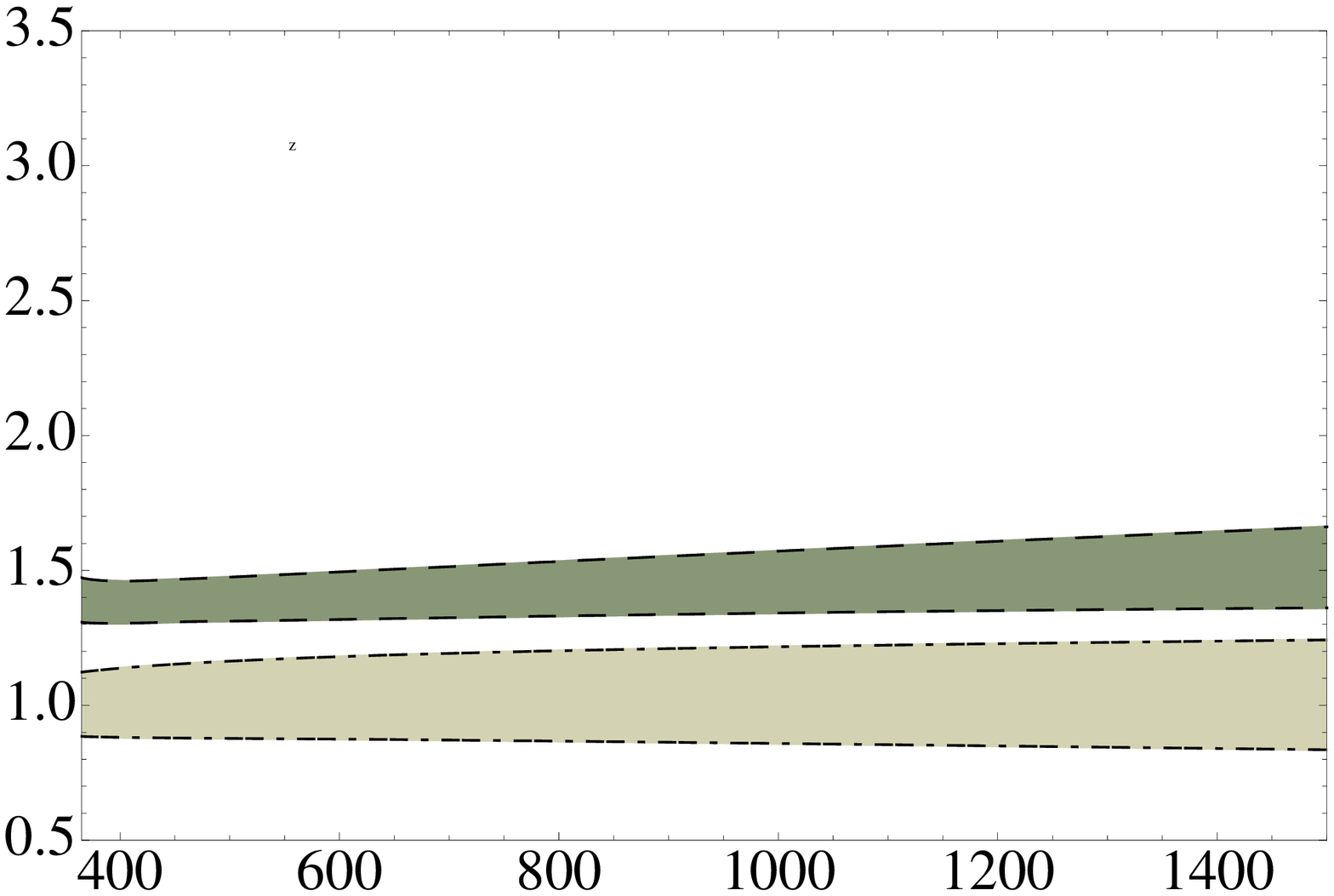}      
\end{tabular}
\end{center}
\vspace{-1ex}
\caption{\label{fig:BandsTeV} $K$ factors $(d\sigma/dM)/(d\sigma^{{\rm LO,\,def}}/dM)$ in
  fixed-order perturbation theory (left) and $(d\sigma/dM)/(d\sigma^{{\rm NLL,\,def}}/dM)$
  in resummed perturbation theory (right), for the Tevatron (top) and LHC (bottom). The
  light bands in the fixed-order (resummed) results show LO (NLL) results, the medium
  bands show NLO leading (NNLL) results, and the dark bands in the fixed-order results
  refer to the approximate threshold expansion (\ref{eq:NNLOthresh}) at NNLO in scheme~A.
  The width of the bands reflects the uncertainties associated with variations of the
  scales, as described in the text.}
\end{figure}
%%%%%%%%%%%%%%%%%%%%%%%%%%%%%%%%%%%%%%%%%%%%%%%%%%%%%%%%%%%%%%%%%%%%%%%%%%%%%%%%

Having chosen default values for the scales, we now discuss in more detail the behavior of
the fixed-order and resummed predictions for the invariant mass distribution. We have
already seen how the results depend on the scale $\mu_f$. In the lower two plots of
Figure~\ref{fig:ScaleVarTeV}, we show the dependence of the resummed results at
$M=400$\,GeV on the scales $\mu_h$ and $\mu_s$. The results as a function of these two
scales are significantly more stable at NNLL than at NLL.

\subsubsection{Convergence of the perturbation series}

An interesting difference between the fixed-order and resummed results is that the
perturbative uncertainties and the size of the higher-order corrections in the resummed
results depend much less on the value of $M$. This is seen in Figure~\ref{fig:BandsTeV},
where we show the $K$ factors and uncertainties in the invariant mass spectrum as a
function of $M$, comparing fixed-order results with the resummed ones. The bands in fixed
order reflect the uncertainty associated with varying the factorization scale around its
default value $\mu_f=M$ by a factor of two. Here and in the following figures, for the
approximate NNLO formulas we show results only in scheme A; those in scheme B look very
similar. To make the bands in resummed perturbation theory, we first obtain uncertainties
associated with the scales $\mu_f$, $\mu_h$, and $\mu_s$ by varying them individually up
and down by a factor of two from their default values at each point in $M$, with the other
two scales held fixed. We then obtain a total error by adding the three uncertainties
obtained this way in quadrature. At the Tevatron, the $K$ factors in resummed perturbation
theory have smaller uncertainties and depend only weakly on $M$, compared to fixed order.
The same is true at the LHC, although at small $M$ the approximate NNLO results and the
resummed ones have comparable uncertainties.

%%%%%%%%%%%%%%%%%%%%%%%%%%%%%%%%%%%%%%%%%%%%%%%%%%%%%%%%%%%%%%%%%%%%%%%%%%%%%%%%
\begin{table}[t]
\begin{center}
\begin{tabular}{|l|c|c|}
  \hline
  Tevatron & $M=400$\,GeV [fb/GeV] &  $M=1$\,TeV [fb/GeV]
  \\
  \hline
  NLL &
  34.1{\footnotesize $^{+2.4}_{-2.4}$}{\footnotesize $^{+1.7}_{-1.2}$} &
  (22.4{\footnotesize $^{+2.5}_{-1.6}$}{\footnotesize $^{+1.7}_{-1.3}$}$)\cdot 10^{-3}$
  \\
  NLO, leading &
  36.5{\footnotesize $^{+5.1}_{-5.0}$}{\footnotesize $^{+1.9}_{-1.4}$} &
  (23.2{\footnotesize $^{+4.2}_{-4.0}$}{\footnotesize $^{+1.9}_{-1.4}$}$)\cdot 10^{-3}$
  \\
  NNLL &
  41.3{\footnotesize $^{+1.3}_{-1.2}$}{\footnotesize $^{+1.9}_{-1.4}$} &
  (29.6{\footnotesize $^{+1.0}_{-1.1}$}{\footnotesize $^{+2.2}_{-1.6}$}$)\cdot 10^{-3}$
  \\
  NNLO, leading (scheme A) &
  40.1{\footnotesize $^{+3.5}_{-3.6}$}{\footnotesize $^{+1.9}_{-1.4}$} &
  (27.1{\footnotesize $^{+2.2}_{-2.7}$}{\footnotesize $^{+2.1}_{-1.6}$}$)\cdot 10^{-3}$
  \\
  NNLO, leading (scheme B) &
  39.3{\footnotesize $^{+3.0}_{-3.3}$}{\footnotesize $^{+1.9}_{-1.4}$} &
  (27.9{\footnotesize $^{+2.7}_{-3.0}$}{\footnotesize $^{+2.2}_{-1.6}$}$)\cdot 10^{-3}$
  \\
  \hline
  \hline
  LHC ($\sqrt{s}=7$\,TeV) & $M=400$\,GeV [fb/GeV] & $M=1$\,TeV [fb/GeV]
  \\
  \hline
  NLL &
  558{\footnotesize $^{+78}_{-68}$}{\footnotesize $^{+20}_{-21}$} &
  7.43{\footnotesize $^{+1.61}_{-1.07}$}{\footnotesize $^{+0.69}_{-0.70}$}
  \\
  NLO, leading & 
  656{\footnotesize $^{+72}_{-76}$}{\footnotesize $^{+26}_{-27}$} & 
  7.17{\footnotesize $^{+1.19}_{-1.10}$}{\footnotesize $^{+0.69}_{-0.69}$}
  \\
  NNLL &
  775{\footnotesize $^{+39}_{-47}$}{\footnotesize $^{+30}_{-31}$} &
  10.83{\footnotesize $^{+0.84}_{-0.87}$}{\footnotesize $^{+1.01}_{-1.03}$}~\,
  \\
  NNLO, leading (scheme A) &
  750{\footnotesize $^{+38}_{-47}$}{\footnotesize $^{+29}_{-30}$} &
  9.38{\footnotesize $^{+0.82}_{-0.87}$}{\footnotesize $^{+0.90}_{-0.90}$}
  \\
  NNLO, leading (scheme B) &
  717{\footnotesize $^{+20}_{-36}$}{\footnotesize $^{+28}_{-29}$} &
  9.11{\footnotesize $^{+0.63}_{-0.78}$}{\footnotesize $^{+0.86}_{-0.86}$}
  \\
  \hline  
\end{tabular}
\end{center}
\vspace{-2mm}
\caption{\label{tab:InvMassTables} 
Values for $d\sigma/dM$ for $M=400$\,GeV and $M=1$\,TeV at the Tevatron and the LHC at $\sqrt{s}=7$\,TeV. The first error refers to perturbative scale uncertainties, the second to PDF uncertainties, see text for a detailed explanation.}
\end{table}
%%%%%%%%%%%%%%%%%%%%%%%%%%%%%%%%%%%%%%%%%%%%%%%%%%%%%%%%%%%%%%%%%%%%%%%%%%%%%%%%

%%%%%%%%%%%%%%%%%%%%%%%%%%%%%%%%%%%%%%%%%%%%%%%%%%%%%%%%%%%%%%%%%%%%%%%%%%%%%%%%
\begin{table}
\vspace{4mm}
\begin{center}
\begin{tabular}{|l|c|c|c|}
\hline
Order & PDF set & $\alpha_s(M_Z)$ \\ 
\hline
LO & MSTW2008LO & 0.139 \\ 
NLO, NLL & MSTW2008NLO & 0.120 \\ 
NNLO approx, NLO+NNLL & MSTW2008NNLO & 0.117 \\ 
\hline
\end{tabular}
\end{center}
\vspace{-2mm}
\caption{\label{tab:PDForder} 
Order of the PDFs \cite{Martin:2009bu} and the corresponding values of the strong coupling used for the different perturbative approximations.}
\end{table}
%%%%%%%%%%%%%%%%%%%%%%%%%%%%%%%%%%%%%%%%%%%%%%%%%%%%%%%%%%%%%%%%%%%%%%%%%%%%%%%%

To illustrate more precisely the quantitative differences between the various perturbative
approximations to the invariant mass spectrum, we show in Table~\ref{tab:InvMassTables}
the exact numerical values of the spectrum at the points $M=400$\,GeV and $M=1000$\,GeV.
We have assigned uncertainties associated with variations of the various scales by factors
of two up and down from their default values. To obtain a total scale uncertainty for the
resummed results, we have added the uncertainties associated with variations of $\mu_h$,
$\mu_s$, and $\mu_f$ in quadrature. We have also included uncertainties associated with
the PDFs, by using the set of MSTW2008NNLO PDFs from \cite{Martin:2009bu} at 90\%
confidence level (CL). The perturbative scale uncertainties are smaller or comparable than
those from the PDFs only once the NNLL or approximate NNLO corrections are taken into
account. For the practical reasons explained earlier, we have not matched the higher-order
results with the fixed-order NLO results. However, the threshold approximation works
rather well. For reference, at the Tevatron the exact NLO results are
$(38.6{}^{+5.1}_{-5.2})$\,fb/GeV for $M=400$\,GeV and $(24.8{}^{+4.5}_{-4.8})\cdot
10^{-3}$\,fb/GeV for $M=1000$\,GeV, while at the LHC they are
$(654{}^{+98}_{-89})$\,fb/GeV for $M=400$\,GeV and $(6.84{}^{+1.40}_{-1.11})$\,fb/GeV for
$M=1000$\,GeV. The deviations from the leading NLO terms shown in the second line in both
parts of the table are smaller than 7\% for the Tevatron and 5\% for the LHC.

%%%%%%%%%%%%%%%%%%%%%%%%%%%%%%%%%%%%%%%%%%%%%%%%%%%%%%%%%%%%%%%%%%%%%%%%%%%%%%%%
\begin{figure}[t]
\begin{center}
\begin{tabular}{lr}
\psfrag{x}[][][1][90]{$d\sigma/dM$ [fb/GeV]}
\psfrag{y}[]{$M$ [GeV]}
\psfrag{z}[][][0.85]{$\sqrt{s}=1.96$\,TeV}
\includegraphics[width=0.43\textwidth]{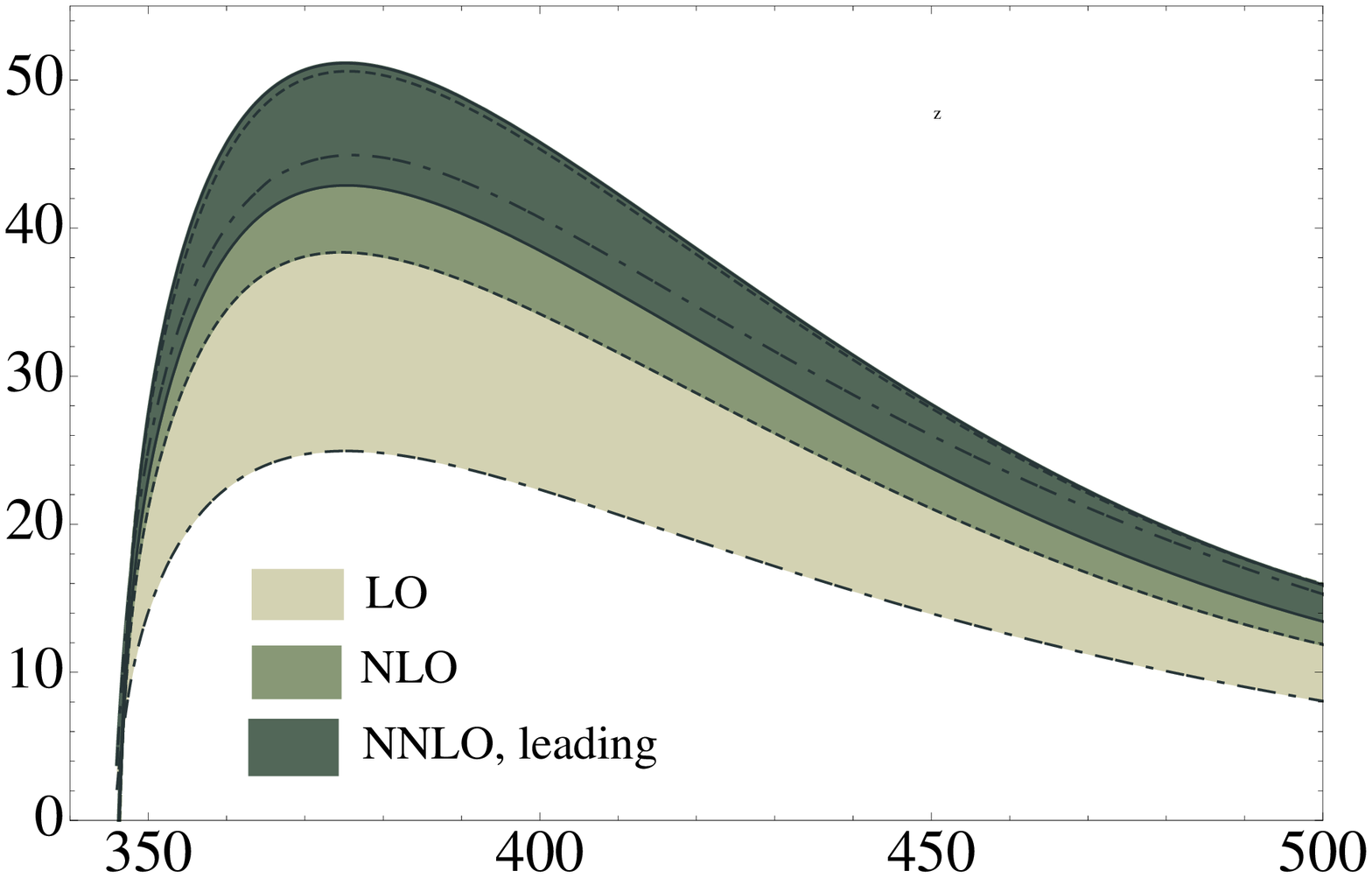}
&
\psfrag{x}[][][1][90]{$d\sigma/dM$ [fb/GeV]}
\psfrag{y}[]{$M$ [GeV]}
\psfrag{z}[][][0.85]{$\sqrt{s}=1.96$\,TeV}
\includegraphics[width=0.43\textwidth]{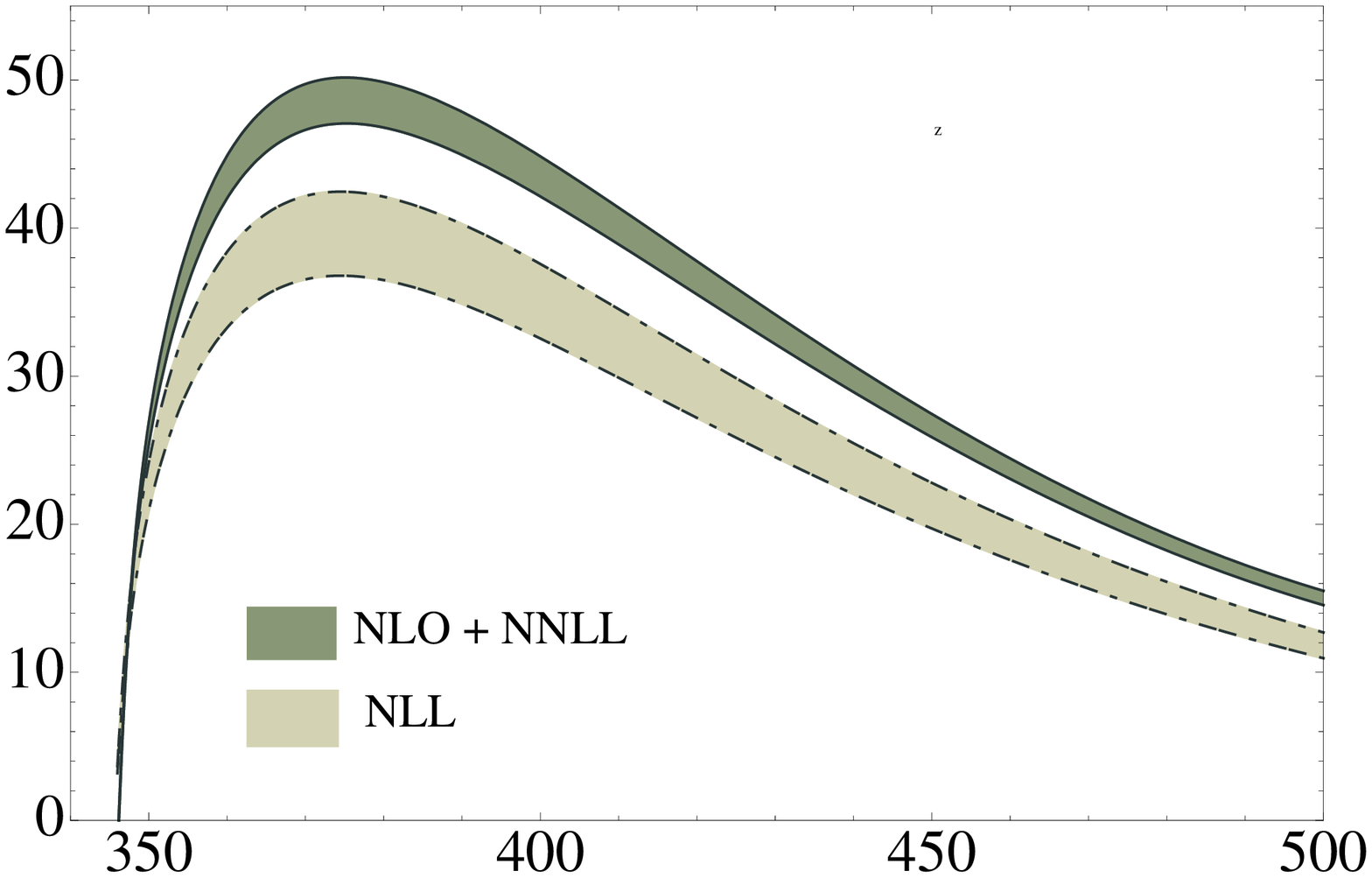}
\\[3mm]
\psfrag{x}[][][1][90]{$d\sigma/dM$ [pb/GeV]}
\psfrag{y}[]{$M$ [GeV]}
\psfrag{z}[][][0.85]{$\sqrt{s}=7$\,TeV}
\includegraphics[width=0.43\textwidth]{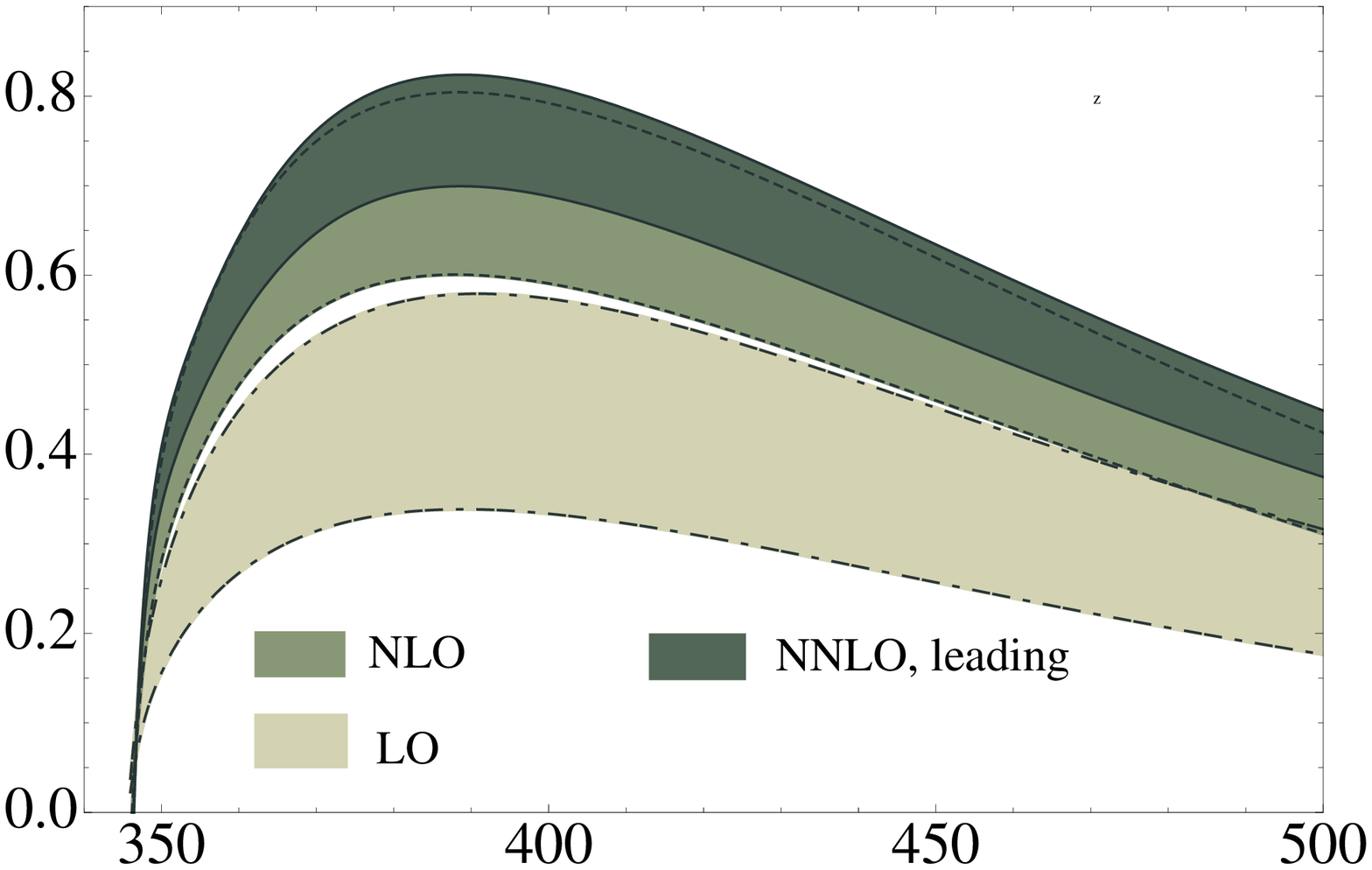}
&
\psfrag{x}[][][1][90]{$d\sigma/dM$ [pb/GeV]}
\psfrag{y}[]{$M$ [GeV]}
\psfrag{z}[][][0.85]{$\sqrt{s}=7$\,TeV}
\includegraphics[width=0.43\textwidth]{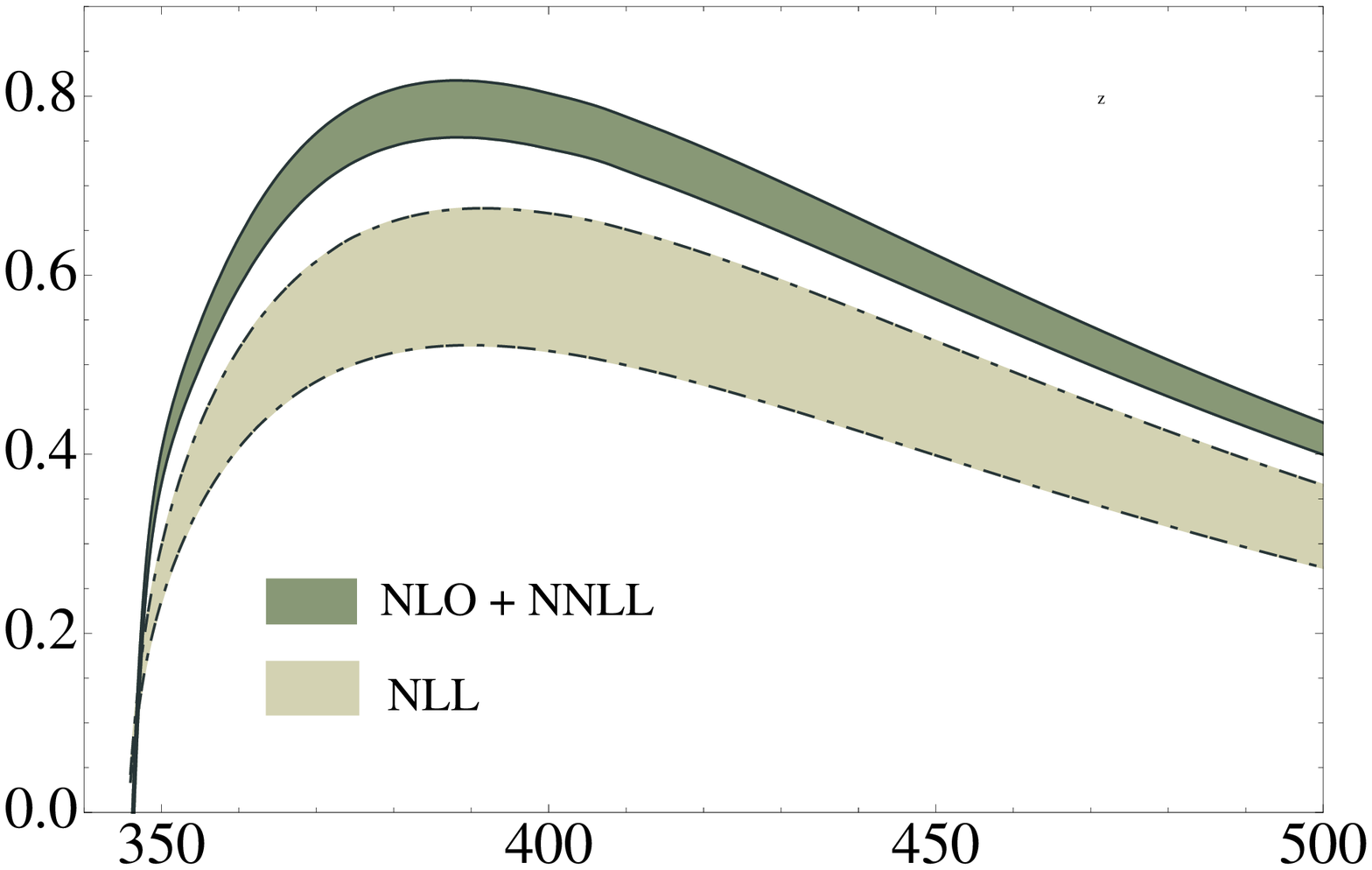}
\end{tabular}
\end{center}
\vspace{-2mm}
\caption{\label{fig:SpectrumPDFS} Left: Fixed-order predictions for the invariant mass
  spectrum at LO (light), NLO (darker), and approximate NNLO (dark bands) for the Tevatron
  (top) and LHC (bottom). Right: Corresponding predictions at NLL (light) and NLO+NNLL
  (darker bands) in resummed perturbation theory. The width of the bands reflects the
  uncertainty of the spectrum under variations of the matching and factorization scales,
  as explained in the text.}
\end{figure}
%%%%%%%%%%%%%%%%%%%%%%%%%%%%%%%%%%%%%%%%%%%%%%%%%%%%%%%%%%%%%%%%%%%%%%%%%%%%%%%%

\subsection{Invariant mass distribution: Phenomenological results}
\label{subsec:InvMassPheno}

After these systematic studies, we now present our final results for the $t\bar t$
invariant mass distributions at the Tevatron and LHC. Here and below, we will use
different sets of PDFs, as appropriate for the order of the perturbative approximation
employed. Strictly speaking, theoretical predictions obtained in resummed perturbation
theory would require PDF sets extracted from data using resummed predictions for the
relevant cross sections; however, such PDF sets do not exist at present. Since our
resummed expressions include the bulk of the perturbative corrections appearing one order
higher in $\alpha_s$, we use NLO parton densities for the NLO and NLL approximations, and
NNLO parton densities for the approximate NNLO and matched NLO+NNLL approximations, as
summarized in Table~\ref{tab:PDForder}. The associated running couplings $\alpha_s(\mu)$
are taken in the $\overline{\rm{MS}}$ scheme with five active flavors, using one-loop
running at LO, two-loop running at NLO, and three-loop running at NNLO.

%%%%%%%%%%%%%%%%%%%%%%%%%%%%%%%%%%%%%%%%%%%%%%%%%%%%%%%%%%%%%%%%%%%%%%%%%%%%%%%%
\begin{figure}[t]
\begin{center}
\begin{tabular}{lr}
\psfrag{x}[][][1][90]{$K$}
\psfrag{y}[]{$M$ [GeV]}
\psfrag{z}[][][0.85]{$\sqrt{s}=1.96$\,TeV}
\includegraphics[width=0.43\textwidth]{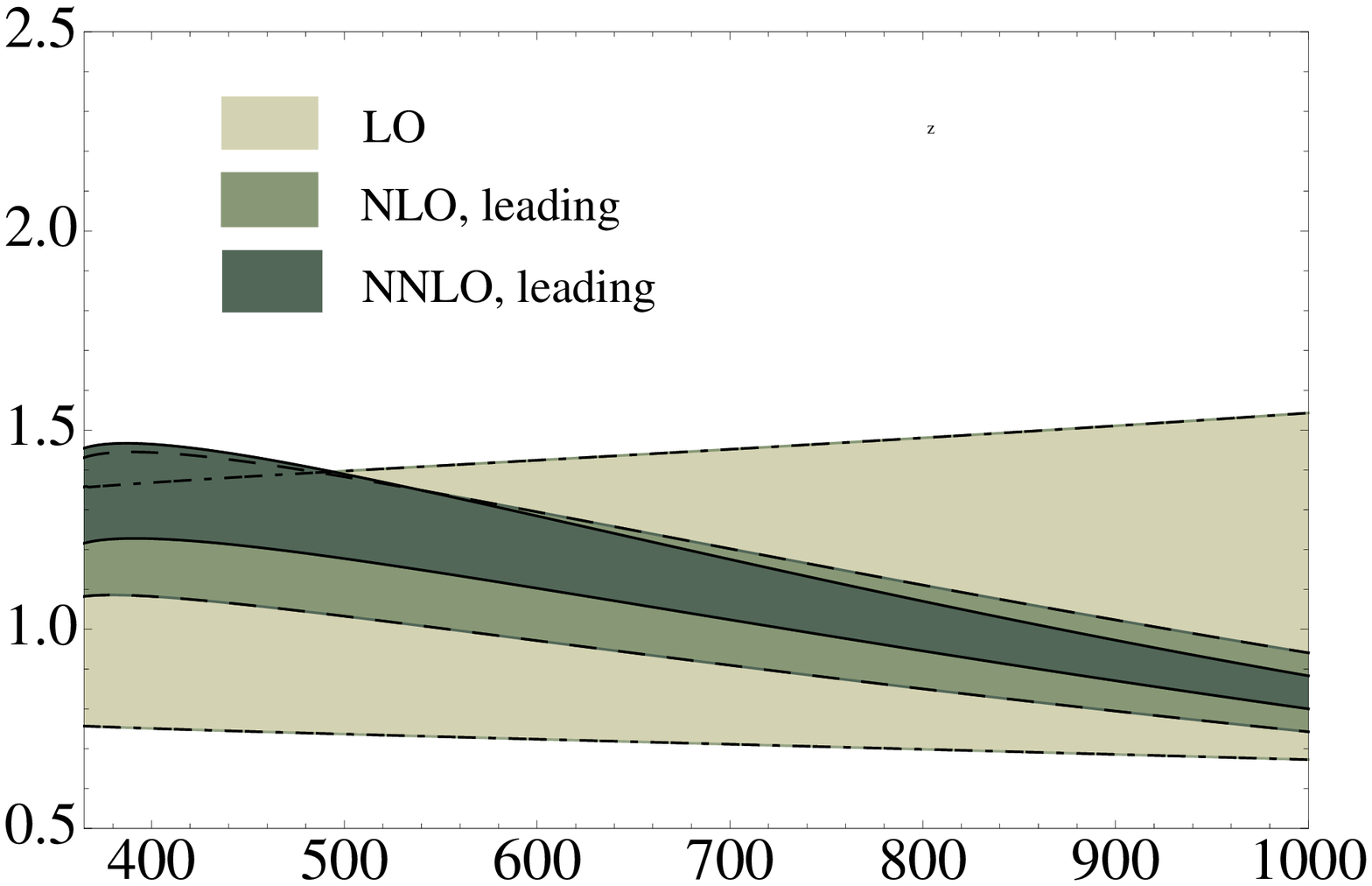} & 
\psfrag{x}[][][1][90]{$K$}
\psfrag{y}[]{$M$ [GeV]}
\psfrag{z}[][][0.85]{$\sqrt{s}=1.96$\,TeV}
\includegraphics[width=0.43\textwidth]{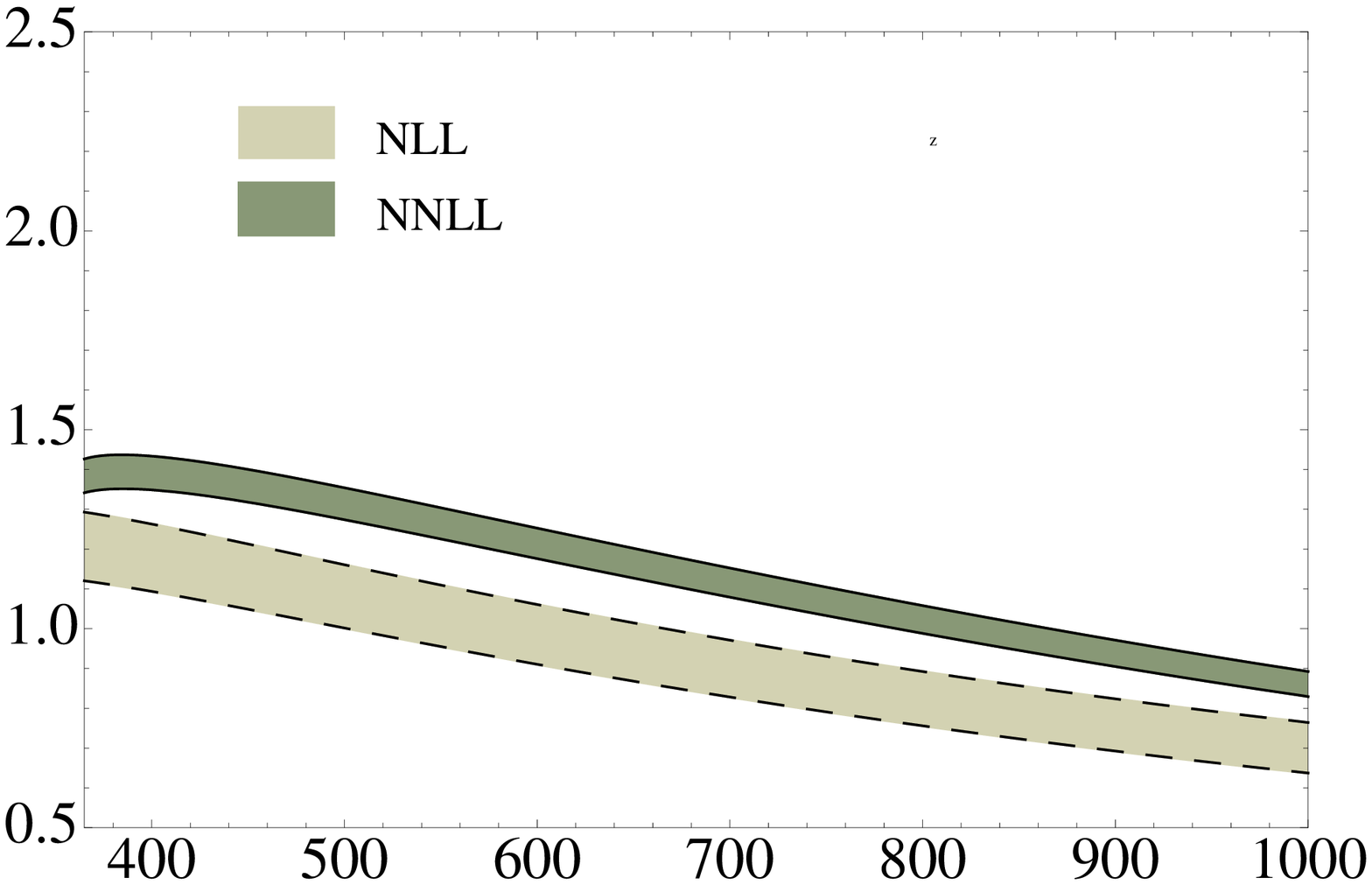} 
\\[3mm]
\psfrag{x}[][][1][90]{$d\sigma/dM$ [fb/GeV]}
\psfrag{y}[]{$M$ [GeV]}
\psfrag{z}[][][0.85]{$\sqrt{s}=1.96$\,TeV}
\includegraphics[width=0.43\textwidth]{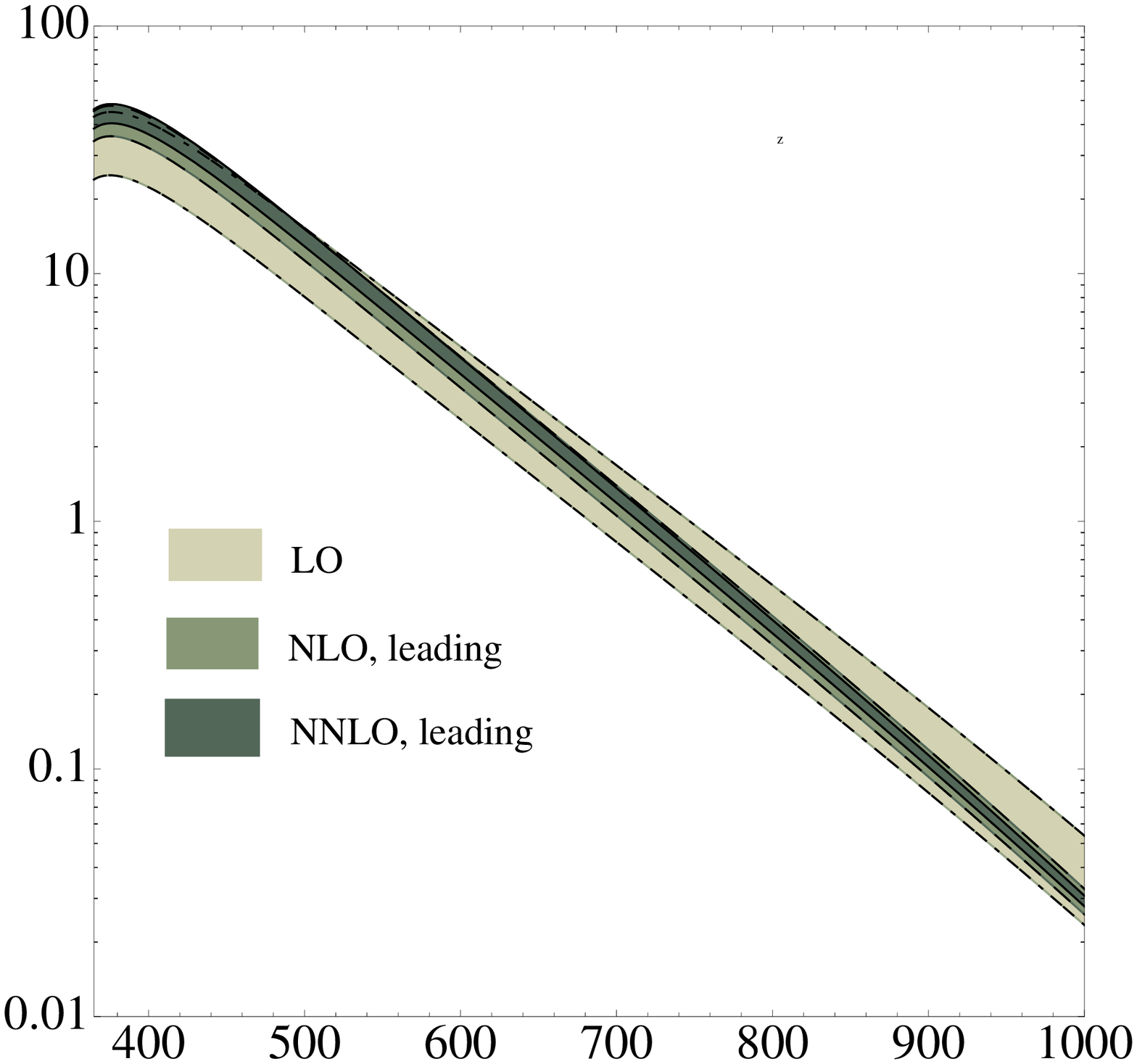} & 
\psfrag{x}[][][1][90]{$d\sigma/dM$ [fb/GeV]}
\psfrag{y}[]{$M$ [GeV]}
\psfrag{z}[][][0.85]{$\sqrt{s}=1.96$\,TeV}
\includegraphics[width=0.43\textwidth]{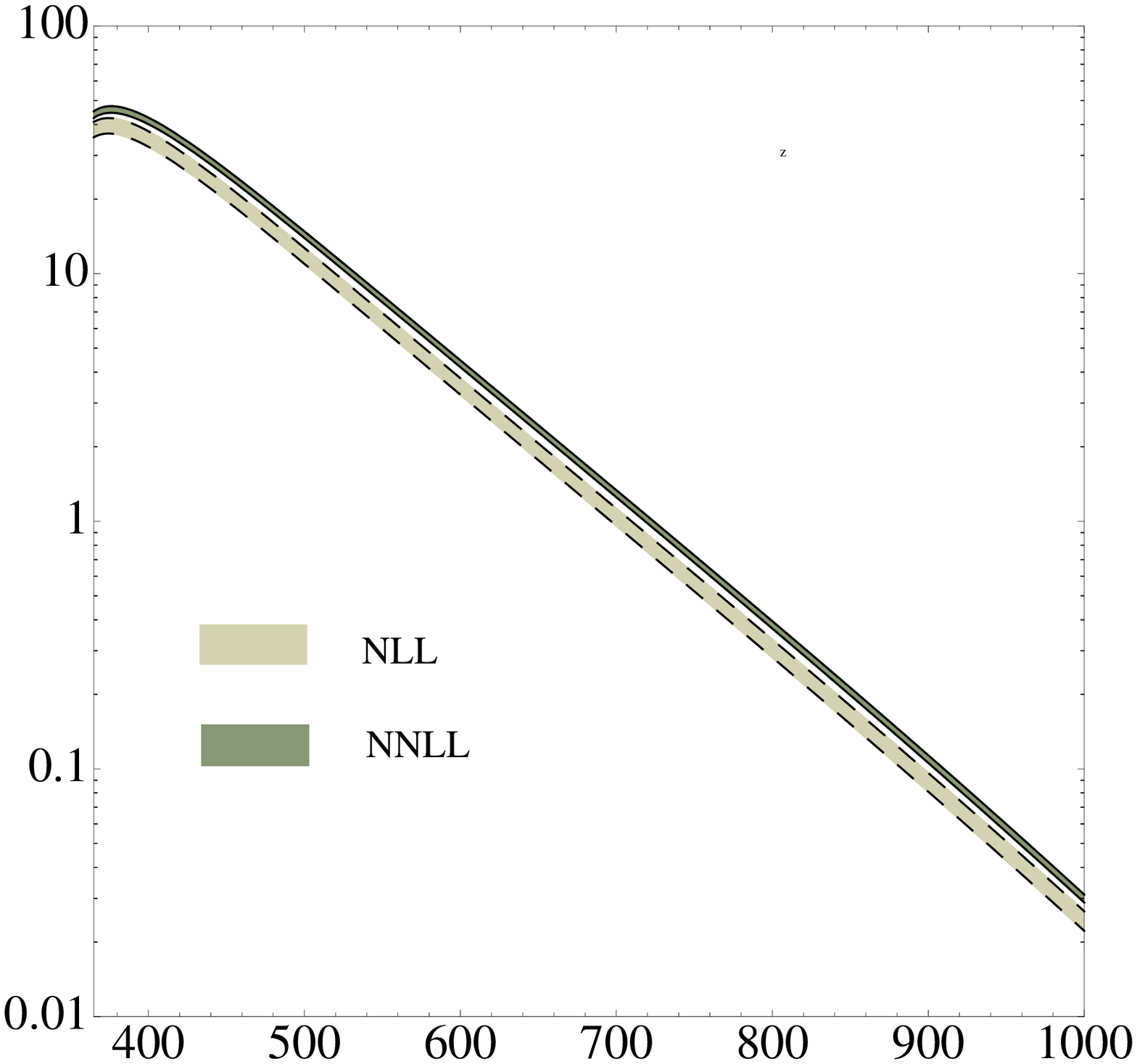}
\end{tabular}
\end{center}
\vspace{-2mm}
\caption{\label{fig:HighMassTev} Left: Fixed-order predictions for the $K$ factor and
  invariant mass spectrum at LO (light), NLO (darker), and approximate NNLO (dark bands)
  for the Tevatron. Right: Corresponding predictions at NLL (light) and NLO+NNLL (darker
  bands) in resummed perturbation theory. The width of the bands reflects the uncertainty
  of the spectrum under variations of the matching and factorization scales, as explained
  in the text.}
\end{figure}
%%%%%%%%%%%%%%%%%%%%%%%%%%%%%%%%%%%%%%%%%%%%%%%%%%%%%%%%%%%%%%%%%%%%%%%%%%%%%%%%

We begin by studying in more detail the invariant mass spectrum at relatively low values
of $M$, where it is the largest, in fixed-order and resummed perturbation theory. Contrary
to the previous section, we now match the results in resummed perturbation theory with the
exact fixed-order results at NLO using the MCFM program, according to
(\ref{eq:FixedMatching}). In this way we obtain state-of-the-art predictions, which
include everything known about the perturbative series for the spectrum. Our results are
shown in Figure~\ref{fig:SpectrumPDFS}. The bands reflect uncertainties in scale
variations according to the same procedure explained in the previous paragraph, but in
this case with $\mu_f=400$\,GeV by default. For the range of $M$ in the plot, this choice
is very close to our preferred scheme $\mu_f=M$, but allows for a simple matching with the
fixed-order results from MCFM. One sees that the perturbative uncertainty estimated by
scale variations is by far the smallest at NLO+NNLL order.

%%%%%%%%%%%%%%%%%%%%%%%%%%%%%%%%%%%%%%%%%%%%%%%%%%%%%%%%%%%%%%%%%%%%%%%%%%%%%%%%
\begin{figure}[t]
\begin{center}
\begin{tabular}{lr}
\psfrag{x}[][][1][90]{$K$}
\psfrag{y}[]{$M$ [GeV]}
\psfrag{z}[][][0.85]{$\sqrt{s}=7$\,TeV}
\includegraphics[width=0.43\textwidth]{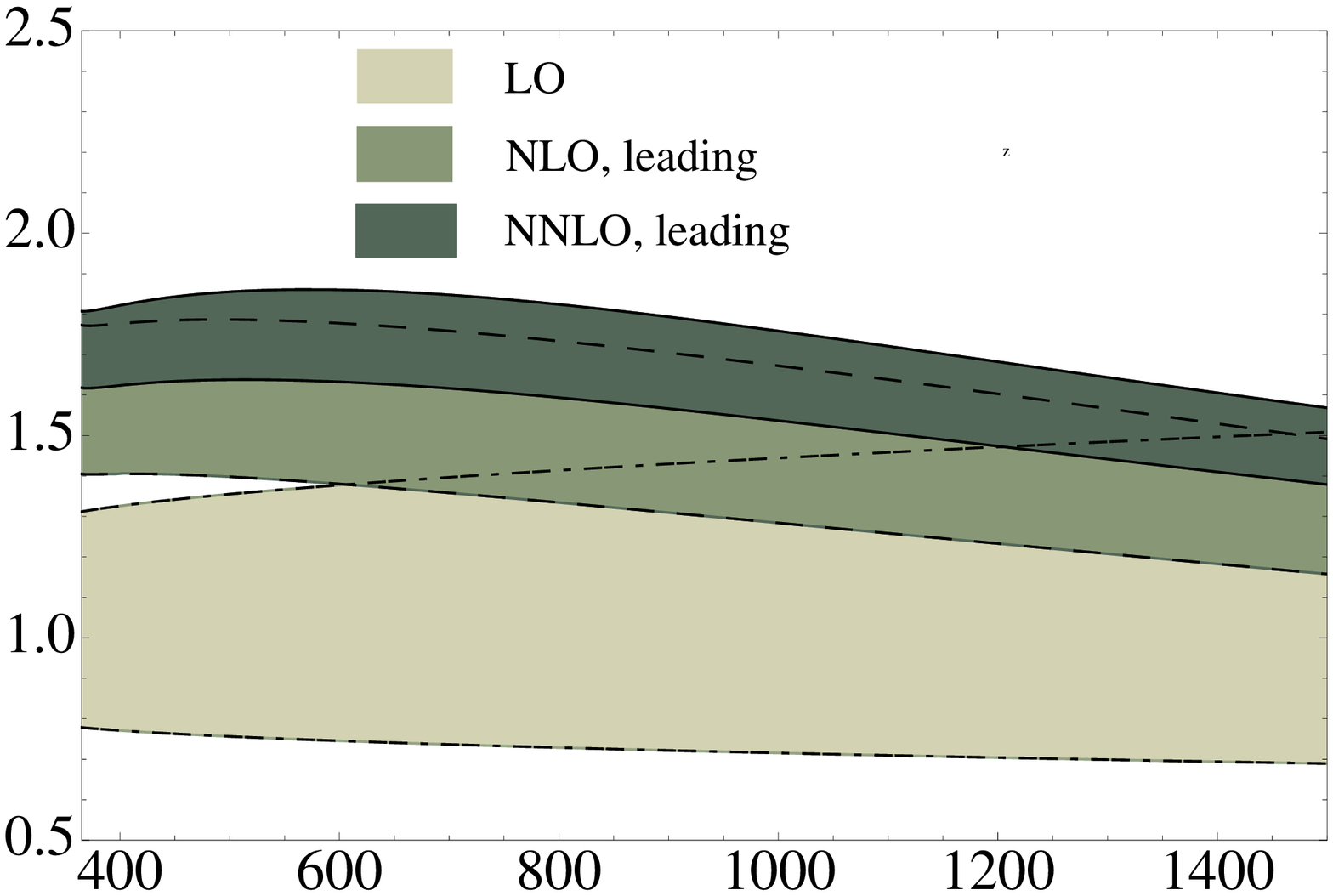} & 
\psfrag{x}[][][1][90]{$K$}
\psfrag{y}[]{$M$ [GeV]}
\psfrag{z}[][][0.85]{$\sqrt{s}=7$\,TeV}
\includegraphics[width=0.43\textwidth]{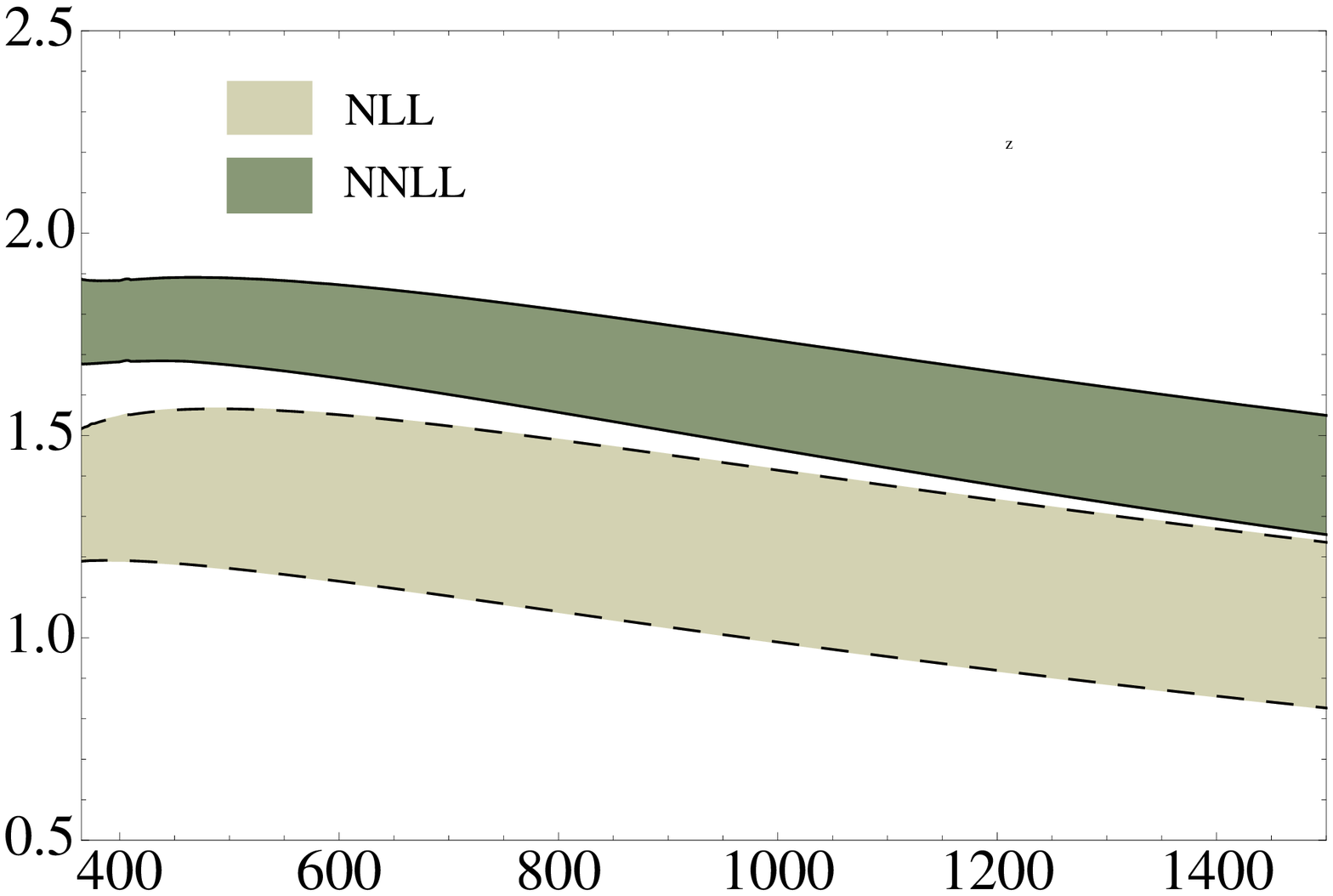} 
\\[3mm]
\psfrag{x}[][][1][90]{$d\sigma/dM$ [fb/GeV]}
\psfrag{y}[]{$M$ [GeV]}
\psfrag{z}[][][0.85]{$\sqrt{s}=7$\,TeV}
\includegraphics[width=0.43\textwidth]{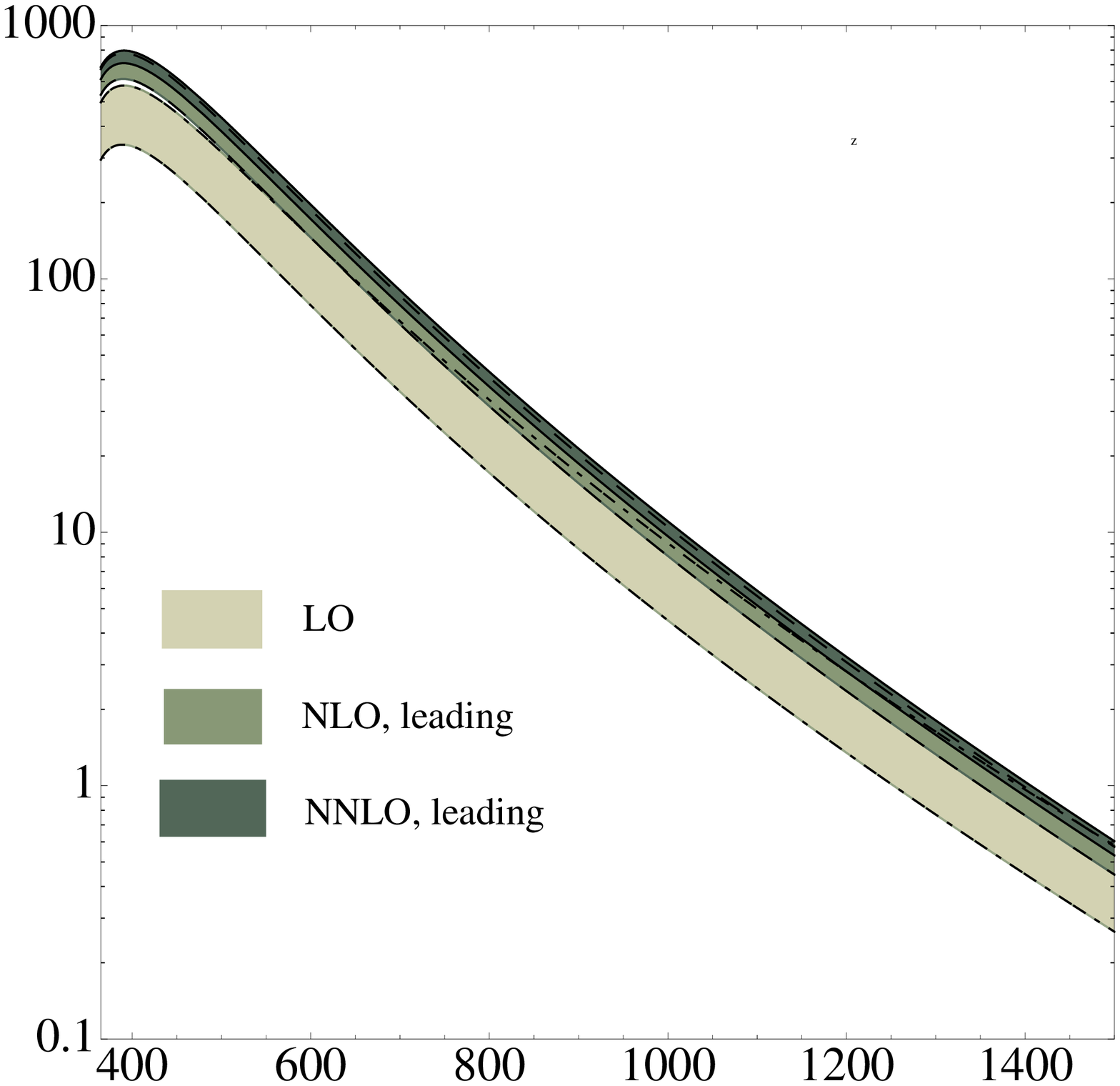} & 
\psfrag{x}[][][1][90]{$d\sigma/dM$ [fb/GeV]}
\psfrag{y}[]{$M$ [GeV]}
\psfrag{z}[][][0.85]{$\sqrt{s}=7$\,TeV}
\includegraphics[width=0.43\textwidth]{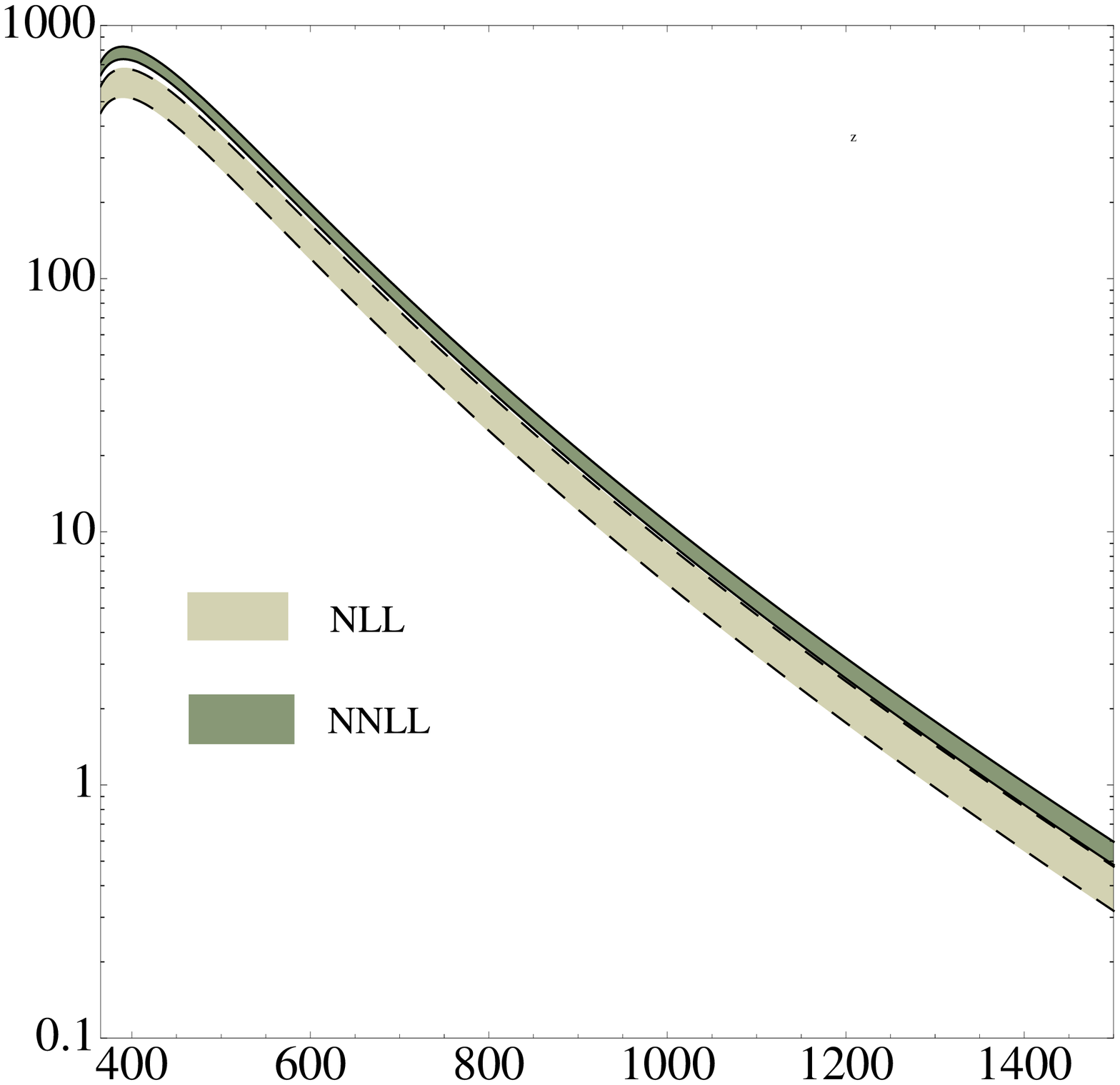}
\end{tabular}
\end{center}
\vspace{-2mm}
\caption{\label{fig:HighMassLHC} Left: Fixed-order predictions for the $K$ factor and
  invariant mass spectrum at LO (light), NLO (darker), and approximate NNLO (dark bands)
  for the LHC. Right: Corresponding predictions at NLL (light) and NLO+NNLL (darker bands)
  in resummed perturbation theory. The width of the bands reflects the uncertainty of the
  spectrum under variations of the matching and factorization scales, as explained in the
  text.}
\end{figure}
%%%%%%%%%%%%%%%%%%%%%%%%%%%%%%%%%%%%%%%%%%%%%%%%%%%%%%%%%%%%%%%%%%%%%%%%%%%%%%%%

We now consider the region of higher invariant masses, for which the dominance of the
threshold terms is even more pronounced, as indicated by the convergence of the dark bands
and dashed lines in Figure~\ref{fig:subl_terms} toward higher $M$ values.
Figure~\ref{fig:HighMassTev} shows our results for the Tevatron, both in fixed-order and
resummed perturbation theory. Figure~\ref{fig:HighMassLHC} shows the corresponding results
for the LHC. It is impractical to match onto fixed-order results obtained using the MCFM
program in this case; however, the differences compared with the shown curves are so small
that they would hardly be visible on the scales of the plots. The upper two plots show $K$
factors, which are defined as the ratio of the cross section to the default lowest-order
prediction $d\sigma^{\rm LO, def}/dM$. Contrary to Figure~\ref{fig:BandsTeV}, we now use
the same normalization in both fixed-order and resummed perturbation theory, so that the
two spectra can more readily be compared to each other. The lower plots show the
corresponding spectra directly. We observe similar behavior as in the low-mass region. The
bands obtained in fixed-order perturbation theory become narrower in higher orders and
overlap. The bands obtained in resummed perturbation theory are narrower than the
corresponding ones at fixed order. The leading-order resummed prediction is already close
to the final result.

%%%%%%%%%%%%%%%%%%%%%%%%%%%%%%%%%%%%%%%%%%%%%%%%%%%%%%%%%%%%%%%%%%%%%%%%%%%%%%%%
\begin{figure}
\begin{center}
\begin{tabular}{lr}
\psfrag{y}[][][1][90]{$d\sigma/d\beta_t$ [pb]}
\psfrag{x}[][][1]{$\beta_t$}
\psfrag{z}[][][0.85]{$\sqrt{s}=1.96$\,TeV}
\includegraphics[width=0.43\textwidth]{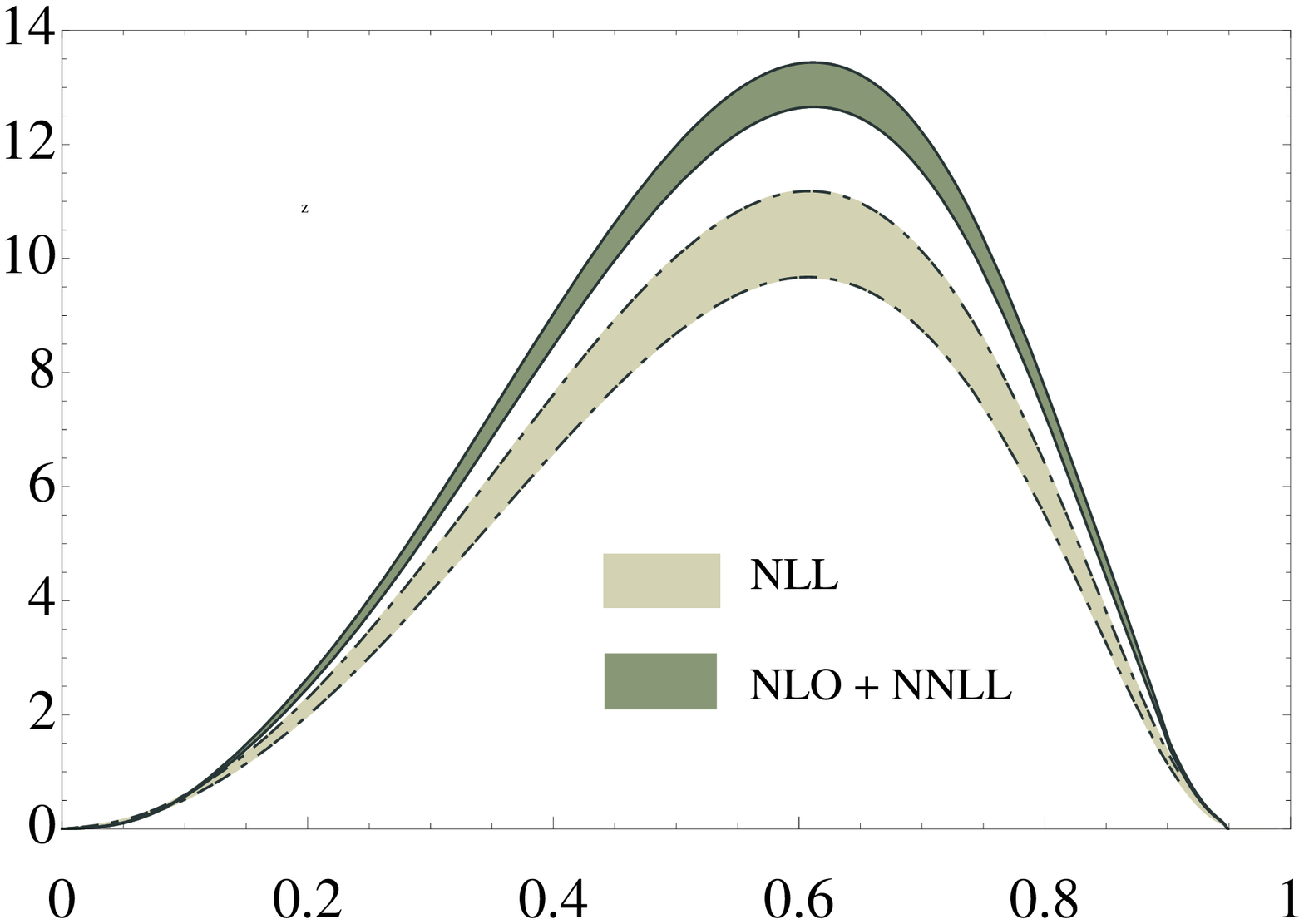} & 
\psfrag{y}[][][1][90]{$d\sigma/d\beta_t$ [pb]}
\psfrag{x}[][][1]{$\beta_t$}
\psfrag{z}[][][0.85]{$\sqrt{s}=7$\,TeV}
\includegraphics[width=0.43\textwidth]{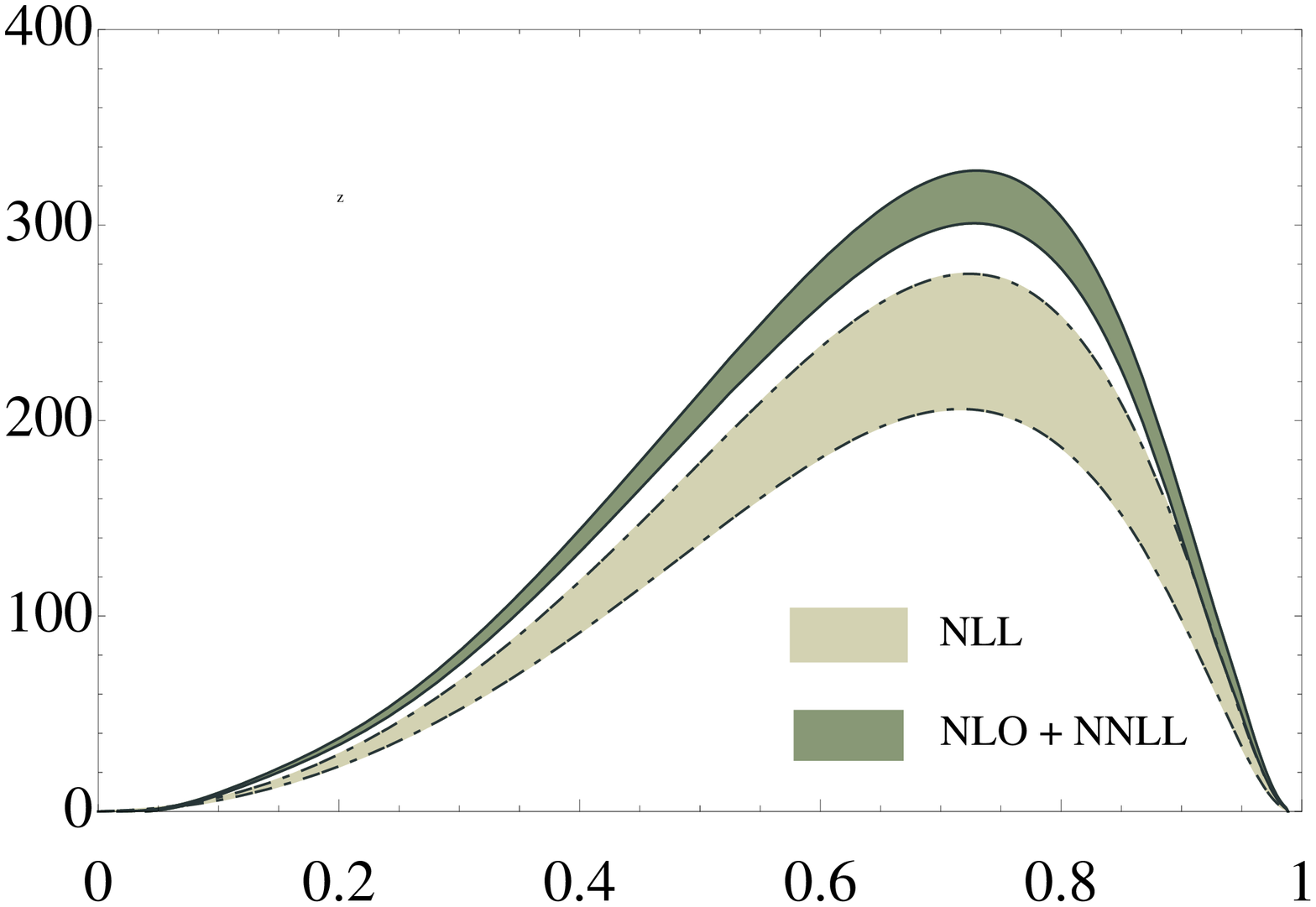}
\end{tabular}
\end{center}
\vspace{-2mm}
\caption{\label{fig:betat} 
Distributions $d\sigma/d\beta_t$ at the Tevatron (left) and LHC (right).}
\end{figure}
%%%%%%%%%%%%%%%%%%%%%%%%%%%%%%%%%%%%%%%%%%%%%%%%%%%%%%%%%%%%%%%%%%%%%%%%%%%%%%%%

The information contained in Figures~\ref{fig:SpectrumPDFS}--\ref{fig:HighMassLHC} can be
represented differently in terms of the very useful distribution $d\sigma/d\beta_t$, with
$\beta_t$ defined as in (\ref{betatdef}). A simple change of variables yields
\begin{equation}
  \frac{d\sigma}{d\beta_t} 
  = \frac{2m_t\beta_t}{(1-\beta_t^2)^{\frac32}}\,\frac{d\sigma}{dM} \,.
\end{equation}
The resulting spectra for the Tevatron and LHC, obtained using RG-improved perturbation
theory, are shown in Figure~\ref{fig:betat}. As before, the distributions are normalized
such that the area under the curves corresponds to the total cross section. Recall that
the physical meaning of the variable $\beta_t$ is that of the 3-velocity of the top quarks
in the $t\bar t$ rest frame. The distributions show that the dominant contributions to the
cross section arise from the region of relativistic top quarks, with velocities of order
0.4--0.8 at the Tevatron and 0.5--0.9 at the LHC. We will come back to the significance of
this observation in the next section.

%%%%%%%%%%%%%%%%%%%%%%%%%%%%%%%%%%%%%%%%%%%%%%%%%%%%%%%%%%%%%%%%%%%%%%%%%%%%%%%%
\begin{figure}
\begin{center}
\begin{tabular}{lr}
\psfrag{x}[][][1]{$M$ [GeV]}
\psfrag{y}[][][1][90]{$d\sigma/dM$ [fb/GeV]}
\psfrag{z}[][][0.85]{$\sqrt{s}=1.96$\,TeV}
\includegraphics[width=0.44\textwidth]{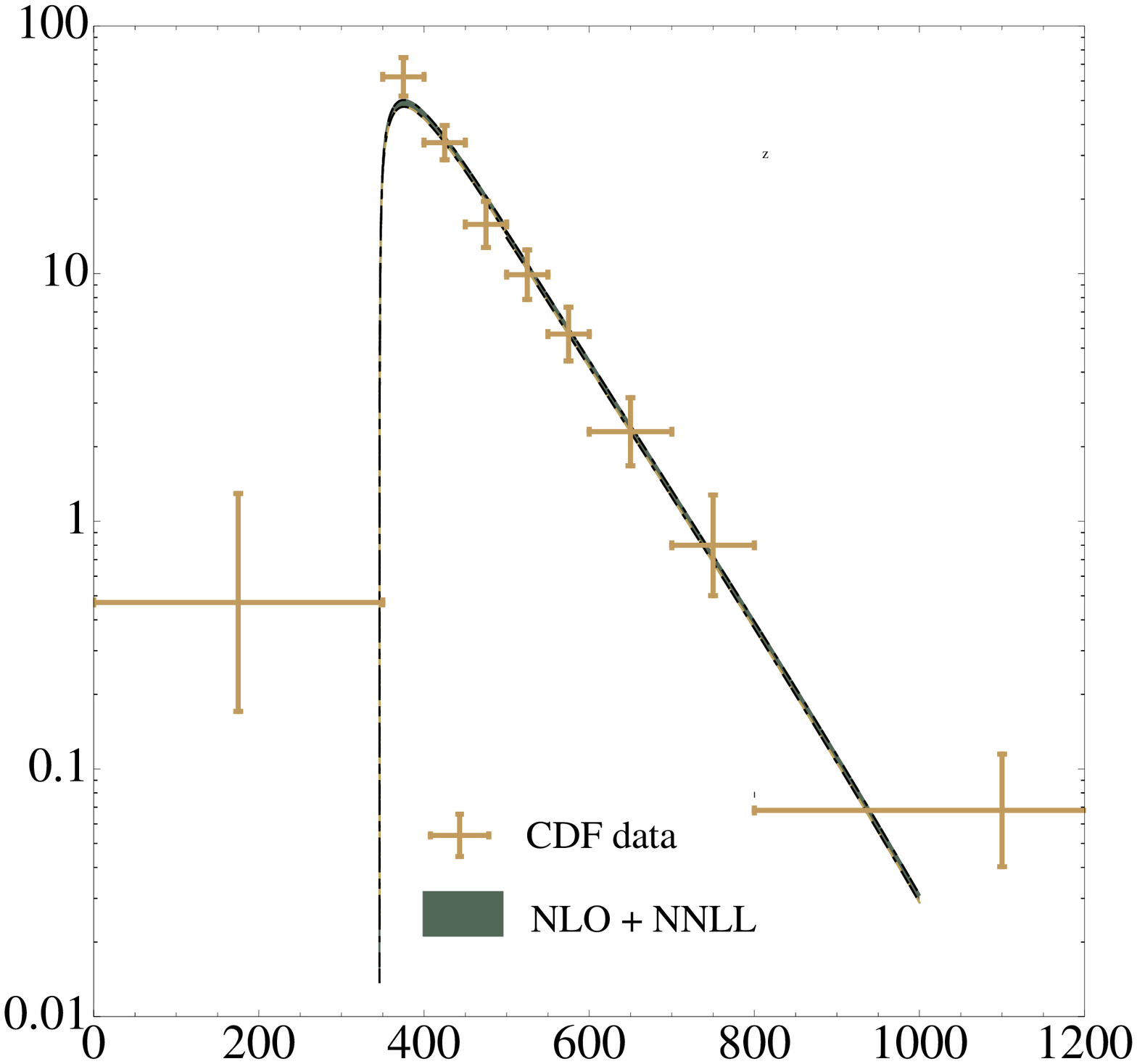} &
\psfrag{x}[][][1]{$M$ [GeV]}
\psfrag{y}[][][1][90]{$d \sigma/dM$ [fb/GeV]}
\psfrag{z}[][][0.85]{$\sqrt{s}=1.96$\,TeV}
\includegraphics[width=0.425\textwidth]{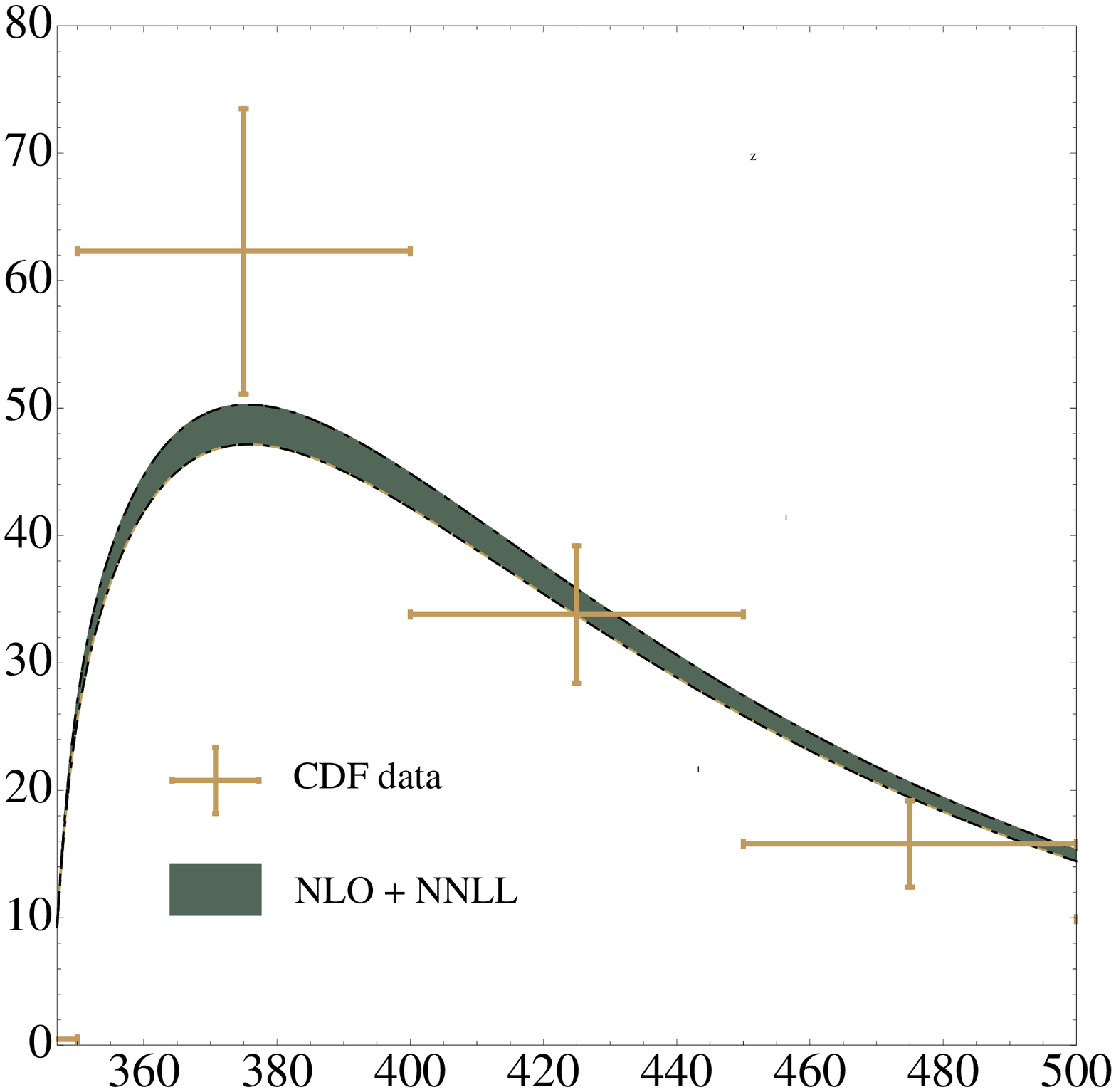}
\end{tabular}
\end{center}
\vspace{-2mm}
\caption{\label{fig:CDF-compare} Comparison of the RG-improved predictions for the
  invariant mass spectrum with CDF data \cite{Aaltonen:2009iz}. The value $m_t=173.1$\,GeV
  has been used. No fit to the data has been performed.}
\end{figure}
%%%%%%%%%%%%%%%%%%%%%%%%%%%%%%%%%%%%%%%%%%%%%%%%%%%%%%%%%%%%%%%%%%%%%%%%%%%%%%%%

In Figure~\ref{fig:CDF-compare}, we compare our RG-improved prediction for the invariant
mass spectrum to a measurement of the CDF collaboration obtained using the ``lepton +
jets'' decay mode of the top quark \cite{Aaltonen:2009iz}. We observe an overall good
agreement between our prediction and the measurement, especially for higher values of $M$.
Apparently, there is no evidence of non-standard resonances in the spectrum. The only
small deviation from our prediction concerns the peak region of the distribution, shown in
more details in the right plot. This deviation has also been observed in
\cite{Aaltonen:2009iz}, where a Monte Carlo study of the SM expectation has been
performed.

\subsection{Total cross section: Phenomenological results}
\label{sec:totcross}

The total cross section is obtained in our approach by integrating numerically the doubly
differential cross section in the ranges $-1<\cos\theta<1$ and $2m_t<M<\sqrt{s}$. In this
case it is a simple matter to match onto NLO in fixed-order perturbation theory, using the
analytic results of \cite{Czakon:2008ii}. To do this, however, we can no longer correlate
the factorization scale $\mu_f$ with $M$, as we did when studying the invariant mass
spectrum. Instead, we should resort to representative average values of $M$, which
characterize the spectrum in the region yielding sizable contributions to the total cross
section. One possibility is to take the location of the peak in the $d\sigma/dM$
distributions, which is $M_{\rm peak}\approx 375$\,GeV for the Tevatron and $M_{\rm
  peak}\approx 388$\,GeV for the LHC (see Figure~\ref{fig:SpectrumPDFS}). Another
possibility is to take the average value $\langle M\rangle$ of the distributions, for
which we find $\langle M\rangle\approx 445$\,GeV for the Tevatron and $\langle
M\rangle\approx 496$\,GeV for the LHC. As previously, we take the fixed value
$\mu_f=400$\,GeV as our default choice. On the other hand, we are still free to choose the
hard and soft scales as we have done so far and match with the fixed-order result as shown
in (\ref{eq:FixedMatching}). We display in Table~\ref{tab:CrossSections400} the central
values and scale uncertainties for the total cross section obtained using this procedure.
The results in resummed perturbation theory use $\mu_h=M$ and $\mu_s$ chosen according to
(\ref{eq:SoftScale}) by default, and the uncertainties are obtained by varying these
scales and the factorization scale $\mu_f$ up and down by a factor of two and adding the
different uncertainties in quadrature. The perturbative uncertainties in the fixed-order
results are obtained by varying the factorization scale up and down by a factor of two
from its default value. In addition to the perturbative uncertainties, we also list the
PDF uncertainties obtained by evaluating the cross section with the appropriate set of
MSTW2008 PDFs at 90\% CL. As shown in Table~\ref{tab:PDForder}, the LO cross sections are
evaluated using LO PDF sets, the NLL and NLO cross sections using NLO PDF sets, and the
NNLL and approximate NNLO cross sections using NNLO PDF sets. In the following tables,
these different classes of predictions are separated by horizontal lines.

%%%%%%%%%%%%%%%%%%%%%%%%%%%%%%%%%%%%%%%%%%%%%%%%%%%%%%%%%%%%%%%%%%%%%%%%%%%%%%%%
\begin{table}[t]
\begin{center}
\begin{tabular}{|l|c|c|c|c|}
  \hline
  & Tevatron &  LHC (7\,TeV) & LHC (10\,TeV) & LHC (14\,TeV)
  \\
  \hline
  $\sigma_{\rm LO}$ &
  4.49{\footnotesize $^{+1.71}_{-1.15}$}{\footnotesize $^{+0.24}_{-0.19}$} &
  84{\footnotesize $^{+29}_{-20}$}{\footnotesize $^{+4}_{-5}$}  &
  217{\footnotesize $^{+70}_{-49}$}{\footnotesize $^{+10}_{-11}$} &
  495{\footnotesize $^{+148}_{-107}$}{\footnotesize $^{+19}_{-24}$}
  \\ 
  \hline
  $\sigma_{\rm NLL}$  &
  5.07{\footnotesize $^{+0.37}_{-0.36}$}{\footnotesize $^{+0.28}_{-0.18}$} &
  112{\footnotesize $^{+18}_{-14}$}{\footnotesize $^{+5}_{-5}$} &
  276{\footnotesize $^{+47}_{-37}$}{\footnotesize $^{+10}_{-11}$} &
  598{\footnotesize $^{+108}_{-94}$}{\footnotesize $^{+19}_{-19}$}
  \\
  $\sigma_{\text{NLO, leading}}$ &
  5.49{\footnotesize $^{+0.78}_{-0.78}$}{\footnotesize $^{+0.31}_{-0.20}$} &
  134{\footnotesize $^{+16}_{-17}$}{\footnotesize $^{+7}_{-7}$} &
  341{\footnotesize $^{+34}_{-38}$}{\footnotesize $^{+14}_{-14}$} &
  761{\footnotesize $^{+64}_{-75}$}{\footnotesize $^{+25}_{-26}$}
  \\
  $\sigma_{\rm NLO}$ &
  5.79{\footnotesize $^{+0.79}_{-0.80}$}{\footnotesize $^{+0.33}_{-0.22}$} &
  133{\footnotesize $^{+21}_{-19}$}{\footnotesize $^{+7}_{-7}$}  &
  341{\footnotesize $^{+50}_{-46}$}{\footnotesize $^{+14}_{-15}$} &
  761{\footnotesize $^{+105}_{-101}$}{\footnotesize $^{+26}_{-27}$}
  \\ 
  \hline 
  &&&& \\[-5.9mm]
  \gr {$\sigma_{\rm NLO+NNLL}$} &
  {6.30{\footnotesize $^{+0.19}_{-0.19}$}{\footnotesize $^{+0.31}_{-0.23}$}} &
  {149{\footnotesize $^{+7}_{-7}$}{\footnotesize $^{+8}_{-8}$}} &
  {373{\footnotesize $^{+17}_{-15}$}{\footnotesize $^{+16} _{-16}$}} &
  {821{\footnotesize $^{+40}_{-42}$}{\footnotesize $^{+24}_{-31}$}}
  \\
  $\sigma_{\rm NNLO,\, approx}$ {\footnotesize (scheme A)} &
  6.14{\footnotesize $^{+0.49}_{-0.53}$}{\footnotesize $^{+0.31}_{-0.23}$} &
  146{\footnotesize $^{+13}_{-12}$}{\footnotesize $^{+8}_{-8}$} &
  369{\footnotesize $^{+34}_{-30}$}{\footnotesize $^{+16}_{-16}$} &
  821{\footnotesize $^{+71}_{-65}$}{\footnotesize $^{+27}_{-29}$}
  \\
  $\sigma_{\rm NNLO,\, approx}$ {\footnotesize (scheme B)} &
  6.05{\footnotesize $^{+0.43}_{-0.50}$}{\footnotesize $^{+0.31}_{-0.23}$} &
  139{\footnotesize $^{+9}_{-9}$}{\footnotesize $^{+7}_{-7}$} &
  349{\footnotesize $^{+23}_{-23}$}{\footnotesize $^{+15}_{-15}$} &
  773{\footnotesize $^{+47}_{-50}$}{\footnotesize $^{+25}_{-27}$}
  \\
  \hline
\end{tabular}
\end{center}
\vspace{-2mm}
\caption{\label{tab:CrossSections400} 
Results for the total cross section in pb, using the default choice $\mu_f=400$\,GeV. The first set of errors refers to perturbative uncertainties associated with scale variations, the second to PDF uncertainties. The most advanced prediction is the NLO+NNLL expansion highlighted in gray.}
\vspace{4mm}
\begin{center}
\begin{tabular}{|l|c|c|c|c|}
  \hline
  & Tevatron &  LHC (7\,TeV) & LHC (10\,TeV) & LHC (14\,TeV)
  \\
  \hline
  $\sigma_{\rm LO}$  & 
  6.66{\footnotesize $^{+2.95}_{-1.87}$}{\footnotesize $^{+0.34}_{-0.27}$} &
  122{\footnotesize $^{+49}_{-32}$}{\footnotesize $^{+6}_{-7}$}  &
  305{\footnotesize $^{+112}_{-76}$}{\footnotesize $^{+14}_{-16}$} &
  681{\footnotesize $^{+228}_{-159}$}{\footnotesize $^{+26}_{-34}$}
  \\ 
  \hline
  $\sigma_{\rm NLL}$  & 
  5.20{\footnotesize $^{+0.40}_{-0.36}$}{\footnotesize $^{+0.29}_{-0.19}$} &
  103{\footnotesize $^{+17}_{-14}$}{\footnotesize $^{+5}_{-5}$} &
  253{\footnotesize $^{+44}_{-36}$}{\footnotesize $^{+10}_{-10}$} &
  543{\footnotesize $^{+101}_{-88}$}{\footnotesize $^{+18}_{-19}$}
  \\
  $\sigma_{\text{NLO, leading}}$  & 
  6.42{\footnotesize $^{+0.42}_{-0.76}$}{\footnotesize $^{+0.35}_{-0.23}$} &
  152{\footnotesize $^{+7}_{-15}$}{\footnotesize $^{+8}_{-8}$} &
  381{\footnotesize $^{+12}_{-32}$}{\footnotesize $^{+16}_{-17}$} &
  835{\footnotesize $^{+18}_{-60}$}{\footnotesize $^{+29}_{-30}$}
  \\
  $\sigma_{\rm NLO}$  & 
  6.72{\footnotesize $^{+0.36}_{-0.76}$}{\footnotesize $^{+0.37}_{-0.24}$} &
  159{\footnotesize $^{+20}_{-21}$}{\footnotesize $^{+8}_{-9}$}  &
  402{\footnotesize $^{+49}_{-51}$}{\footnotesize $^{+17}_{-18}$} &
  889{\footnotesize $^{+107}_{-106}$}{\footnotesize $^{+31}_{-32}$}
  \\ 
  \hline 
  &&&& \\[-5.9mm] 
  \gr {$\sigma_{\rm NLO+NNLL}$}  & 
  {6.48{\footnotesize $^{+0.17}_{-0.21}$}{\footnotesize $^{+0.32}_{-0.25}$}} & 
  {146{\footnotesize $^{+7}_{-7}$}{\footnotesize $^{+8}_{-8}$}} &
  {368{\footnotesize $^{+20} _{-14}$}{\footnotesize $^{+19} _{-15}$}} &
  {813{\footnotesize $^{+50}_{-36}$}{\footnotesize $^{+30}_{-35}$}}
  \\
  $\sigma_{\rm NNLO,\, approx}$ {\footnotesize (scheme A)} & 
  6.72{\footnotesize $^{+0.45}_{-0.47}$}{\footnotesize $^{+0.33}_{-0.24}$} & 
  162{\footnotesize $^{+19}_{-14}$}{\footnotesize $^{+9}_{-9}$} &
  411{\footnotesize $^{+49}_{-35}$}{\footnotesize $^{+17}_{-20}$  } &
  911{\footnotesize $^{+111}_{-77}$}{\footnotesize $^{+35}_{-32}$}
  \\
  $\sigma_{\rm NNLO,\, approx}$ {\footnotesize (scheme B)} &
  6.55{\footnotesize $^{+0.32}_{-0.41}$}{\footnotesize $^{+0.33}_{-0.24}$} & 
  149{\footnotesize $^{+10}_{-9}$}{\footnotesize $^{+8}_{-8}$}&
  377{\footnotesize $^{+28}_{-23}$}{\footnotesize $^{+16}_{-18}$} &
  832{\footnotesize $^{+65}_{-50}$}{\footnotesize $^{+31}_{-29}$} 
  \\
  \hline
\end{tabular}
\end{center}
\vspace{-2mm}
\caption{\label{tab:CrossSections} 
Same as Table~\ref{tab:CrossSections400}, but with the ``educated'' scale choice $\mu_f=m_t$.}
\end{table}
%%%%%%%%%%%%%%%%%%%%%%%%%%%%%%%%%%%%%%%%%%%%%%%%%%%%%%%%%%%%%%%%%%%%%%%%%%%%%%%%

A few comments are in order concerning the results shown in the table. At NLO the cross
sections $\sigma_{\text{NLO, leading}}$ evaluated using only the leading singular terms
from the threshold expansion reproduce between 95\% (for the Tevatron) to almost 100\%
(for the LHC) of the exact fixed-order result at the default values of the factorization
scale. The subleading terms in $(1-z)$, obtained by integrating $d\sigma_{\text{NLO,
    subleading}}$, contribute the remaining few percent. In other words, the singular
terms capture about 85\% of the NLO correction at the Tevatron and practically 100\% of it
at the LHC. We cannot say whether the threshold expansion works so well also at higher
orders in perturbation theory, although this does not seem unreasonable. Our best
prediction is obtained by matching the fixed-order result with the resummed result at
NLO+NNLL accuracy and is highlight in gray. The effect of resummation is roughly a
10--15\% enhancement over the fixed-order NLO result. A more important effect is that the
resummation stabilizes the scale dependence significantly. Concerning the approximate NNLO
schemes, the results from scheme~A are noticeably higher than those from scheme~B, but
these differences are well inside the quoted errors. Since the two schemes differ only by
terms proportional to $\delta(1-z)$, this gives an indication of the size of the unknown
constant terms.

To some extent, the enhancement effect resulting from the resummation of the leading
threshold terms can be mimicked using fixed-order results evaluated at a significantly
lower factorization and renormalization scale $\mu_f$. Such an ``educated'' scale choice,
which is often adopted in the literature on fixed-order calculations, is $\mu_f=m_t$.
Table~\ref{tab:CrossSections} shows the cross-section predictions obtained in this case.
The fixed-order results are indeed significantly enhanced with this scale choice. The
resummed predictions, on the other hand, do not change much compared to those shown in
Table~\ref{tab:CrossSections400}.

\subsection{Total cross section: Comparison with previous calculations}

\subsubsection{Small-$\bm\beta$ expansion}

The approach pursued here offers an alternative to the direct threshold expansion of the
total partonic cross section in the limit $\beta\to 0$, corresponding to $\hat s\to
4m_t^2$. In this case not only the phase-space for real gluon emissions shrinks to zero,
but in addition the top and anti-top quarks are produced at rest in the partonic
center-of-mass frame, which implies that in addition to soft-gluon singularities one
encounters Coulomb singularities. The leading terms in the $\beta\to 0$ limit were first
calculated at NNLO in \cite{Moch:2008qy, Langenfeld:2009wd}, and later corrected in
\cite{Beneke:2009ye}.

It is important to emphasize that the leading singular contributions to the total cross
section arising from the $\beta\to 0$ limit do not coincide with those arising from the
limit $z\to 1$, even after integrating over all kinematic variables. The reason is that
after convolution with the PDFs there are no truly small scale ratios left in the process
(the total center-of-mass energy $\sqrt{s}$ at the Tevatron or LHC are so large that they
can be taken to infinity compared with the scale $m_t$). The large perturbative
corrections to the cross section arise dynamically, because of the relatively strong
fall-off of the parton luminosities combined with the fact that the partonic cross
sections receive their dominant contributions from the region near Born-level kinematics
\cite{Becher:2007ty, Ahrens:2008nc}. One would then expect that the most accurate account
of enhanced perturbative corrections should be the one that captures enhanced
contributions in all relevant regions of phase space.

%%%%%%%%%%%%%%%%%%%%%%%%%%%%%%%%%%%%%%%%%%%%%%%%%%%%%%%%%%%%%%%%%%%%%%%%%%%%%%%%
\begin{figure}
\begin{center}
\includegraphics[width=0.5\textwidth]{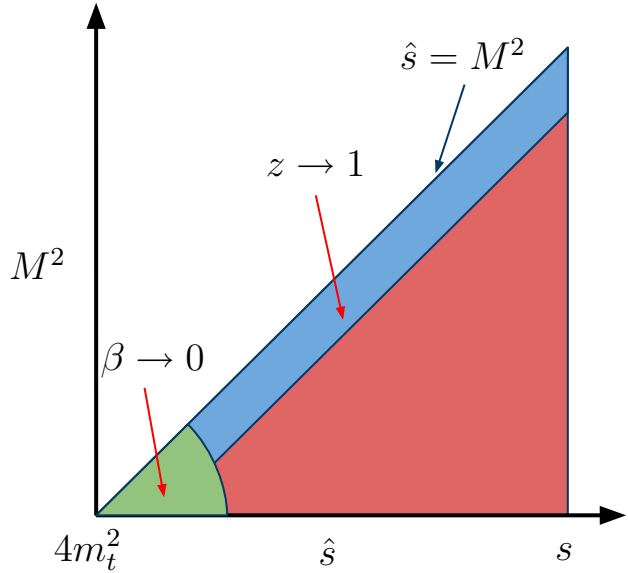} 
\end{center}
\vspace{-2mm}
\caption{\label{fig:phasespace} Phase space in the $(\hat s,M^2)$ plane. In the blue
  region along the diagonal threshold singularities arise, and the cross sections receives
  its main contributions. In the small green region near the origin Coulomb singularities
  appear and the small-$\beta$ expansion applies.}
\end{figure}
%%%%%%%%%%%%%%%%%%%%%%%%%%%%%%%%%%%%%%%%%%%%%%%%%%%%%%%%%%%%%%%%%%%%%%%%%%%%%%%%

In terms of the variables $\hat s$ and $M^2$, the phase space is given by the triangular
region $4m_t^2\le M^2\le\hat s\le s$, as illustrated in Figure~\ref{fig:phasespace}. The
large threshold terms considered in the present paper and in our previous work
\cite{Ahrens:2009uz} are located along the diagonal, where the partonic cross sections are
largest. The large corrections arising in the $\beta\to 0$ limit are located near the
origin of the diagram, where both $\hat s$ and $M^2$ approach $4m_t^2$. The parton
luminosities are largest for small values of $\hat s$. It is obvious from the figure, and
also by considering the invariant mass distributions shown in
Figure~\ref{fig:SpectrumPDFS}, that the region near the origin gives only a very small
contribution to the total cross section. This fact is most clearly demonstrated by the
distributions in the variable $\beta_t$ shown in Figure~\ref{fig:betat}, which peak at
$\beta_t\approx 0.6$ (Tevatron) and $\beta_t\approx 0.7$ (LHC). The region of small
velocity, say below $\beta_t=0.2$, obviously yields very small contributions to the total
cross sections. Since the variable $\beta$ is always larger than $\beta_t$, this
conclusion is even more true for the small-$\beta$ region. On the contrary, the approach
pursued in the present work accounts for enhanced perturbative contributions in all
regions of phase space giving rise to large contributions to the total cross section. It
is completely analogous to threshold (or soft-gluon) resummation for Drell-Yan or Higgs
production at fixed value of the lepton pair or Higgs boson mass. Even though one can
never be sure how accurately the full NNLO correction to a cross sections is approximated
by a subset of calculable terms, we strongly believe that our treatment provides an
approximation that captures more physics than that based on the $\beta\to 0$ limit. We
therefore expect our predictions for the invariant mass distribution, total cross section,
and forward-backward asymmetry to be the most reliable available at present.

Having just argued that the $\beta\to 0$ limit is not of much relevance for the total
cross section, it is nevertheless interesting to study how well our predictions fare in
this region. Since in our case the top quarks are generically not at rest in their
center-of-mass frame, we are not dealing with Coulomb singularities, and hence our
approximate prediction for the NNLO corrections to the cross section misses a subset of
terms involving potential-gluon exchange. We will now study in more detail which of the
singular terms in the $\beta\to 0$ limit can be recovered in our approach. To this end, we
write the total cross section in the form
\begin{align}
  \sigma(s,m^2_t) = \frac{\alpha_s^2}{m_t^2} \sum_{ij} \int_{4 m_t^2}^s \frac{d
    \hat{s}}{s} \, \ff_{ij}\bigg(\frac{\hat{s}}{s},\mu\bigg)\,f_{ij} \bigg( \frac{4
    m_t^2}{\hat{s}} ,\mu \bigg) \,.
\end{align}
We can obtain an expression for the perturbative functions $f_{ij}$ by integrating
(\ref{eq:genfact}) over all of phase-space, in which case we find
\begin{align}
  \alpha_s^2\,f_{ij}\bigg(\frac{4m_t^2}{\hat s},\mu \bigg)=\frac{8\pi m_t^2}{3\hat s}
  \int_{2m_t}^{\sqrt{\hat s}} \frac{dM}{M}\int_{-1}^{1} d\cos\theta
  \sqrt{1-\frac{4m_t^2}{M^2}}\,C_{ij} \bigg(\frac{M^2}{\hat s},M,m_t,\cos\theta,\mu\bigg)
  \, .
\end{align}
We can now evaluate the above formula in the limit $\hat s \to 4m_t^2$. Defining expansion
coefficients for the functions $f_{ij}$ as
\begin{align}
  \label{eq:fexp}
  f_{ij} &= f_{ij}^{(0)} + 4\pi \alpha_s f_{ij}^{(1)} + \left(4\pi\alpha_s\right)^2 \left[
    f_{ij}^{(2,0)} + f_{ij}^{(2,1)} \ln\left(\frac{\mu_f^2}{m_t^2}\right) + f_{ij}^{(2,2)}
    \ln^2\left(\frac{\mu_f^2}{m_t^2}\right) \right] + \dots ,
\end{align}
the answer for the scale-independent pieces with $n_h=1$, $n_l=5$, and $N=3$ can be
written as
\begin{align}
  \label{eq:betaexpans}
  f_{q\bar q}^{(2,0)} &= \frac{1}{(16\pi^2)^2} \frac{\pi\beta}{9} \Big[ 910.22 \ln^4\beta
  - 1315.5 \ln^3\beta + 592.29\ln^2\beta + 452.52 \ln\beta \nonumber
  \\
  &\quad -\frac{1}{\beta} \left( 140.37 \ln^2\beta + 18.339 \ln\beta - 72.225 \right) +
  f_{q\bar q}^{{\rm potential}} \Big] + \ldots , \nonumber
  \\
  f_{gg}^{(2,0)} &= \frac{1}{(16\pi^2)^2} \frac{7\pi\beta}{192} \Big[ 4608.0 \ln^4\beta -
  1894.9 \ln^3\beta - 912.35\ln^2\beta + 2747.5 \ln\beta \nonumber
  \\
  &\quad + \frac{1}{\beta} \left( 496.30 \ln^2\beta + 400.41 \ln\beta - 236.22 \right) +
  f_{gg}^{{\rm potential}} \Big] + \ldots ,
\end{align}
where the dots refer to ${\cal O}(\beta\ln^0\!\beta)$ terms, which are yet unknown. We
have split the answer into the piece recovered from the expansion of our results, which we
have written explicitly, and a piece related to NNLO effects from potential gluons, which
would be recovered from the small-$\beta$ expansion of the as yet unknown
$\mu$-independent part of the NNLO coefficient $C_0$ in (\ref{eq:C2}). Such
potential-gluon contributions were obtained in \cite{Beneke:2009ye} by using the two-loop
calculations of \cite{Czarnecki:1997vz, Beneke:1999qg, Czarnecki:2001gi}, and lead to the
additional terms
\begin{align}
  \label{eq:potential}
  f_{q\bar q}^{\rm potential} &= \frac{3.6077}{\beta^2} + \frac{1}{\beta} \left(
    50.445\ln\beta - 68.274 \right) + 76.033 \ln\beta \, , \nonumber
  \\
  f_{gg}^{\rm potential} &= \frac{68.547}{\beta^2} + \frac{1}{\beta} \left(
    -79.270\ln\beta + 227.59 \right) - 290.76 \ln\beta \, .
\end{align}
The coefficients $f_{ij}^{(2,1)}$ and $f_{ij}^{(2,2)}$ (with $ij=q\bar q,gg$) in
(\ref{eq:fexp}) are also recovered from the expansion of our results, up to terms of
${\cal O}(\beta^2)$. We have checked that they agree with the results given in
\cite{Langenfeld:2009wd} when expanded to that order.

%%%%%%%%%%%%%%%%%%%%%%%%%%%%%%%%%%%%%%%%%%%%%%%%%%%%%%%%%%%%%%%%%%%%%%%%%%%%%%%%
\begin{table}[t]
\begin{center}
\begin{tabular}{|l|c|c|c|c|}
  \hline
  & Tevatron &  LHC (7\,TeV) & LHC (10\,TeV) & LHC (14\,TeV)
  \\
  \hline
  $\sigma_{\rm NLO}$ &
  5.79{\footnotesize $^{+0.79}_{-0.80}$}{\footnotesize $^{+0.33}_{-0.22}$} &
  133{\footnotesize $^{+21}_{-19}$}{\footnotesize $^{+7}_{-7}$}  &
  341{\footnotesize $^{+50}_{-46}$}{\footnotesize $^{+14}_{-15}$} &
  761{\footnotesize $^{+105}_{-101}$}{\footnotesize $^{+26}_{-27}$}
  \\ 
  $\sigma_{\text{NLO, leading}}$ &
  5.49{\footnotesize $^{+0.78}_{-0.78}$}{\footnotesize $^{+0.31}_{-0.20}$} &
  134{\footnotesize $^{+16}_{-17}$}{\footnotesize $^{+7}_{-7}$} &
  341{\footnotesize $^{+34}_{-38}$}{\footnotesize $^{+14}_{-14}$} &
  761{\footnotesize $^{+64}_{-75}$}{\footnotesize $^{+25}_{-26}$}
  \\
  $\sigma_{\text{NLO, $\beta$-exp.\ v1}}$ &
  8.22{\footnotesize $^{+0.54}_{-0.88}$}{\footnotesize $^{+0.49}_{-0.33}$} &
  157{\footnotesize $^{+12}_{-16}$}{\footnotesize $^{+8}_{-8}$} &
  395{\footnotesize $^{+24}_{-36}$}{\footnotesize $^{+14}_{-15}$} &
  877{\footnotesize $^{+49}_{-73}$}{\footnotesize $^{+29}_{-30}$}
  \\
  $\sigma_{\text{NLO, $\beta$-exp.\ v2}}$ &
  6.59{\footnotesize $^{+0.96}_{-0.95}$}{\footnotesize $^{+0.38}_{-0.25}$} &
  151{\footnotesize $^{+15}_{-18}$}{\footnotesize $^{+8}_{-8}$} &
  386{\footnotesize $^{+30}_{-39}$}{\footnotesize $^{+15}_{-16}$} &
  863{\footnotesize $^{+49}_{-73}$}{\footnotesize $^{+29}_{-30}$}
  \\
  \hline 
  &&&& \\[-5.9mm] 
  \gr {$\sigma_{\rm NLO+NNLL}$} &
  6.30{\footnotesize $^{+0.19}_{-0.19}$}{\footnotesize $^{+0.31}_{-0.23}$} &
  149{\footnotesize $^{+7}_{-7}$}{\footnotesize $^{+8}_{-8}$} &
  373{\footnotesize $^{+17}_{-15}$}{\footnotesize $^{+16} _{-16}$} &
  821{\footnotesize $^{+40}_{-42}$}{\footnotesize $^{+24}_{-31}$}
  \\
  $\sigma_{\text{NNLO, $\beta$-exp. v1}}$ &
  7.37{\footnotesize $^{+0.01}_{-0.20}$}{\footnotesize $^{+0.39}_{-0.29}$} &
  156{\footnotesize $^{+2}_{-5}$}{\footnotesize $^{+8}_{-8}$} &
  392{\footnotesize $^{+4}_{-11}$}{\footnotesize $^{+16}_{-17}$} &
  865{\footnotesize $^{+5}_{-17}$}{\footnotesize $^{+29}_{-30}$}
  \\
  $\sigma_{\text{NNLO, $\beta$-exp.+potential v1}}$ &
  7.30{\footnotesize $^{+0.01}_{-0.18}$}{\footnotesize $^{+0.39}_{-0.28}$} &
  158{\footnotesize $^{+3}_{-6}$}{\footnotesize $^{+8}_{-8}$} &
  398{\footnotesize $^{+7}_{-13}$}{\footnotesize $^{+16}_{-17}$} &
  880{\footnotesize $^{+12}_{-22}$}{\footnotesize $^{+29}_{-31}$}
  \\
  $\sigma_{\text{NNLO, $\beta$-exp.\ v2}}$ &
  6.98{\footnotesize $^{+0.17}_{-0.40}$}{\footnotesize $^{+0.37}_{-0.27}$} &
  156{\footnotesize $^{+2}_{-6}$}{\footnotesize $^{+8}_{-8}$} &
  394{\footnotesize $^{+2}_{-10}$}{\footnotesize $^{+16}_{-17}$} &
  871{\footnotesize $^{+0}_{-14}$}{\footnotesize $^{+29}_{-31}$}
  \\
  $\sigma_{\text{NNLO, $\beta$-exp.+potential v2}}$ &
  6.95{\footnotesize $^{+0.16}_{-0.39}$}{\footnotesize $^{+0.36}_{-0.26}$} &
  159{\footnotesize $^{+3}_{-7}$}{\footnotesize $^{+8}_{-8}$} &
  401{\footnotesize $^{+6}_{-12}$}{\footnotesize $^{+17}_{-17}$} &
  888{\footnotesize $^{+7}_{-19}$}{\footnotesize $^{+30}_{-32}$}
  \\
  \hline
\end{tabular}
\end{center}
\vspace{-2mm}
\caption{\label{tab:CrossSections400beta} 
Results for the total cross section in pb, using the default choice $\mu_f=400$\,GeV. Some numbers from Table~\ref{tab:CrossSections400} are compared with results obtained from different implementations of the small-$\beta$ expansion (see text for explanation). The errors have the same meaning as before.}
\end{table}
%%%%%%%%%%%%%%%%%%%%%%%%%%%%%%%%%%%%%%%%%%%%%%%%%%%%%%%%%%%%%%%%%%%%%%%%%%%%%%%%

It is worth noting that after obtaining the small-$\beta$ expansion as in
(\ref{eq:betaexpans}), one can replace the approximated Born prefactors, $\pi\beta/9$ in
the $q\bar{q}$ case and $7\pi\beta/192$ in the $gg$ case, with the exact Born-level
results. This procedure has been adopted in the recent literature on the small-$\beta$
expansion \cite{Moch:2008qy, Langenfeld:2009wd, Beneke:2009ye}. Therefore, we will
differentiate the two versions of the small-$\beta$ expansion: version~1 refers to the
version where a complete expansion is carried out as in (\ref{eq:betaexpans}), while
version~2 refers to the version with the exact Born-level results as prefactors.

\begin{figure}[t!]
\begin{center}
\begin{tabular}{lr}
\psfrag{t}[][][0.75]{$q\bar q$ channel}
\psfrag{y}[][][1][90]{$\alpha_s$ correction [pb]}
\psfrag{x}[][][1]{$\beta$}
\psfrag{z}[][][0.85]{$\sqrt{s}=1.96$\,TeV}
\includegraphics[width=0.43\textwidth]{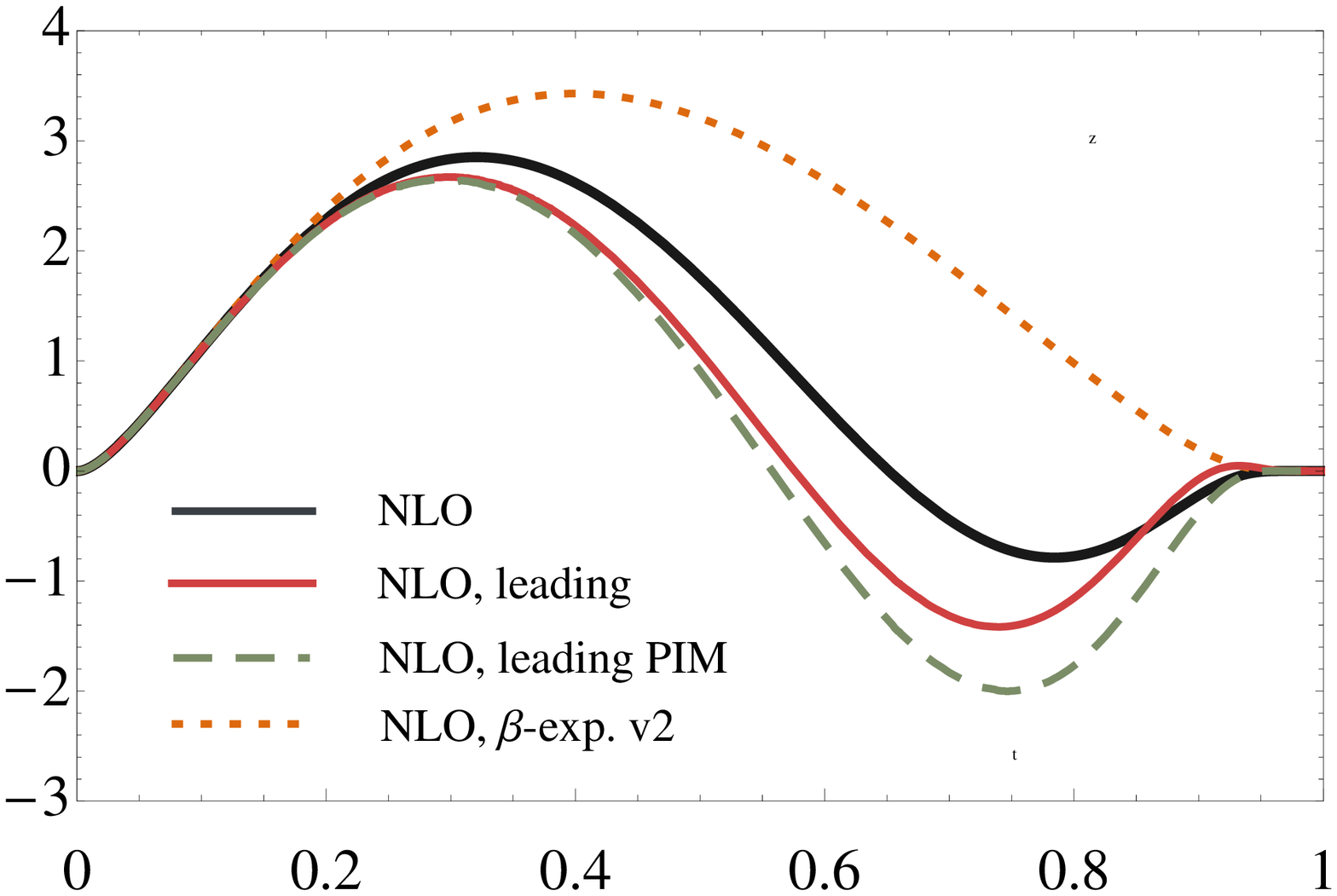} & 
\psfrag{t}[][][0.75]{$gg$ channel}
\psfrag{y}[][][1][90]{$\alpha_s$ correction [pb]}
\psfrag{x}[][][1]{$\beta$}
\psfrag{z}[][][0.85]{$\sqrt{s}=1.96$\,TeV}
\includegraphics[width=0.445\textwidth]{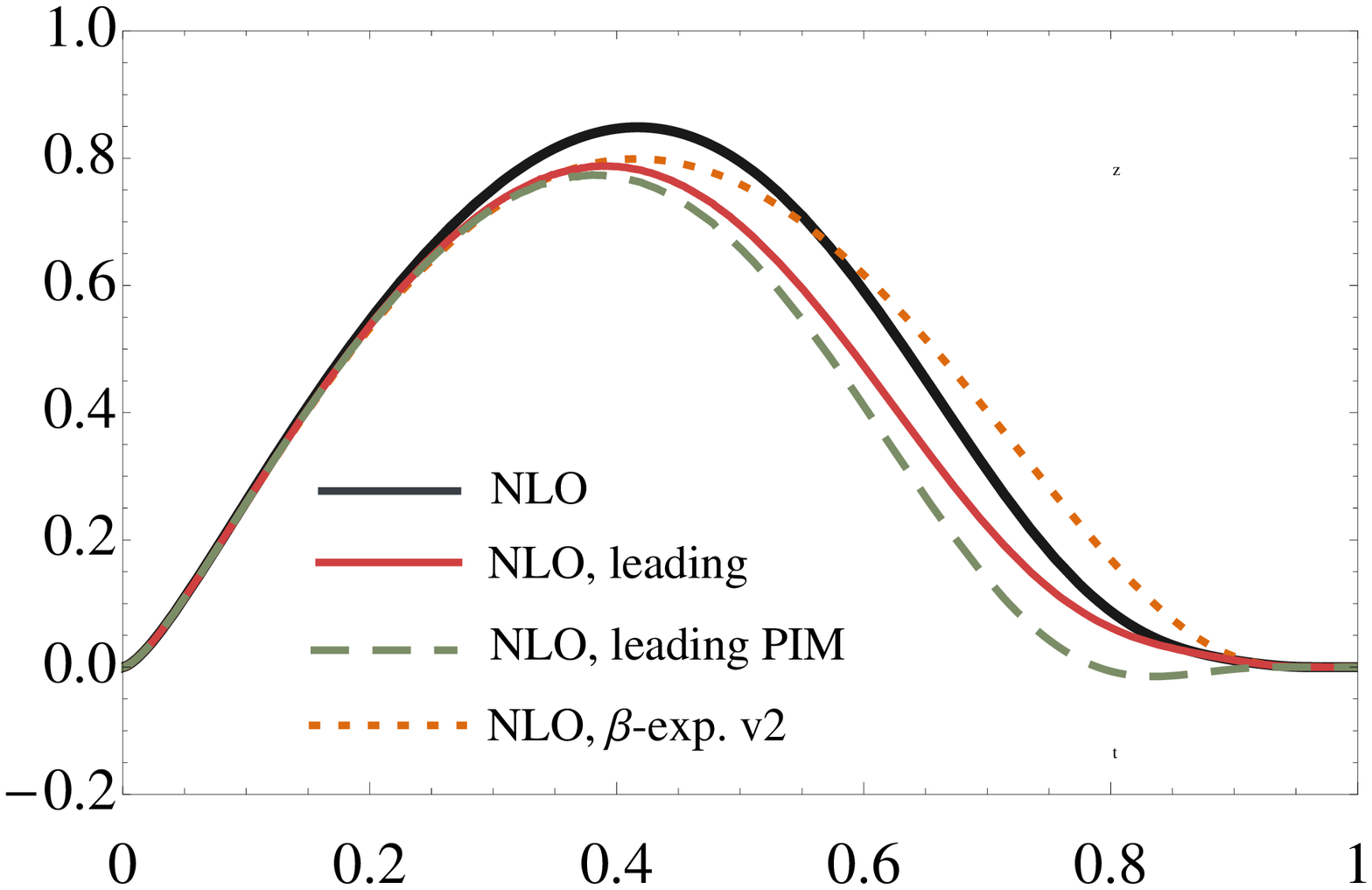} \\
& \\
\psfrag{t}[][][0.75]{$q\bar q$ channel}
\psfrag{y}[][][1][90]{$\alpha_s$ correction [pb]}
\psfrag{x}[][][1]{$\beta$}
\psfrag{z}[][][0.85]{$\sqrt{s}=7$\,TeV}
\includegraphics[width=0.43\textwidth]{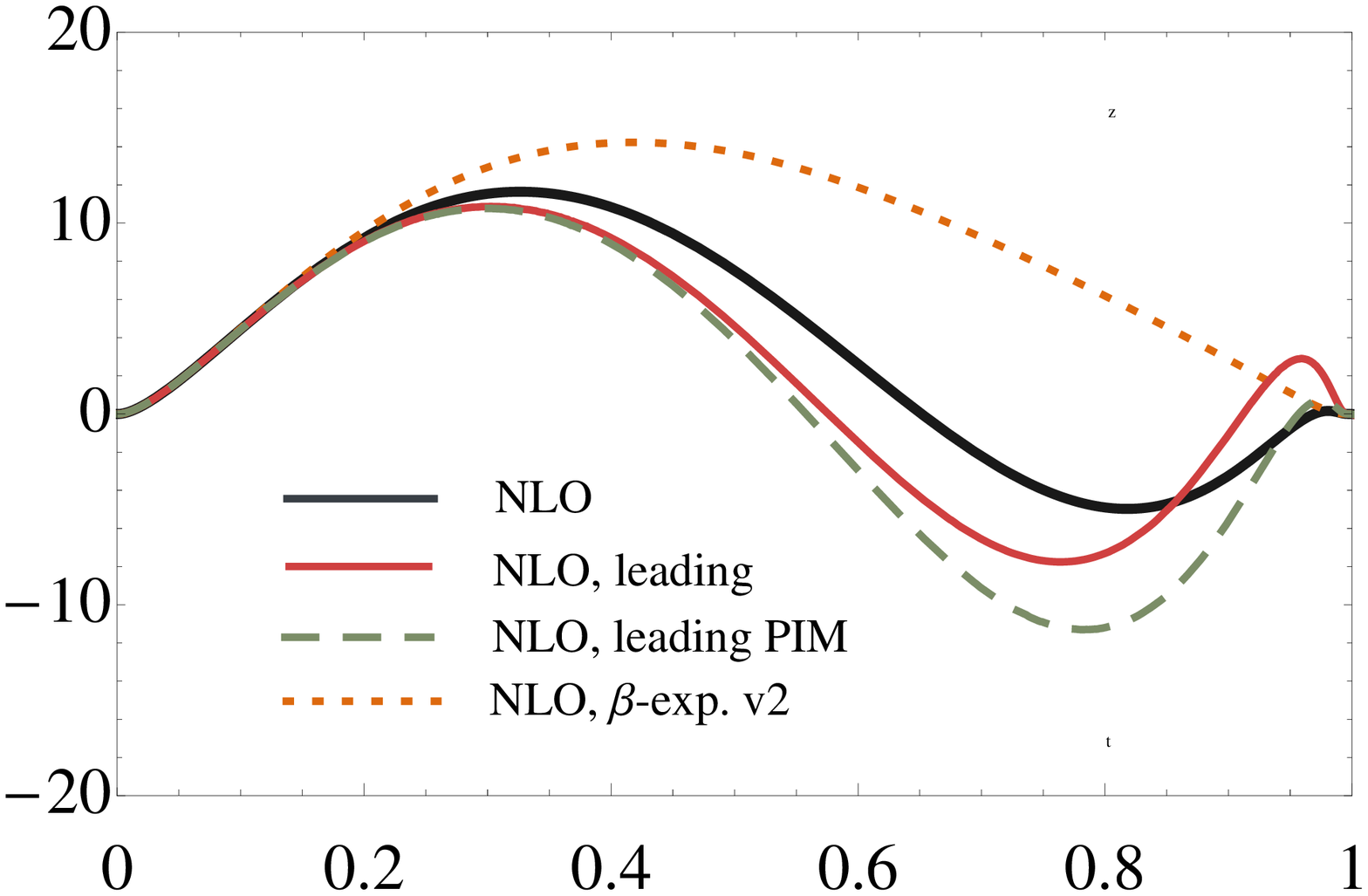} & 
\psfrag{t}[][][0.75]{$gg$ channel}
\psfrag{y}[][][1][90]{$\alpha_s$ correction [pb]}
\psfrag{x}[][][1]{$\beta$}
\psfrag{z}[][][0.85]{$\sqrt{s}=7$\,TeV}
\includegraphics[width=0.435\textwidth]{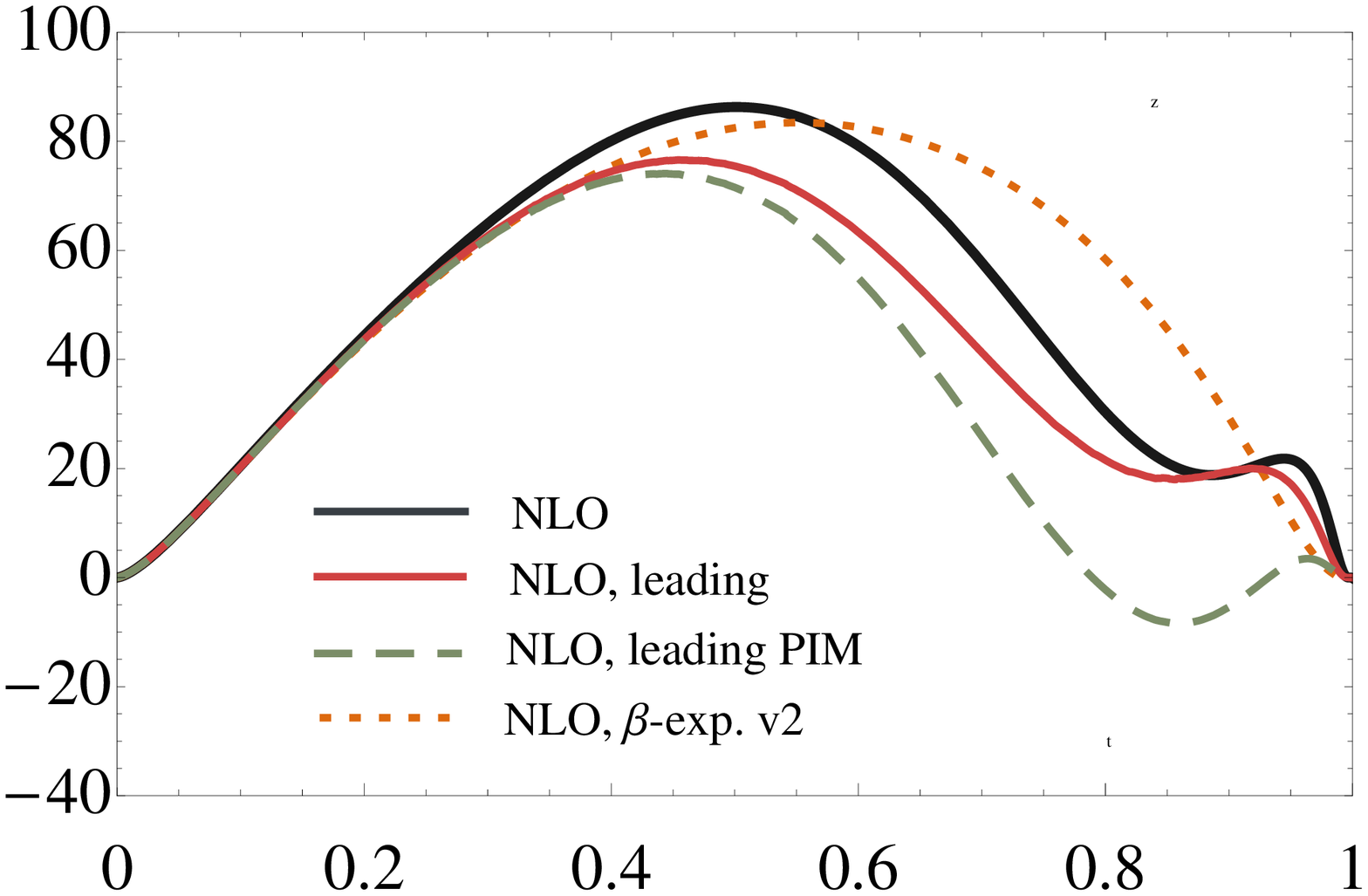} \\
& \\
\psfrag{t}[][][0.75]{$q\bar q$ channel}
\psfrag{y}[][][1][90]{$\alpha_s$ correction [pb]}
\psfrag{x}[][][1]{$\beta$}
\psfrag{z}[][][0.85]{$\sqrt{s}=14$\,TeV~~~}
\includegraphics[width=0.43\textwidth]{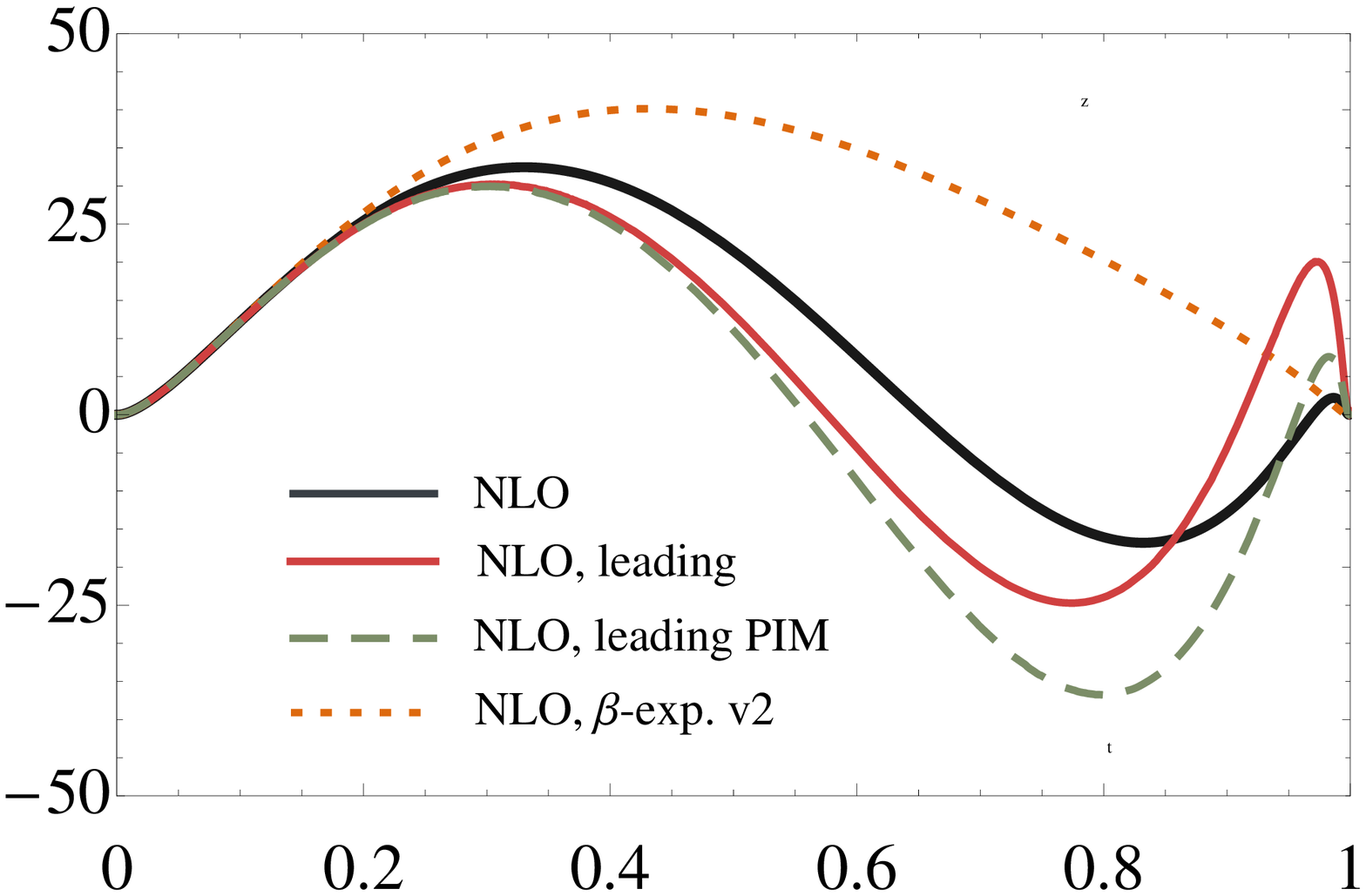} & 
\psfrag{t}[][][0.75]{$gg$ channel}
\psfrag{y}[][][1][90]{$\alpha_s$ correction [pb]}
\psfrag{x}[][][1]{$\beta$}
\psfrag{z}[][][0.85]{~~~~$\sqrt{s}=14$\,TeV}
\includegraphics[width=0.44\textwidth]{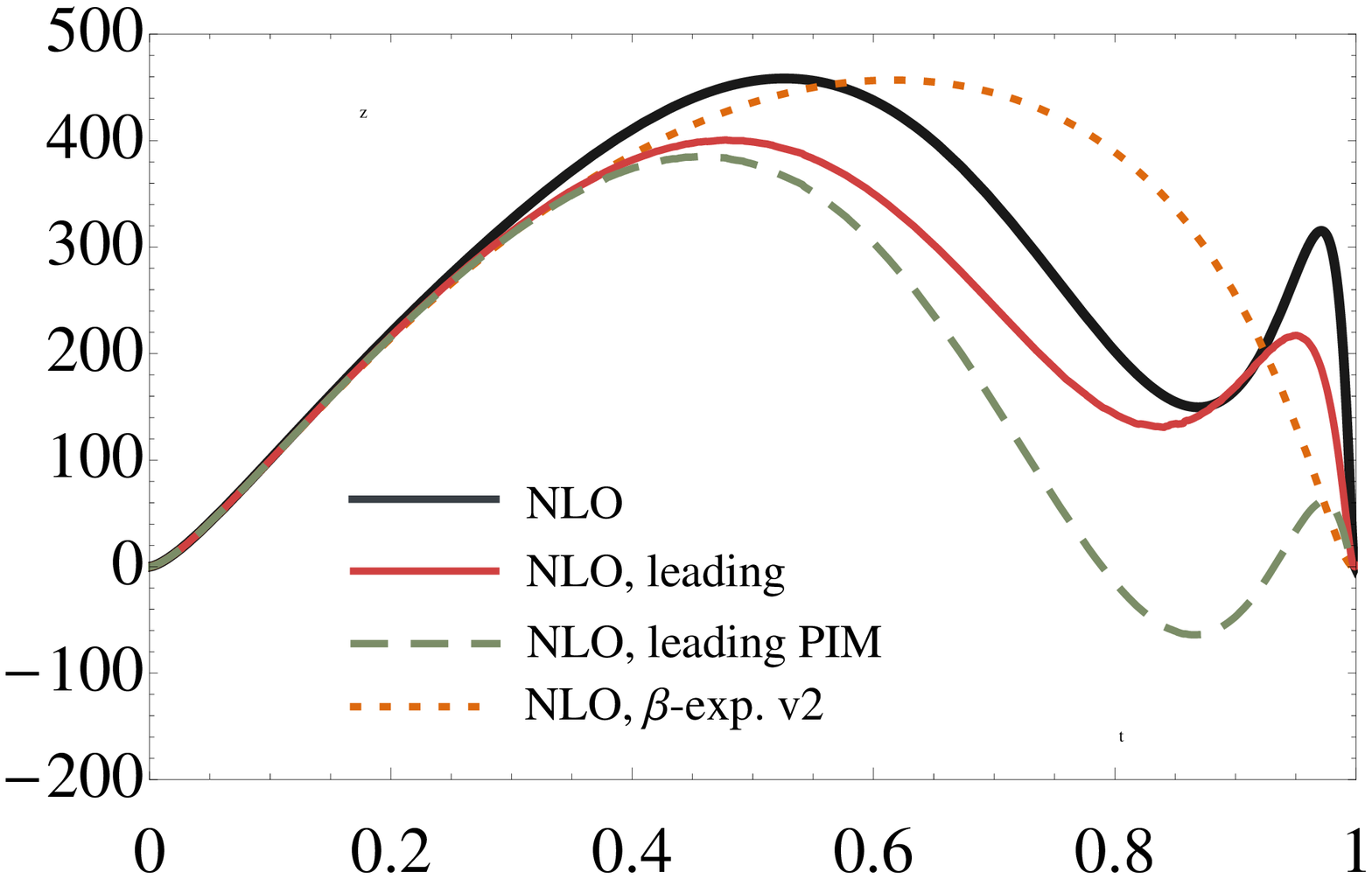} 
\end{tabular}
\end{center}
\vspace{-2mm}
\caption{\label{fig:beta-plot} The $\alpha_s$ corrections to $d\sigma/d\beta$ for the
  different approximations mentioned in the text at the Tevatron and LHC, with
  $\mu_f=m_t$. The plots on the left side show the $q\bar q$ channel, those on the right
  the $gg$ channel.}
\end{figure}

We are now ready to compare our results to those obtained using the small-$\beta$
expansion. We focus first on a comparison at NLO, where the exact answers are known. In
the upper portion of Table~\ref{tab:CrossSections400beta}, we show the results of the
different approximations as well as the exact NLO cross sections. As we have mentioned in
the last section, our NLO leading singular terms are always a good approximation to the
exact NLO results, at both the Tevatron and the LHC. On the other hand, the small-$\beta$
expansion version~1 tends to overestimate the cross section at the Tevatron by more than
2\,pb. By incorporating the exact Born prefactors (version~2) the small-$\beta$ expansion
works better, but still it overestimates the exact results by about 1\,pb. At the LHC, the
small-$\beta$ expansion happens to give results closer to the exact answers. However, as
we will now explain, this is a coincidence. In Figure~\ref{fig:beta-plot} we plot the NLO
corrections to the cross sections, including the parton luminosities, as functions of
$\beta$. The black solid curves show the exact results, the red solid curves our leading
singular terms, and the dotted curves the results obtained using version~2 of the
small-$\beta$ expansion (the dashed curves will be explained in the next subsection). In
the small-$\beta$ region, all the approximations work rather well as expected. With
increasing $\beta$, the different approximations start to deviate from one another. We
observe that, at both the Tevatron and the LHC, our approximations always reproduce the
shapes of the exact results quite well, which is not at all achieved by the small-$\beta$
expansion. The fact that the small-$\beta$ expansion overestimates the cross section at
the Tevatron, where the $q\bar{q}$ channel dominates, is evident from the left plots in
Figure~\ref{fig:beta-plot}. At the LHC, where the $gg$ channel dominates, the
small-$\beta$ expansion does not reproduce the shapes of the exact results, even though it
happens that the integrated cross sections are close to the exact ones due to a
coincidental cancellation. However, it is unlikely that a similar cancellation will happen
at NNLO.

We next compare our best prediction, NLO+NNLL, to the best prediction obtained using the
small-$\beta$ expansion at NNLO, which is obtained by adding the approximate NNLO
corrections derived using the small-$\beta$ expansion to the exact NLO cross sections.
Without knowing the exact expression for the NNLO corrections, it is hard to tell which
one is closer to the true answer, but we can study the validity of small-$\beta$ expansion
by investigating the effects of the subleading terms in $\beta$ that are contained in our
results. We have included in Table~\ref{tab:CrossSections400beta} the numerical results
for the cross section obtained by evaluating the small-$\beta$ expansion
(\ref{eq:betaexpans}) of our approximate NNLO formula, without including the extra
potential terms (labeled ``$\beta$-exp.'' in the table), and that obtained using this
expansion plus the potential terms in (\ref{eq:potential}) (labeled
``$\beta$-exp.+potential''). We notice that the NLO+NNLL resummed results and the
small-$\beta$ expansion differ by about 10--15\% at the Tevatron and about 6\% at the LHC.
On the other hand, the effect of adding the potential-gluon contributions to the
small-$\beta$ expansion, which cannot be reproduced in our approach, is always smaller
than 2\%. We conclude that the bulk of the terms that become singular in the $\beta\to 0$
limit are accounted for in our approach. The reverse statement is not true. A resummation
based on the $\beta\to 0$ expansion does {\em not\/} account for the bulk of the terms
that become singular in the $z\to 1$ limit, and our analysis suggests that subleading
terms in $\beta$ are by no means generically small.

The CDF collaboration has recently published a combined value for the $t\bar t$ cross
section, finding $\sigma_{\rm exp}=(7.50\pm 0.48)$\,pb \cite{cdf9913}. They have compared
their result with the theoretical prediction $\sigma_{\rm th}=(7.46^{+0.66}_{-0.80})$\,pb
derived using the small-$\beta$ expansion approach from \cite{Langenfeld:2009wd}. This
number is significantly higher than our best prediction
$\sigma_{\text{NLO+NNLL}}=(6.30^{+0.19+0.31}_{-0.19-0.23})$\,pb given in
Table~\ref{tab:CrossSections400}, but it is also higher than the value
$\sigma_{\text{NNLO,~$\beta$-exp.+potential v2}}=(6.95^{+0.16+0.36}_{-0.39-0.26})$\,pb
shown in Table~\ref{tab:CrossSections400beta}, which is our own result obtained using the
small-$\beta$ expansion. This latter difference is partly due to a smaller top-quark mass
(172.5\,GeV instead of 173.1\,GeV), a different PDF set (CTEQ6.6 instead of MSTW2008NNLO),
and a different value of $\alpha_s(M_Z)$ (0.118 instead of 0.117), and partly due to a
lower factorization scale ($m_t$ instead of 400~GeV). Changing our input values to those
used in \cite{Langenfeld:2009wd}, we would obtain
$\sigma_{\text{NLO+NNLL}}=(6.86^{+0.19+0.44}_{-0.20-0.35})$\,pb and
$\sigma_{\text{NNLO,~$\beta$-exp.+potential v2}}=(7.55^{+0.01+0.49}_{-0.27-0.38})$\,pb. We
observe that with these settings the predictions are indeed higher,\footnote{The
  difference between the central values of our NLO+NNLL predictions (6.86\,pb vs.\
  6.30\,pb) is 0.56\,pb, of which about 0.18\,pb is due to the scale choice, about
  0.10\,pb is due to the smaller $m_t$, and another 0.28\,pb is due to the different PDF
  set and $\alpha_s(M_Z)$ value.} and our $\sigma_{\text{NNLO,~$\beta$-exp.+potential
    v2}}$ is now very close to the result from \cite{Langenfeld:2009wd}. The remaining
slight difference can be attributed to the use of an incorrect two-loop coefficient in
\cite{Langenfeld:2009wd}, which was later corrected in \cite{Beneke:2009ye}. Our best
prediction $\sigma_{\text{NLO+NNLL}}$ is still about 10\% lower than the result obtained
using the small-$\beta$ expansion, which can be understood from the discussions in the
previous paragraphs, but is consistent with the experimental result within errors.

\subsubsection{Other approaches and kinematics}

Since in this paper we are interested in the invariant mass distribution of the $t\bar{t}$
pair, we have adopted the so-called pair-invariant mass (PIM) kinematics, which is defined
by
\begin{align}
  N_1(P_1) + N_2(P_2) \to t\bar{t}(p_3+p_4) + X(k) \, .
\end{align}
When calculating the transverse-momentum and rapidity distributions of the top quark, with
the momentum of the anti-top quark integrated over, one instead considers the one-particle
inclusive (1PI) kinematics, which can be written as
\begin{align}
  N_1(P_1) + N_2(P_2) \to t(p_3) + X'[\bar{t}](p_4') \, ,
\end{align}
where here the final state $X'$ contains the $\bar t$ quark accompanied by additional
emissions (see e.g.~\cite{Kidonakis:2000ui, Kidonakis:2001nj}). In both cases, the
threshold limit corresponds to the limit in which these extra emissions are soft, implying
$k^0 \to 0$ and $p_4'^2 \to m_t^2$, respectively. In the PIM case, this implies $M^2 \to
\hat{s}$, where $M^2=(p_3+p_4)^2$ and $\hat s=(p_1+p_2)^2$ can be calculated without
reference to the unobserved momentum $k$. In the 1PI case, it implies $s_4/\hat
s=(1-2E_3/\sqrt{\hat s})\to 0$, where $s_4=\hat s+t_1+u_1$ can be calculated from
(\ref{mandelstam}) without reference to the unobserved momentum $p_4'$, and $E_3$ denotes
the energy of the top quark in the partonic center-of-mass frame.

Although the PIM and 1PI kinematics are applicable in different differential
distributions, they can both be integrated over to obtain the total cross section. In the
sense that they are both applicable in the limit where the extra emissions are soft,
threshold resummation for the total cross section based on PIM or 1PI kinematics amounts
to resumming the same leading contributions, but differs by subleading corrections.
Several authors have found significant numerical differences between the results obtained
using the two kinematic schemes (see e.g.~\cite{Kidonakis:2000ui, Kidonakis:2001nj,
  Kidonakis:2003qe}). Therefore, it is interesting to work out the 1PI kinematics also in
our approach.

Before moving onto 1PI kinematics, it is however necessary to point out an important
difference between our results and previous ones obtained using PIM kinematics. In the
traditional approach, the leading singular terms in $(1-z)$ are written in terms of the
distributions $P_n(z)=[\ln^n(1-z)/(1-z)]_+$, while in our approach they are more naturally
written in terms of the distributions
\begin{align}
  P'_n(z) = \left[ \frac{1}{1-z} \ln^n \left( \frac{M^2(1-z)^2}{\mu^2z} \right) 
  \right]_+ .
\end{align}
The additional factor of $z$ in the logarithms is a subleading effect, but it is relevant
in practice. This has been studied in detail for the simpler cases of Drell-Yan
\cite{Becher:2007ty} and Higgs production \cite{Ahrens:2008nc}, but analogous remarks hold
also in the present case. Of crucial importance in this context is the fact that our
resummation method works directly in momentum space \cite{Becher:2006nr}, and as a result
the matching onto analytical fixed-order expressions is particularly transparent. We have
observed in previous work \cite{Becher:2007ty, Ahrens:2008nc} that at NLO in perturbation
theory {\em all\/} logarithmic corrections to Drell-Yan and Higgs production, even those
entering at subleading orders in $(1-z)$, can be expressed in terms of the ``hard'' and
``soft'' logarithms $\ln(M^2/\mu_f^2)$ and $\ln(\hat s\,(1-z)^2/\mu_f^2)$, where
$M^2=M_{l^+l^-}^2$ or $m_H^2$ as appropriate, and $\hat s=M^2/z$. Choosing $\hat
s\,(1-z)^2=M^2(1-z)^2/z$ as the argument of the soft functions in the SCET approach thus
provides an ``optimal'' matching onto fixed-order expressions, which automatically
accounts for subleading logarithms $\ln z$. We have observed that, with this choice, the
leading singular terms near partonic threshold account for the bulk of the NLO and NNLO
corrections to the Drell-Yan and Higgs production cross sections. In the present paper, we
have shown that the same is true for the NLO corrections to the $t\bar t$ production cross
sections at the Tevatron and LHC. Our resummation procedure is thus different, in the way
we treat subleading terms, from the traditional PIM scheme, and for the reason just
explained it is bound to be more accurate. This can be clearly seen from
Figure~\ref{fig:beta-plot}, where we have plotted the NLO leading singular terms both in
our scheme (red solid curves) and in the traditional PIM scheme (dashed curves). One can
find that our results always provide a better approximation to the exact NLO correction
than the traditional PIM results. Especially for the $gg$ channel at the LHC, where the
traditional PIM results were argued to be a bad approximation \cite{Kidonakis:2001nj,
  Kidonakis:2003qe}, our result still provides a decent approximation.

We now turn to the 1PI kinematics, in which case the relevant differential cross section is
\begin{align}
  \frac{d^2\sigma}{dT_1\,dU_1} = \frac{8\pi}{3s^2} \sum_{i,j} \int \frac{dx_1}{x_1}\,
  \frac{dx_2}{x_2} f_{i/N_1}(x_1,\mu_f) \, f_{j/N_2}(x_2,\mu_f) \,
  C'_{ij}(s_4,t_1,u_1,m_t,\mu_f) \, .
\end{align}
Here $T_1=(P_1-p_3)^2-m_t^2$ and $U_1=(P_2-p_3)^2-m_t^2$ are the hadronic counterparts of
$t_1$ and $u_1$. Near the partonic threshold $s_4 \to 0$, the hard-scattering kernels
$C'_{ij}$ can again be factorized as
\begin{align}
  C'_{ij}(s_4,t_1,u_1,m_t,\mu_f) = \Tr \left[ \bm{H}_{ij}(t_1,u_1,m_t,\mu_f) \,
    \bm{S}'_{ij}(s_4,t_1,u_1,m_t,\mu_f) \right] ,
\end{align}
where the hard functions $\bm{H}_{ij}$ are the same as in the PIM case, once we re-express
$M^2$ and $\cos\theta$ in terms of $t_1$ and $u_1$. The new soft functions $\bm{S}'_{ij}$
can be calculated from (\ref{eq:softintegrals}) with the replacement $k^0 \to v_4 \cdot
k$. After these changes, all techniques we have developed in this paper can be applied to
the 1PI case, too. In particular, we can derive a resummed expression for the
hard-scattering kernels
\begin{align}
  C'(s_4,t_1,u_1,m_t,\mu_f) &= \exp \left[ 2a_{\Gamma}(\mu_s,\mu_f)
    \ln\frac{m_t^2\mu_s^2}{t_1u_1} + 4a_{\gamma^{\phi}}(\mu_s,\mu_f) \right] \nonumber
  \\
  &\quad \times \Tr \Bigg[ \bm{U}(t_1,u_1,m_t,\mu_h,\mu_s) \, \bm{H}(t_1,u_1,m_t,\mu_h) \,
  \bm{U}^\dagger(t_1,u_1,m_t,\mu_h,\mu_s) \nonumber
  \\
  &\hspace{15mm} \times \tilde{\bm{s}}'(\partial_\eta,t_1,u_1,m_t,\mu_s) \Bigg]
  \frac{1}{s_4} \left(\frac{s_4}{m_t\mu_s}\right)^{2\eta}
  \frac{e^{-2\gamma_E\eta}}{\Gamma(2\eta)} \, .
\end{align}
By setting $\mu_h=\mu_s=\mu_f$ in the above formula, we can derive the leading singular
terms in 1PI kinematics. We have checked that the leading terms at NLO agree with those in
the literature \cite{Beenakker:1988bq, Laenen:1998qw}. A numerical analysis of the
resummation in 1PI kinematics is better to be done in the context of transverse-momentum
and rapidity distributions of the top quark, where a systematic study as we have done for
PIM kinematics can be carried out to determine the optimal choices of the scales $\mu_h$,
$\mu_s$ and $\mu_f$. Therefore we will leave it for a future analysis. Nevertheless, given
the arguments in the previous paragraphs, we do not expect significant differences in the
results for the total cross sections obtained using the two kinematic schemes.

\subsection{Charge and forward-backward asymmetries}
\label{sec:Afb}

The top-quark pair charge asymmetry is an important observable at the Tevatron, which
originates from the difference in the production rates for top and anti-top quarks at
fixed scattering angle or rapidity \cite{Kuhn:1998jr, Kuhn:1998kw}. To describe the total
charge asymmetry, we first introduce a charge-asymmetric cross section as the difference
between the production of top and anti-top quarks in the forward direction,
\begin{align}
  \label{eq:casym}
  \Delta\sigma_c \equiv \int_0^1 d\cos\theta \int_{2m_t}^{\sqrt{s}} dM \left[ \frac{d^2
      \sigma^{p\bar{p} \to t\bar{t} X}}{dMd\cos\theta} - \frac{d^2 \sigma^{p\bar{p} \to
        \bar{t} t X}}{dMd\cos\theta} \right] .
\end{align}
The total charge asymmetry is then defined by $A_c = \Delta\sigma_c/\sigma$, where
$\sigma$ is the total cross section. As a consequence of the charge conjugation invariance
of QCD, $\Delta\sigma_c$ is the same as the difference between the cross sections for the
production of top quarks in the forward and backward directions,
\begin{align}
  \label{eq:fbasym}
  \Delta\sigma_c = \Delta\sigma_{\text{FB}} \equiv \int_{2m_t}^{\sqrt{s}} dM \left[
    \int_0^1 d\cos\theta\,\frac{d^2 \sigma^{p\bar{p} \to t\bar{t} X}}{dMd\cos\theta} -
    \int_{-1}^0 d\cos\theta\,\frac{d^2 \sigma^{p\bar{p} \to t\bar{t} X}}{dMd\cos\theta}
  \right] .
\end{align}
Therefore, the charge asymmetry can also be interpreted as a forward-backward asymmetry
for the top quark, defined as $A^t_{\text{FB}} = \Delta\sigma_{\text{FB}}/\sigma$.

%%%%%%%%%%%%%%%%%%%%%%%%%%%%%%%%%%%%%%%%%%%%%%%%%%%%%%%%%%%%%%%%%%%%%%%%%%%%%%%%
\begin{figure}[t]
\begin{center}
\includegraphics[width=0.45\textwidth]{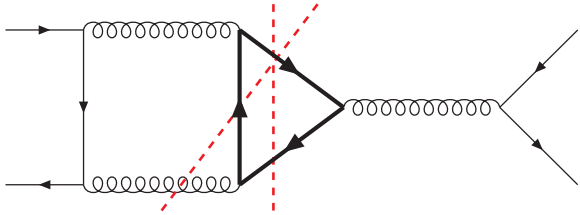}
\end{center}
\vspace{-2mm}
\caption{\label{fig:feyn} Examples of interferences contributing to the charge-asymmetric
  cross section. The two-particle cut corresponds to the interference of a one-loop box
  with the tree-level diagram, while the three-particle cut corresponds to the
  interference of tree-level diagrams with a $t\bar{t} g$ final state.}
\end{figure}
%%%%%%%%%%%%%%%%%%%%%%%%%%%%%%%%%%%%%%%%%%%%%%%%%%%%%%%%%%%%%%%%%%%%%%%%%%%%%%%%

At leading order in QCD ($\mathcal{O}(\alpha_s^2)$), the charge-asymmetric cross section
$\Delta\sigma_c$ vanishes. This quantity receives non-zero contributions starting at
$\mathcal{O}(\alpha_s^3)$. These arise if, in the interference of one-loop and tree-level
diagrams, the top-quark fermionic line and the light-quark fermionic line are connected by
three gluons. The same observation applies also to the interference of two tree-level
diagrams with three particles in the final state. In Figure~\ref{fig:feyn} we show the
interference of the planar box with the tree-level diagram and the corresponding
interference of real emission diagrams. The other contribution to the asymmetry at
${\mathcal O}(\alpha_s^3)$ in the quark-annihilation channel originates from the
interference of the crossed box and tree-level diagram (or from the corresponding real
emission case). This can be visualized by imagining to cross the two gluons on the left
side of the heavy-quark triangle in Figure~\ref{fig:feyn}. The color factors multiplying
the structure in Figure~\ref{fig:feyn} or its crossed counterpart are respectively
\begin{align}
  \mathcal{C}_{\text{planar}} = \frac{1}{16 N^2} \left( f_{abc}^2 + d_{abc}^2 \right) ,
  \qquad \mathcal{C}_{\text{crossed}} = \frac{1}{16 N^2} \left( d_{abc}^2 - f_{abc}^2
  \right) ,
\end{align}
where $f_{abc}^2=(N^2-1)N$ and $d_{abc}^2=(N^2-1)(N^2-4)/N$. When the color factors are
stripped off, the interference in Figure~\ref{fig:feyn} and its crossed counterpart
satisfy the relation
\begin{align}
  d \sigma_{\text{planar}}^{p\bar{p} \to t \bar{t} X} = -d
  \sigma_{\text{crossed}}^{p\bar{p} \to \bar{t} t X} \, .
  \label{eq:gfer}
\end{align}
This relation holds both for the three-particle and the two-particle cuts. Therefore, the
charge-asymmetric cross section is proportional to $d_{abc}^2$. The interference of the
one-loop box diagrams with the tree-level diagram gives a positive contribution to the
asymmetry, which is partially canceled by the asymmetry originating from the interference
of initial- and final-state radiation diagrams. An additional small contribution to the
asymmetry at $\mathcal{O}(\alpha_s^3)$ originates from the flavor excitation channel
$gq(\bar{q}) \to t \bar{t} X$ at tree level. The gluon-fusion channel does not contribute
to the charge-asymmetric cross section. The study of the charge-asymmetric cross section
at $\mathcal{O}(\alpha_s^3)$ shows that top quarks (anti-top quarks) are preferably
emitted of the direction of the incoming quark (anti-quark); consequently, in $p\bar{p}$
collisions top quarks are preferably emitted in the direction of the incoming proton
\cite{Kuhn:1998jr, Kuhn:1998kw}.

In the asymmetric cross section (\ref{eq:casym}), the definition of the scattering angle
$\theta$ depends on the reference frame, therefore the value of the asymmetry is also
frame-dependent. Experimentally the forward-backward asymmetry has been measured both in
the laboratory frame and in the $t\bar{t}$ rest frame \cite{:2007qb, Aaltonen:2008hc},
where the result in the former frame is about 70\% of the result in the latter. Near the
partonic threshold the $t\bar{t}$ rest frame and the partonic center-of-mass frame
coincide. Using the complete $\mathcal{O}(\alpha_s^3)$ expressions for $\Delta\sigma_{\rm
  FB}$ in \cite{Kuhn:1998kw}, we have checked explicitly that the results in these two
frames only differ slightly. It is therefore interesting to employ the formulas derived in
the previous sections to calculate the forward-backward asymmetry in the partonic
center-of-mass frame. The resummation of partonic threshold effects in the
forward-backward asymmetry at NLL order in Mellin space was first considered in
\cite{Almeida:2008ug}. In what follows we extend these calculations to NNLL order, and
also study the effect of the approximate NNLO corrections in fixed order.

%%%%%%%%%%%%%%%%%%%%%%%%%%%%%%%%%%%%%%%%%%%%%%%%%%%%%%%%%%%%%%%%%%%%%%%%%%%%%%%%
\begin{table}
\begin{center}
\begin{tabular}{|l|c|c|c|c|}
  \hline
  & \multicolumn{2}{c|}{$0.2<\mu_f/{\rm TeV}<0.8$}
  & \multicolumn{2}{c|}{$m_t/2<\mu_f<2m_t$} 
  \\
  \cline{2-5} & $\Delta\sigma_{\rm FB}$ [pb] &  $A^t_{\text{FB}}$ [\%]
  & $\Delta\sigma_{\rm FB}$ [pb] & $A^t_{\text{FB}}$ [\%]
  \\
  \hline
  NLL 
  & 0.29{\footnotesize $^{+0.16}_{-0.16}$} 
  & 5.8{\footnotesize $^{+3.3}_{-3.2}$} 
  & 0.31{\footnotesize $^{+0.16}_{-0.17}$} 
  & 5.9{\footnotesize $^{+3.4}_{-3.3}$}
  \\
  NLO, leading 
  & 0.19{\footnotesize $^{+0.09}_{-0.06}$} 
  & 5.2{\footnotesize $^{+0.4}_{-0.4}$} 
  & 0.31{\footnotesize $^{+0.16}_{-0.10}$} 
  & 5.7{\footnotesize $^{+0.5}_{-0.4}$}
  \\
  NLO 
  & 0.25{\footnotesize $^{+0.12}_{-0.07}$} 
  & 6.7{\footnotesize $^{+0.6}_{-0.4}$} 
  & 0.40{\footnotesize $^{+0.21}_{-0.13}$} 
  & 7.4{\footnotesize $^{+0.7}_{-0.6}$}
  \\ 
  \hline 
  &&&& \\[-5.9mm] 
  \gr NLO+NNLL
  & 0.40{\footnotesize $^{+0.06}_{-0.06}$}
  & 6.6{\footnotesize $^{+0.6}_{-0.5}$}
  & 0.45{\footnotesize $^{+0.08}_{-0.07}$} 
  & 7.3{\footnotesize $^{+1.1}_{-0.7}$}
  \\
  NNLO, approx (scheme A)
  & 0.37{\footnotesize $^{+0.10}_{-0.08}$} 
  & 6.4{\footnotesize $^{+0.9}_{-0.7}$}
  & 0.48{\footnotesize $^{+0.11}_{-0.10}$} 
  & 7.5{\footnotesize $^{+1.3}_{-0.9}$} 
  \\
  NNLO, approx (scheme B)
  & 0.34{\footnotesize $^{+0.08}_{-0.07}$}
  & 5.8{\footnotesize $^{+0.8}_{-0.6}$}
  & 0.45{\footnotesize $^{+0.09}_{-0.09}$}
  & 6.8{\footnotesize $^{+1.1}_{-0.8}$} 
  \\
  \hline  
\end{tabular}
\end{center}
\vspace{-2mm}
\caption{\label{tab:FB} 
The asymmetric cross section and forward-backward asymmetry at the Tevatron, evaluated at different orders in perturbation theory in the partonic center-of-mass frame. The errors refer to perturbative uncertainties related to scale variations, as explained in the text.}
\end{table}
%%%%%%%%%%%%%%%%%%%%%%%%%%%%%%%%%%%%%%%%%%%%%%%%%%%%%%%%%%%%%%%%%%%%%%%%%%%%%%%%

The results of our analysis are summarized in Table~\ref{tab:FB}. In the column labeled
$\Delta\sigma_{\rm FB}$, we show the total charge-asymmetric cross section evaluated at
different orders in perturbation theory. As in the previous sections, we have used
$m_t=173.1$\,GeV and adapted the order of the PDFs as described in
Table~\ref{tab:PDForder}. For the factorization scale, we consider our standard choice
$\mu_f=400$\,GeV as well as $\mu_f=m_t$. The results in resummed perturbation theory use
$\mu_h=M$ and $\mu_s$ chosen according to (\ref{eq:SoftScale}) by default, and the
uncertainties are obtained by varying these up and down by a factor of two and adding the
different uncertainties in quadrature. The uncertainties in the fixed-order results are
obtained by varying $\mu_f$ up and down by a factor of two. The counting used in the table
refers to the order at which the differential cross section itself is needed, relative to
$\alpha_s^2$; this differs slightly from the counting in fixed order used in
\cite{Kuhn:1998kw}, which would count, for instance, our NLO as LO. To obtain the result
in fixed order at NLO, we have used the formulas in Appendix A of \cite{Kuhn:1998kw}. At
both $\mu_f=400$\,GeV and $\mu_f=m_t$, the NLO threshold terms recover about 80\% of the
full result in fixed order, which is roughly in line with our findings for the NLO
corrections to the cross section. The table also includes our results for the
forward-backward asymmetry. In calculating the asymmetry, we first evaluate the numerator
and denominator of the ratio $A^t_{\text{FB}} = \Delta\sigma_{\text{FB}}/\sigma$ to a
given order in RG-improved or fixed-order perturbation theory, and then further expand the
ratio itself. When performing the calculation in this way, the errors in the asymmetry at
NLO are actually smaller than those at NLO+NNLL order, even though the scale variations in
the numerator and denominator of the ratio are much larger. We note, however, that if we
chose instead to not further expand the ratio, the NLO result would be decreased by about
20\%, while the NLO+NNLL result changes only by about 5\% and should therefore be
considered the more reliable prediction. In \cite{Kuhn:1998kw}, an overall factor of 1.09
due to electroweak corrections is included for the asymmetry. However, a smaller
correction was recently obtained in \cite{Bernreuther:2010ny}, and it is also
scale-dependent. Therefore we have chosen not to include these corrections in our results.

Our results are in good agreement with the previous findings reported in
\cite{Antunano:2007da, Almeida:2008ug}. We have not performed the calculation in the
$p\bar p$ frame, but expect that the boost to this frame decreases the asymmetry by
roughly 30\%, as found in \cite{Antunano:2007da}. On the other hand, the CDF collaboration
at the Tevatron recently reported the value $A_{\text{FB}}^t(\text{exp}) = (19.3 \pm
6.9)\, \%$ for the asymmetry in the $p\bar{p}$ frame \cite{cdf9724}. The measured
asymmetry exceeds the predicted one by about two standard deviations. In light of our
results, we conclude that higher-order QCD effects are not sufficient to explain the large
experimental value. Possible explanations of this fact in the framework of several new
physics scenarios were recently investigated in \cite{Antunano:2007da, Frampton:2009rk,
  Shu:2009xf, Dorsner:2009mq, Cao:2009uz, Cao:2010zb}.

\subsection{Sensitivity to the top-quark mass}
\label{sec:topmass}

%%%%%%%%%%%%%%%%%%%%%%%%%%%%%%%%%%%%%%%%%%%%%%%%%%%%%%%%%%%%%%%%%%%%%%%%%%%%%%%%
\begin{figure}[t!]
\begin{center}
\psfrag{x}[][][1][90]{$\sigma$ [pb]}
\psfrag{y}[][][1]{$m_t$ [GeV]}
\psfrag{z}[][][0.85]{$\sqrt{s}=1.96$\,TeV}
\includegraphics[width=0.5\textwidth]{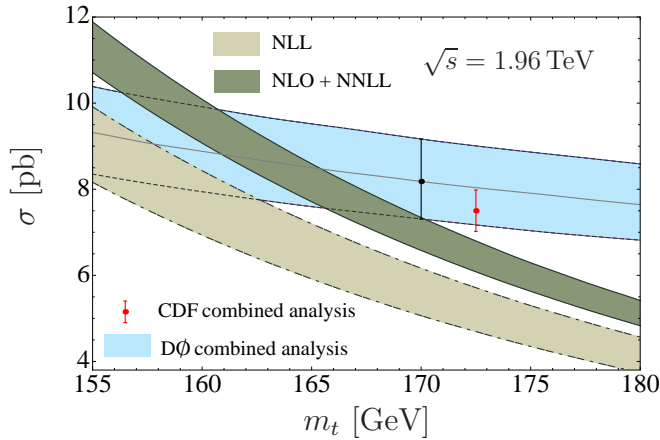}
\end{center}
\vspace{-2mm}
\caption{\label{fig:mtdepenTC} Dependence of the total cross section on the top-quark mass
  defined in the pole scheme. The NLL and NLO+NNLL bands indicate the scale uncertainties
  and PDF uncertainties combined linearly, while the blue band represents the dependence
  of the D0 measurement of the cross section on $m_t$, as described in
  \cite{Abazov:2009ae}. For comparison, a CDF combined measurement \cite{cdf9913} is also
  shown.}
\end{figure}
%%%%%%%%%%%%%%%%%%%%%%%%%%%%%%%%%%%%%%%%%%%%%%%%%%%%%%%%%%%%%%%%%%%%%%%%%%%%%%%%

Up to this point our results were obtained using the default value $m_t=173.1$\,GeV for
the top-quark mass defined in the pole scheme. Figure~\ref{fig:mtdepenTC} shows the
dependence of the total cross section on the value of $m_t$. We show our NLL and NLO+NNLL
results as bands representing the linearly combined errors from the scale uncertainties
and PDF uncertainties, which has been converted to 68\% CL (1$\sigma$) to match the
confidence level of the experimental errors. We compare our result with a recent combined
measurement of the total cross section performed by the D0 collaboration
\cite{Abazov:2009ae}. We also show a combined measurement by the CDF collaboration
\cite{cdf9913}; however, since they did not provide the mass dependence of the cross
section, we cannot use their results to extract the top-quark mass. From the overlap
region of the bands we determine the top-quark mass to be
$m_t=(163.0^{+7.2}_{-6.3})$\,GeV, which is lower than the latest world average
$m_t=(173.1^{+1.3}_{-1.3})$\,GeV \cite{:2009ec}. One should, however, interpret this
``discrepancy'' with care. It is unclear if the quoted world-average top-quark mass really
refers to the pole scheme, and even if it does, to which order of perturbation theory it
corresponds. Since the pole mass of a quark is not well-defined beyond perturbation
theory, the value of $m_t$ in the pole scheme changes significantly in different orders.
It would therefore be preferable to perform both the theoretical and experimental analyses
with a more physical mass parameter such as that defined in the $\overline{\text{MS}}$
scheme (see below).
  
\begin{figure}[t!]
\begin{center}
\begin{tabular}{lr}
\psfrag{y}[][][1][90]{$M_{\rm peak}$}
\psfrag{x}[][][1]{$m_t$ [GeV]}
\psfrag{z}[][][0.85]{$\sqrt{s}=1.96$\,TeV}
\includegraphics[width=0.41\textwidth]{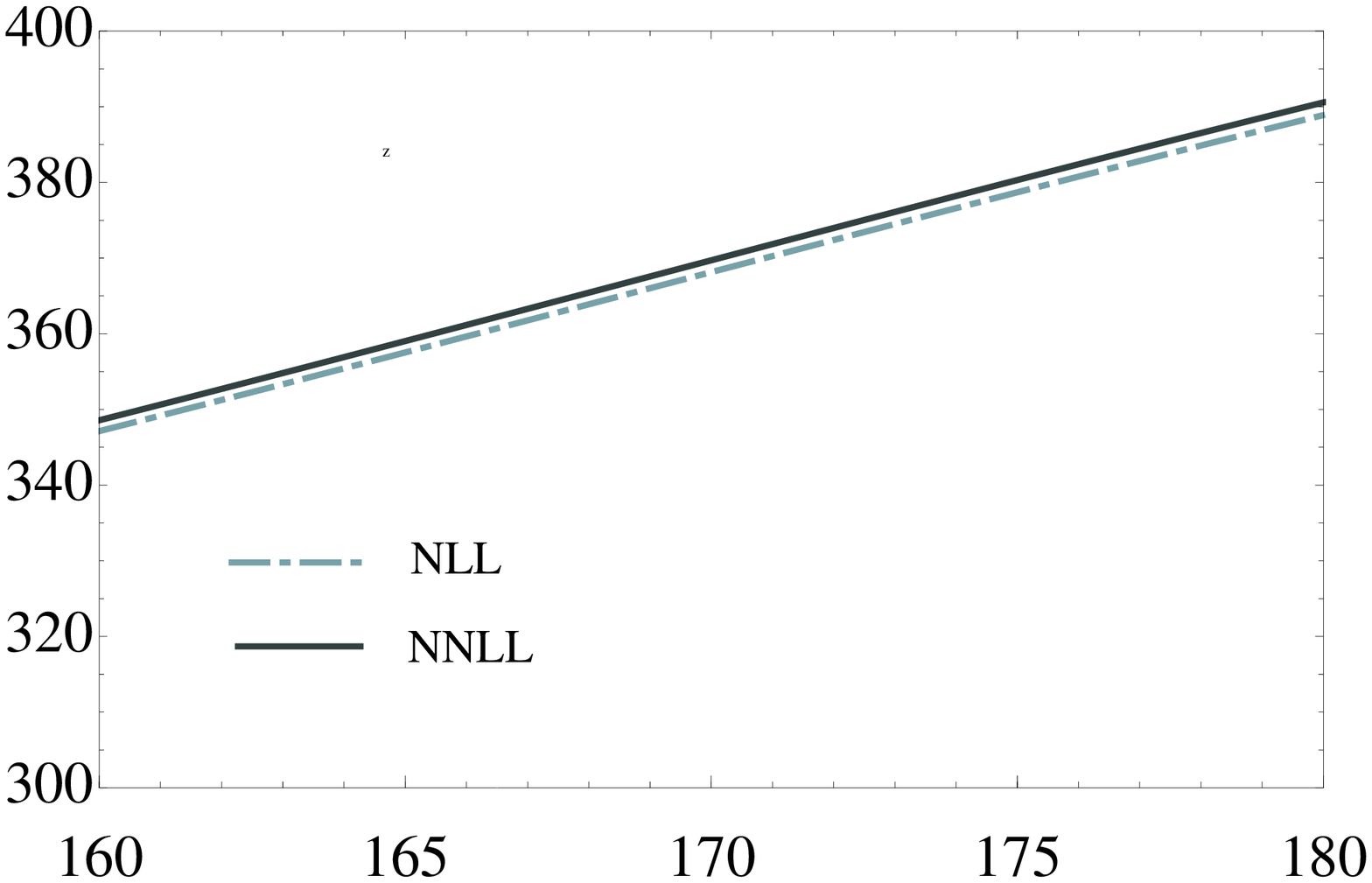} &
\psfrag{y}[][][1][90]{$\langle M\rangle$}
\psfrag{x}[][][1]{$m_t$ [GeV]}
\psfrag{z}[][][0.85]{$\sqrt{s}=1.96$\,TeV}
\includegraphics[width=0.41\textwidth]{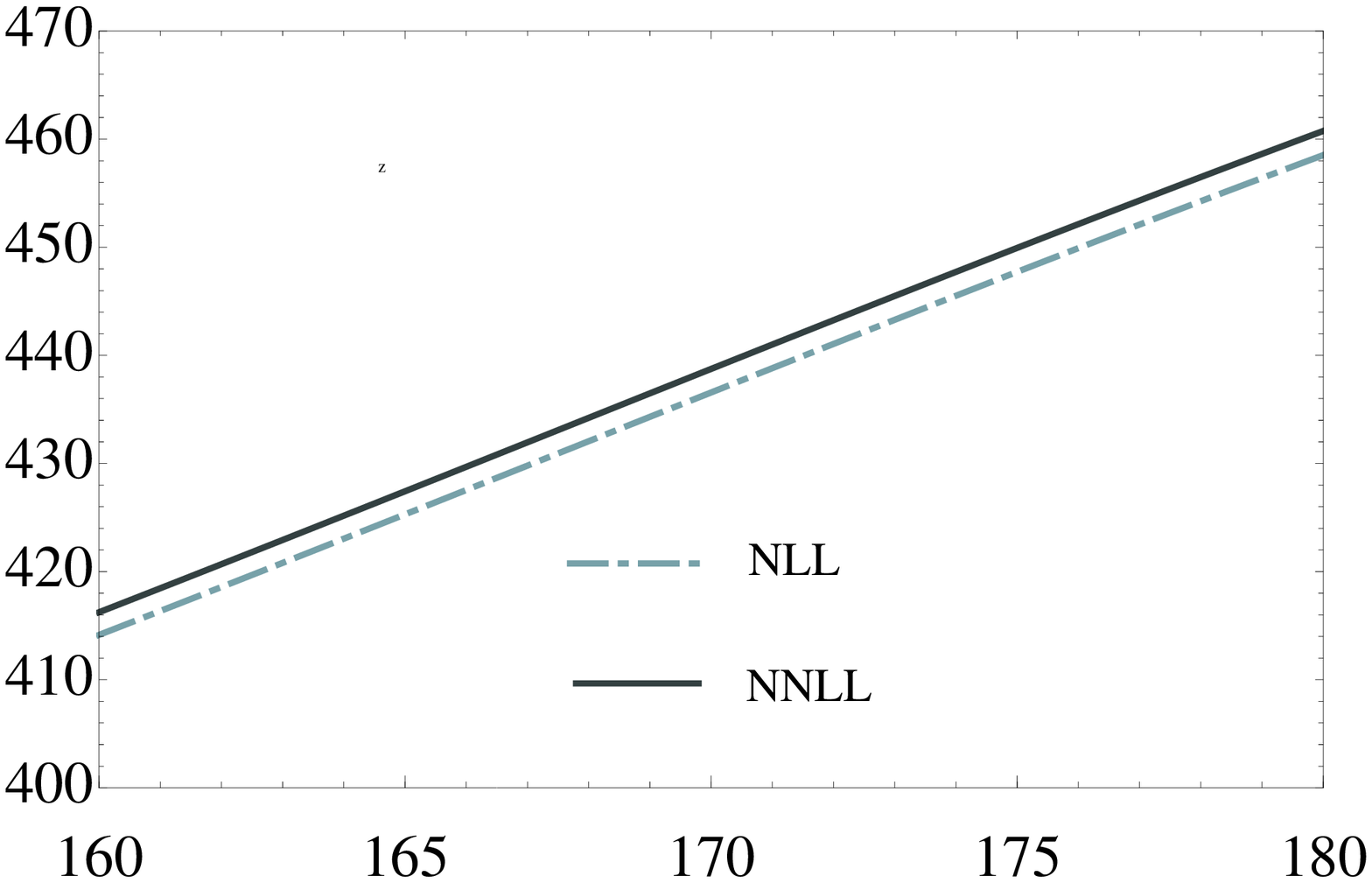}
\end{tabular}
\end{center}
\vspace{-2mm}
\caption{\label{fig:mt-peak} Position of the peak of the invariant mass distribution
  (left) and of the mean invariant mass $\langle M\rangle$ (right) as functions of the
  top-quark mass.}
\end{figure}

As we have seen, there are still rather large uncertainties in using the total cross
section as a mean to extract the top-quark mass. We therefore investigate what additional
information can be gained from the invariant mass distribution. On the left in
Figure~\ref{fig:mt-peak}, we show the position of the peak of the invariant mass
distribution, $M_{\rm peak}$, as a function of $m_t$. On the right we show the
corresponding dependence of the mean invariant mass $\langle M\rangle$. We do not show the
errors arising from scale variations, as they would be too small to be visible on the
scales of the plots. As illustrated in Table~\ref{tab:peak}, both observables turn out to
be very stable under the change of the order of perturbation theory. (Observing that the
shifts between different orders are larger than those indicated by the scale variations,
we however note that scale variation might not be a good estimator of the uncertainty.)
They show an almost linear dependence on the mass of the top quark, which leads to a
nearly linear translation of errors. A precise measurement of one of these observables
would thus lead to a direct measurement of $m_t$ with about the same precision. Of course
this can only be taken as a qualitative remark at the moment, but with increasing data
this might become an important strategy for a precision measurement of $m_t$.

%%%%%%%%%%%%%%%%%%%%%%%%%%%%%%%%%%%%%%%%%%%%%%%%%%%%%%%%%%%%%%%%%%%%%%%%%%%%%%%%
%%%%%%%%%%%%%%%%%%%%%%%%%%%%%%%%%%%%%%%%%%%%%%%%%%%%%%%%%%%%%%%%%%%%%%%%%%%%%%%%

%%%%%%%%%%%%%%%%%%%%%%%%%%%%%%%%%%%%%%%%%%%%%%%%%%%%%%%%%%%%%%%%%%%%%%%%%%%%%%%%
\begin{table}[t!]
\vspace{4mm}
\begin{center}
\begin{tabular}{|r|r|r|}
  \hline
  & $M_{\rm peak}$ [GeV]  & $\langle M\rangle$ [GeV] \\ \hline
  LO         & $375.6^{+0.5}_{-0.4}$ & $450.2^{+1.7}_{-1.5}$ \\ \hline
  NLL        & $374.5^{+0.2}_{-0.2} $  & $443.3^{+0.6}_{-0.8}$ \\ \hline
  NNLL       & $376.1^{+0.2}_{-0.2}$  & $445.5^{+0.4}_{-0.3}$ \\ \hline
  NLO+NNLL   & $375.2^{+0.3}_{-0.2}$  & $445.0^{+0.4}_{-0.4}$ \\ \hline
\end{tabular}
\end{center}
\vspace{-2mm}
\caption{\label{tab:peak} 
Values of the peak position of the mean invariant mass, obtained with $m_t= 173.1$GeV.}
\end{table}
%%%%%%%%%%%%%%%%%%%%%%%%%%%%%%%%%%%%%%%%%%%%%%%%%%%%%%%%%%%%%%%%%%%%%%%%%%%%%%%%

\begin{figure}
\begin{center}
\begin{tabular}{lr}
\psfrag{y}[][][1][90]{$d\sigma/dM$ [fb/GeV]}
\psfrag{x}[][][1]{$M$ [GeV]}
\psfrag{z}[][][0.85]{$\sqrt{s}=1.96$\,TeV}
\psfrag{t}[][][0.7]{}
\includegraphics[width=0.42\textwidth]{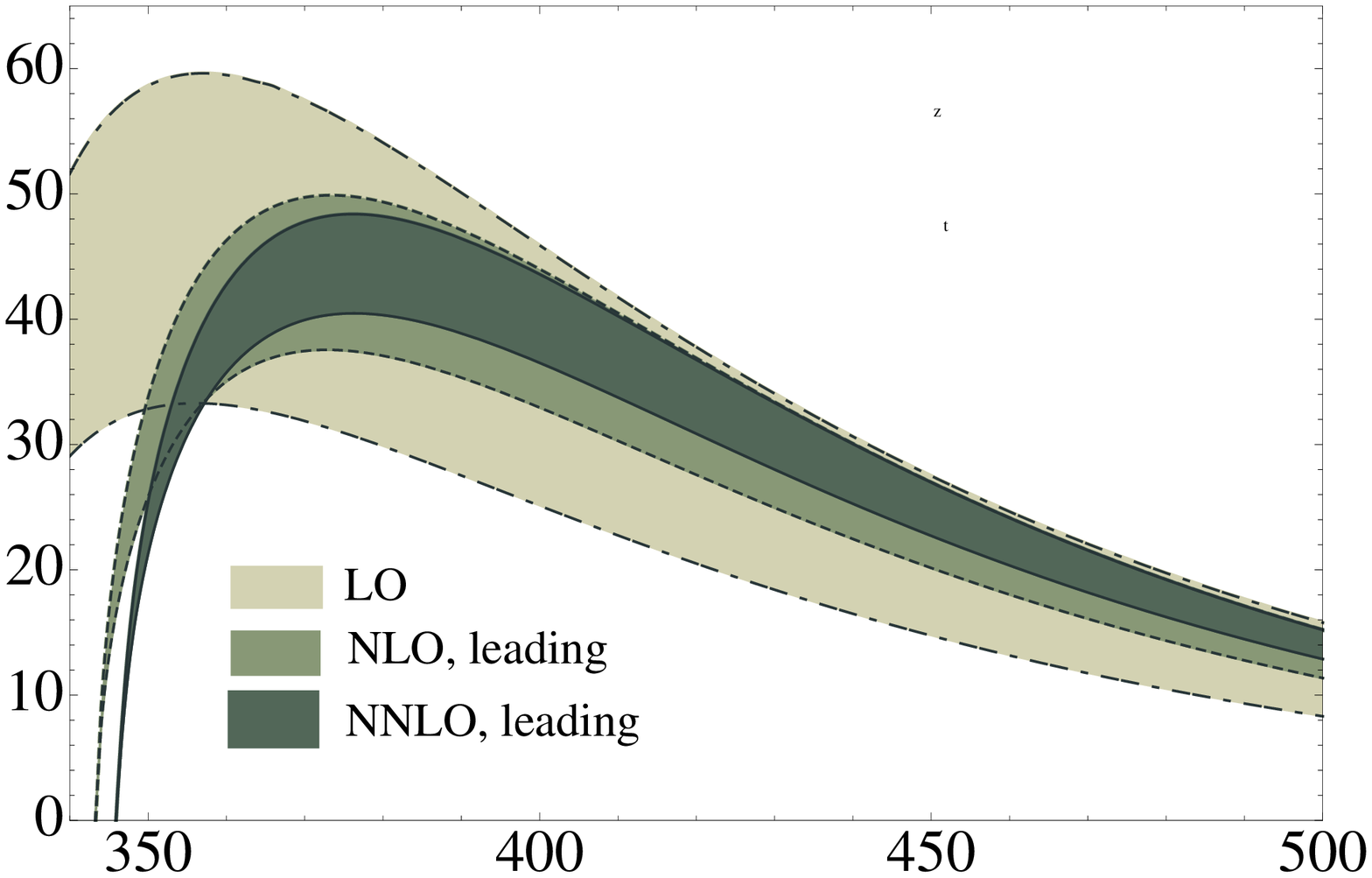}
&
\psfrag{y}[][][1][90]{$d\sigma/dM$ [fb/GeV]}
\psfrag{x}[][][1]{$M$ [GeV]}
\psfrag{z}[][][0.85]{$\sqrt{s}=1.96$\,TeV}
\psfrag{t}[][][0.7]{}
\includegraphics[width=0.42\textwidth]{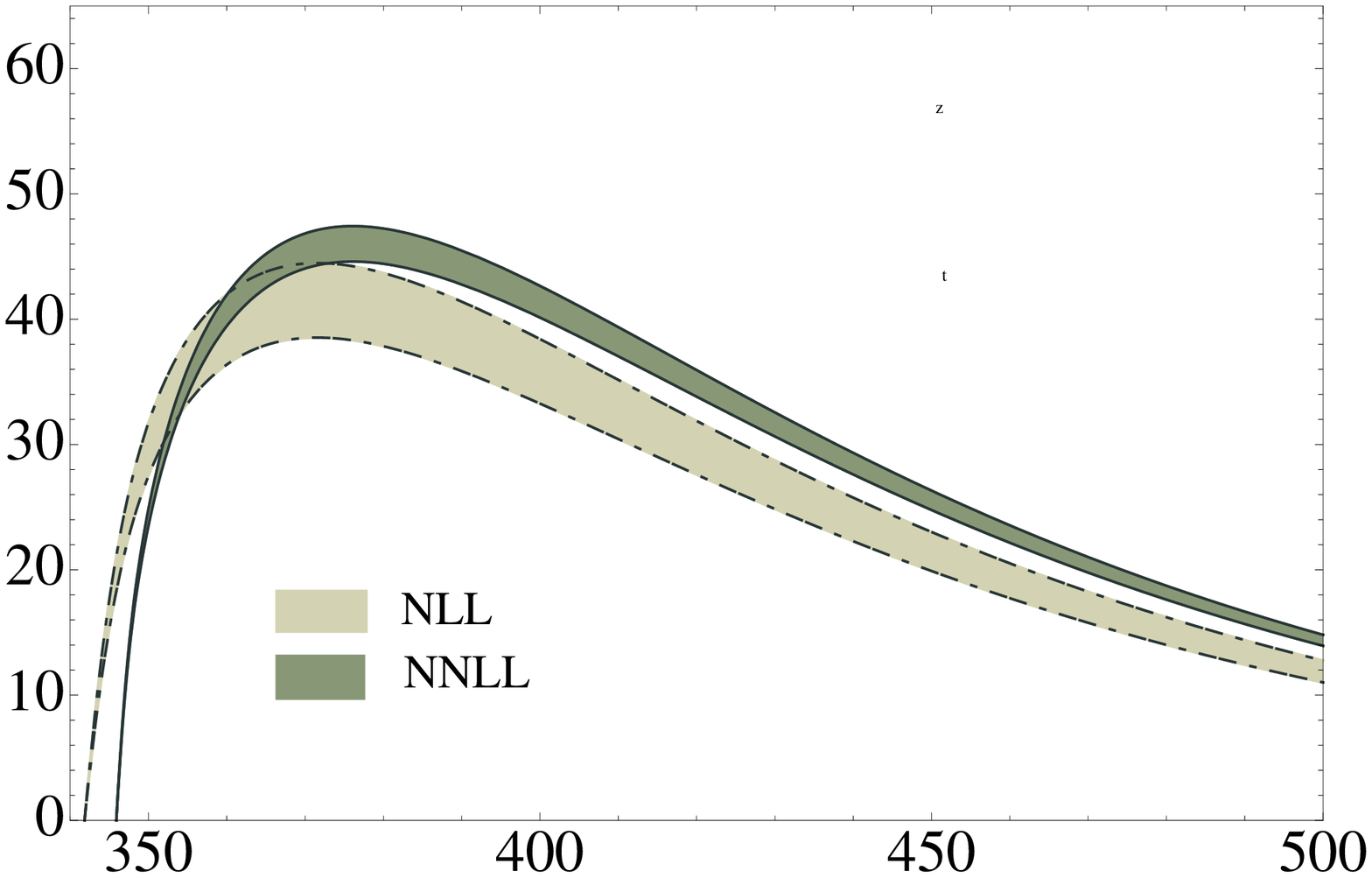}
\end{tabular}
\end{center}
\vspace{-2mm}
\caption{\label{fig:msbarplot} Fixed order (left) and resummed (right) results for
  $d\sigma/dM$, changing $m_t$ at different orders in perturbation theory as shown in
  Table~\ref{tab:mtop}.}
\end{figure}

It is a well-known fact that the pole mass defined in perturbation theory is an
ill-defined concept, as it is plagued by renormalon ambiguities in higher orders of
perturbation theory \cite{Bigi:1994em, Beneke:1994sw}. While the corresponding intrinsic
uncertainty in $m_t$ is of order $\Lambda_{\rm QCD}$ and thus smaller than the present
experimental error in this parameter, it is nevertheless useful to consider predictions
for the cross section (and distributions) parameterized in terms of a more suitable mass
parameter. For the purpose of illustration, we investigate in the following the impact of
using the \msbar\ mass $\overline{m}_t(\mu)$. We do this by changing the value of the pole
mass $m_t$ at different orders in perturbation theory, such that the ``physical'' mass
$\overline{m}_t(\mu)$ remains the same. This is implemented through the relation
\begin{equation}\label{eq:msbar1}
   m_t = \overline{m}_t(\mu)\,\bigg[ 1 + \frac{\alpha_s(\mu)}{\pi}\,d^{(1)} 
    + \left( \frac{\alpha_s(\mu)}{\pi} \right)^2 d^{(2)} + \dots \bigg] \,, 
\end{equation}
where for QCD ($N=3$) with $n_f=5$ active, massless flavors \cite{Gray:1990yh,
  Fleischer:1998dw, Chetyrkin:1999qi}
\begin{equation}
   d^{(1)} = \frac43 + L_m \,, \qquad
   d^{(2)} = \frac{2053}{288} + \frac{\pi^2}{18} + \frac{\pi^2}{9}\,\ln 2 
    - \frac{\zeta_3}{6} + \frac{379}{72}\,L_m + \frac{37}{24}\,L_m^2 \,,
\end{equation}
with $L_m=\ln(\mu^2/\overline{m}_t^2(\mu))$. We use a fixed input value
$\overline{m}_t\equiv\overline{m}_t(\overline{m}_t)=164.0$\,GeV, chosen such that at
two-loop order the corresponding pole-scheme parameter $m_t$ in (\ref{eq:msbar1})
coincides with our default value $m_t=173.1$\,GeV. In lower orders we adjust $m_t$ such
that the \msbar\ mass stays unchanged, which leads to the values collected in
Table~\ref{tab:mtop}. In Figure~\ref{fig:msbarplot}, we show the impact on the invariant
mass distribution of choosing $m_t$ in the way described above. Compared with
Figure~\ref{fig:SpectrumPDFS}, we observe an improved convergence of the perturbation
theory, both in fixed order and after threshold resummation. This finding, which has
previously been made in \cite{Langenfeld:2009wd}, could have been anticipated by looking
at Figure~\ref{fig:mtdepenTC} and Table~\ref{tab:mtop}. The increase of the cross section
and invariant mass distributions in higher orders is to a large extent compensated by the
effect of increasing the value of the pole mass.

%%%%%%%%%%%%%%%%%%%%%%%%%%%%%%%%%%%%%%%%%%%%%%%%%%%%%%%%%%%%%%%%%%%%%%%%%%%%%%%%
\begin{table}[t!]
\begin{center}
\begin{tabular}{|c|c|}
\hline
           & $m_t$ [GeV] \\ \hline
LO         & 164.0       \\ \hline
NLO, NLL   & 171.7       \\ \hline
NNLO, NNLL & 173.1       \\ \hline
\end{tabular}
\end{center}
\vspace{-2mm}
\caption{\label{tab:mtop} 
Values of the pole mass $m_t$ in different orders of perturbation theory, corresponding to a fixed value $\overline{m}_t(\overline{m}_t)=164.0$\,GeV.}
\vspace{4mm}
\end{table}
%%%%%%%%%%%%%%%%%%%%%%%%%%%%%%%%%%%%%%%%%%%%%%%%%%%%%%%%%%%%%%%%%%%%%%%%%%%%%%%%

%%%%%%%%%%%%%%%%%%%%%%%%%%%%%%%%%%%%%%%%%%%%%%%%%%%%%%%%%%%%%%%%%%%%%%%%%%%%%%%%
%%%%%%%%%%%%%%%%%%%%%%%%%%%%%%%%%%%%%%%%%%%%%%%%%%%%%%%%%%%%%%%%%%%%%%%%%%%%%%%%

\section{Conclusions}
\label{sec:Conclusions}

We have studied higher-order perturbative corrections to the differential cross section
for $t\bar t$ hadroproduction in the partonic threshold region, where the invariant mass
of the $t\bar t$ pair approaches the partonic center-of-mass energy. This involved using
SCET to derive a factorization formula expressing the hard-scattering kernels as products
of matrix-valued hard and soft matching coefficients in this kinematic regime, calculating
these coefficients to NLO in perturbation theory, and solving the RG equations for these
functions directly in momentum space to achieve NNLL accuracy for the resummed
differential cross section. We also presented an approximate NNLO formula, which includes
all terms proportional to singular plus distributions in the variable $(1-z)$, as well the
scale-dependent part of the coefficient of the $\delta(1-z)$ term. The momentum-space
resummation techniques allow for a straightforward matching of the resummed results with
the exact results at NLO in fixed-order perturbation theory. We used this fact to perform
numerical studies of the invariant mass distribution, the total cross section, and the
forward backward asymmetry at NLO+NNLL order.

Our results are expected to work progressively better when the invariant mass $M$ becomes
large, since in that case the partonic threshold region becomes more important due to
kinematics. However, our numerical results show that the full result at NLO in fixed-order
perturbation theory is dominated by the leading terms in the threshold expansion, even at
low values of invariant mass near the peak of the distribution. This provides evidence for
a dynamical enhancement of the partonic threshold region. Assuming this to be the case
also at higher orders in perturbation theory, we have argued that our results obtained at
NLO+NNLL order provide the best available predictions for the $t\bar{t}$ invariant mass
spectrum in the whole range of $M$. The benefits of performing the resummation are best
seen through the reduction of scale dependence and the improvement of the convergence
shown in Figure~\ref{fig:BandsTeV}--\ref{fig:HighMassLHC}. Compared to the fixed-order NLO
results, the RG resummation reduces the scale uncertainties in the differential cross
sections and therefore provide the more reliable predictions.

In view of the accuracy we can achieve for the differential cross section, we have also
made predictions for the total cross section and the forward-backward asymmetry by
integrating the differential distribution over the appropriate regions of phase-space. We
believe that these results, summarized in Tables~\ref{tab:CrossSections400},
\ref{tab:CrossSections} and \ref{tab:FB}, are currently the best predictions for these
observables. In particular, we have compared our results for the total cross section with
those obtained in the conventional approach of expanding around $\beta \to 0$, where the
top-quark pair is produced nearly at rest. By expanding our approximate NNLO formulas in
the limit $\beta \to 0$ and dropping subleading terms in $\beta$, we recover the results
recently obtained in \cite{Beneke:2009ye}, up to terms at NNLO related to potential-gluon
exchange. A numerical analysis showed that the contributions of these extra
potential-gluon corrections are actually smaller than the subleading terms in $\beta$
contained in our formulas, whose contribution is about 6--15\%. This is not surprising,
because the cross section receives the dominant contributions from regions with $\beta\sim
0.6$, which is not especially small. For this reason, it seems to us that the results
obtained here offer a legitimate alternative for including higher-order corrections to the
total cross section. As for the forward-backward asymmetry, our results are in
quantitative agreement with previous findings, and thus cannot explain the large value
measured by the CDF collaboration.

\section*{Acknowledgments}

We would like to thank Martin Beneke, Ulrich Haisch, Nikolaos Kidonakis, Kirill Melnikov,
Sven Moch, Markus Schulze, and Berthold Stech for useful discussions. A.~F. is grateful to
the members of the Center for Cosmology and Particle Physics of New York University for
their kind hospitality. This research was supported in part by the State of
Rhineland-Palatinate via the Research Centre {\em Elementary Forces and Mathematical
  Foundations}, by the Helmholtz Institute Mainz, and by the German Federal Ministry for
Education and Research under grant 05H09UME: {\em Precision Calculations for Collider and
  Flavour Physics at the LHC}.

%\newpage
\appendix

\section{RG-evolution factors and anomalous dimensions}
\label{sec:appendix}
\renewcommand{\theequation}{A.\arabic{equation}}
\setcounter{equation}{0}

Here we collect the expressions of the perturbative solutions to NNLL order for the RG
factors $S$, $a_\gamma$, and $\bm{u}$ in (\ref{eq:Sa}) and (\ref{eq:offdiagu}), which
appear in the solution of the evolution matrix $\bm{U}$ in the final resummed result
(\ref{eq:MasterFormula}). We also collect the expansion coefficients of the anomalous
dimensions and QCD $\beta$-function needed at this order.

We first define expansion coefficients of the anomalous dimensions and QCD
$\beta$-function as
\begin{align}
  \label{eq:gammaexp}
  \Gamma_{\text{cusp}}(\alpha_s) &= \Gamma_0 \, \frac{\alpha_s}{4\pi} + \Gamma_1 \left(
    \frac{\alpha_s}{4\pi} \right)^2 + \Gamma_2 \left( \frac{\alpha_s}{4\pi} \right)^3 +
  \ldots \, , \nonumber
  \\
  \beta(\alpha_s) &= -2\alpha_s \left[ \beta_0 \, \frac{\alpha_s}{4\pi} + \beta_1 \left(
      \frac{\alpha_s}{4\pi} \right)^2 + \beta_2 \left( \frac{\alpha_s}{4\pi} \right)^3 +
    \ldots \right] ,
\end{align}
and similarly for the other anomalous dimensions (recall that $\Gamma_{\rm
  cusp}=C_F\gamma_{\rm cusp}$ for the $q\bar q$ channel, and $\Gamma_{\rm
  cusp}=C_A\gamma_{\rm cusp}$ for the $gg$ channel). In terms of these quantities, the
function $a_\Gamma$ is given by \cite{Neubert:2004dd, Becher:2006mr}
\begin{align}
  \label{asol}
  a_\Gamma(\nu,\mu) = \frac{\Gamma_0}{2\beta_0} \, \Bigg\{
  \ln\frac{\alpha_s(\mu)}{\alpha_s(\nu)} + \left( \frac{\Gamma_1}{\Gamma_0} -
    \frac{\beta_1}{\beta_0} \right) \frac{\alpha_s(\mu)-\alpha_s(\nu)}{4\pi} \Bigg\} \, ,
\end{align}
and the result for the Sudakov factor $S$ reads
\begin{multline}
  \label{eq:Ssol}
  S(\nu,\mu) = \frac{\Gamma_0}{4\beta_0^2} \, \Bigg\{ \frac{4\pi}{\alpha_s(\nu)} \left( 1
    - \frac{1}{r} - \ln r \right) + \left( \frac{\Gamma_1}{\Gamma_0} -
    \frac{\beta_1}{\beta_0} \right) (1 - r + \ln r) + \frac{\beta_1}{2\beta_0} \ln^2 r
  \\
  + \frac{\alpha_s(\nu)}{4\pi} \Bigg[ \left( \frac{\beta_1\Gamma_1}{\beta_0\Gamma_0} -
    \frac{\beta_2}{\beta_0} \right) (1 - r + r\ln r) + \left( \frac{\beta_1^2}{\beta_0^2}
    - \frac{\beta_2}{\beta_0} \right) (1-r) \ln r
  \\
  - \left( \frac{\beta_1^2}{\beta_0^2} - \frac{\beta_2}{\beta_0} -
    \frac{\beta_1\Gamma_1}{\beta_0\Gamma_0} + \frac{\Gamma_2}{\Gamma_0} \right)
  \frac{(1-r)^2}{2} \Bigg] \Bigg\} \, ,
\end{multline}
where $r=\alpha_s(\mu)/\alpha_s(\nu)$.

The matrix equation for $\bm{u}$ can be evaluated using the techniques described in
\cite{Buras:1991jm, Buchalla:1995vs}. To this end, we first define the matrix $\bm{V}$
which diagonalizes $\bm{\gamma}^{h(0)}$ as
\begin{align}
  \bm{\gamma}_D^{h(0)} = \bm{V}^{-1} \bm{\gamma}^{h(0)} \bm{V} \, .
\end{align}
We also define the vector $\vec{\gamma}^{h(0)}$ consisting of the diagonal elements of
$\bm{\gamma}_D^{h(0)}$. Then the solution 
at NNLL reads
\begin{align}
  \label{eq:usol}
  \bm{u}(M,m_t,\cos\theta,\mu_h,\mu)
  = \bm{V} \left( 1 + \frac{\alpha_s(\mu)}{4\pi}\,\bm{K} \right)
  \Bigg( \! \left[ \frac{\alpha_s(\mu_h)}{\alpha_s(\mu)} 
  \right]^{\frac{\vec{\gamma}^{h(0)}}{2\beta_0}} \! \Bigg)_D \!
  \left( 1 - \frac{\alpha_s(\mu_h)}{4\pi}\,\bm{K} \right) \bm{V}^{-1} ,
\end{align}
where the matrix elements of $\bm{K}$ are
\begin{align}
  K_{IJ} = \delta_{IJ} \vec{\gamma}_I^{h(0)} \frac{\beta_1}{2\beta_0^2} -
  \frac{\left[\bm{V}^{-1}\bm{\gamma}^{h(1)}\bm{V}\right]_{IJ}}
  {2\beta_0+\vec{\gamma}_I^{h(0)}-\vec{\gamma}_J^{h(0)}} \, .
\end{align}

Finally, we collect the expansion coefficients of the anomalous dimensions and QCD
$\beta$-function needed in this work. The cusp anomalous dimension to three-loop order is
\cite{Moch:2004pa}
\begin{align}
  \gamma^{\text{cusp}}_0 &= 4 \, ,
  \nonumber
  \\
  \gamma^{\text{cusp}}_1 &= \left( \frac{268}{9} - \frac{4\pi^2}{3} \right) C_A -
  \frac{80}{9} T_F n_f \, ,
  \nonumber
  \\
  \gamma^{\text{cusp}}_2 &= C_A^2 \left( \frac{490}{3} - \frac{536\pi^2}{27} +
    \frac{44\pi^4}{45} + \frac{88}{3} \zeta_3 \right) + C_A T_F n_f \left( -
    \frac{1672}{27} + \frac{160\pi^2}{27} - \frac{224}{3} \zeta_3 \right) \nonumber
  \\
  &\quad + C_F T_F n_f \left( - \frac{220}{3} + 64 \zeta_3 \right) - \frac{64}{27} T_F^2
  n_f^2 \, .
\end{align}
For the anomalous dimensions entering $\bm{\gamma}^h$, we have \cite{Becher:2009qa,
  Becher:2009kw}
\begin{align}
  \gamma^q_0 &= - 3 C_F \, ,
  \nonumber
  \\
  \gamma^q_1 &= C_F^2 \left( -\frac{3}{2} + 2 \pi^2 - 24 \zeta_3 \right) + C_F C_A \left(
    - \frac{961}{54} - \frac{11\pi^2}{6} + 26 \zeta_3 \right) 
    + C_F T_F n_f \left( \frac{130}{27} + \frac{2\pi^2}{3} \right) ,
  \nonumber
  \\
  \gamma^g_0 &= - \frac{11}{3} C_A + \frac{4}{3} T_F n_f \, ,
  \nonumber
  \\
  \gamma^g_1 &= C_A^2 \left( - \frac{692}{27} + \frac{11\pi^2}{18} + 2 \zeta_3 \right) +
  C_A T_F n_f \left( \frac{256}{27} - \frac{2\pi^2}{9} \right) + 4 C_F T_F n_f \, ,
  \nonumber
  \\
  \gamma^Q_0 &= - 2 C_F \, ,
  \nonumber
  \\
  \gamma^Q_1 &= C_F C_A \left( -\frac{98}{9} + \frac{2\pi^2}{3} - 4 \zeta_3 \right) +
  \frac{40}{9} C_F T_F n_f \, ,
\end{align}
as well as \cite{Becher:2009kw, Ferroglia:2009ep, Ferroglia:2009ii, Korchemsky:1987wg,
  Korchemsky:1991zp, Kidonakis:2009ev}
\begin{align}
  \gamma^{\text{cusp}}_0(\beta) &= \gamma^{\text{cusp}}_0 \beta \coth \beta
  \, ,
  \nonumber
  \\
  \gamma^{\text{cusp}}_1(\beta) &= \gamma^{\text{cusp}}_1 \beta \coth \beta
  + 8 C_A \Bigg\{ \frac{\pi^2}{6} + \zeta_3 + \beta^2 \nonumber
  \\
  &\quad + \coth^2\beta \bigg[ \Li_3(e^{-2\beta}) + \beta
  \Li_2(e^{-2\beta}) - \zeta_3 + \frac{\pi^2}{6} \beta + \frac{\beta^3}{3}
  \bigg]
  \nonumber
  \\
  &\quad + \coth\beta \bigg[ \Li_2(e^{-2\beta}) - 2\beta
  \ln(1-e^{-2\beta}) - \frac{\pi^2}{6} (1+\beta) - \beta^2 -
  \frac{\beta^3}{3} \bigg] \Bigg\} \, , \nonumber
  \\ 
  g_0(\beta)&=0, \nonumber \\
  g_1(\beta) &= \coth\beta \bigg[ \beta^2 + 2\beta
  \ln(1-e^{-2\beta}) - \Li_2(e^{-2\beta}) + \frac{\pi^2}{6} \bigg] -
  \beta^2 - \frac{\pi^2}{6} \, .
\end{align}
The anomalous dimensions of the PDFs are given by
\begin{align}
  \gamma^{\phi_q}_0 &= 3C_F \, ,
  \nonumber\\
  \gamma^{\phi_q}_1 &= C_F^2 \left( \frac{3}{2} - 2\pi^2 + 24\zeta_3 \right) + C_F C_A
  \left( \frac{17}{6} + \frac{22\pi^2}{9} - 12\zeta_3 \right) - C_F T_F n_f \left(
    \frac{2}{3} + \frac{8\pi^2}{9} \right) ,
  \nonumber\\
  \gamma^{\phi_g}_0 &= \frac{11}{3} C_A - \frac{4}{3} T_F n_f \, ,
  \nonumber\\
  \gamma^{\phi_g}_0 &= C_A^2 \left( \frac{32}{3} + 12\zeta_3 \right) - \frac{16}{3} C_A
  T_F n_f - 4 C_F T_F n_f \, ,
\end{align}
and for the QCD $\beta$-function to three-loop order we have
\begin{align}
  \beta_0 &= \frac{11}{3} C_A - \frac{4}{3} T_F n_f \, ,
  \nonumber\\
  \beta_1 &= \frac{34}{3} C_A^2 - \frac{20}{3} C_A T_F n_f - 4 C_F T_F n_f \, ,
  \nonumber\\
  \beta_2 &= \frac{2857}{54} C_A^3 + \left( 2 C_F^2 - \frac{205}{9} C_F C_A -
    \frac{1415}{27} C_A^2 \right) T_F n_f + \left( \frac{44}{9} C_F + \frac{158}{27} C_A
  \right) T_F^2 n_f^2 \, .
\end{align}

\newpage

%%%%%%%%%% BEGIN BIBLIOGRAPHY %%%%%%%%%%

%%%%%%%%%% END BIBLIOGRAPHY %%%%%%%%%%

\end{document}